\documentclass[aps,prx,reprint,superscriptaddress,nofootinbib,floatfix,longbibliography]{revtex4-2}
\usepackage{amsmath,amssymb,amsthm,mathtools,bm}
\usepackage{times,txfonts}
\usepackage{graphicx,microtype,booktabs}
\graphicspath{{figures/}{../figures/}}
\usepackage[colorlinks=true,linkcolor=blue,citecolor=blue,urlcolor=blue]{hyperref}
\usepackage{orcidlink}
\newtheorem{theorem}{Theorem}
\newtheorem{lemma}[theorem]{Lemma}
\newtheorem{proposition}[theorem]{Proposition}
\newtheorem{corollary}[theorem]{Corollary}
\theoremstyle{remark}
\newcommand{\Tr}{\operatorname{Tr}}
\newcommand{\diag}{\operatorname{diag}}
\newcommand{\Id}{\mathbb I}
\newcommand{\Om}{\Omega}
\newcommand{\EF}{E_{\mathrm F}}
\newcommand{\EG}{E_{\mathrm F}^{\mathrm G}}
\newcommand{\EC}{E_{\mathrm C}}

\AtBeginDocument{\lefthyphenmin=2\righthyphenmin=3}
\hypersetup{pdftitle={Entanglement cost and classical capacity of multimode Gaussian states and channels},pdfauthor={Yuan Gao and Gerardo Adesso}}

\begin{document}
\title{Entanglement cost and classical capacity of multimode Gaussian states and channels}
\author{Yuan Gao}
\email{willgao@alumni.harvard.edu}
\affiliation{First Principle Science, San Francisco, CA, United States}

\author{Gerardo Adesso\orcidlink{0000-0001-7136-3755}}
\email{gerardo.adesso@nottingham.ac.uk}
\affiliation{Algorithmiq S.r.L., Via della Chiusa 15, 20123 Milano, Italy}
\affiliation{School of Mathematical Sciences and Centre for the Mathematics and Theoretical Physics of Quantum Non-Equilibrium Systems, University of Nottingham, University Park, Nottingham NG7 2RD, United Kingdom}
\date{\today}

\begin{abstract}
Gaussian states and operations are central to continuous-variable quantum information and communication. For systems with any finite number of modes, we prove that entanglement of formation, Gaussian entanglement of formation, and entanglement cost coincide for every finite-energy bipartite Gaussian state. Their common value is strongly superadditive and lower-bounds entanglement of formation and entanglement cost for arbitrary non-Gaussian states with the same covariance matrix. We then determine the entanglement cost of every multimode Gaussian channel as well as its classical capacity under a mean energy constraint. These quantities reduce to Gaussian covariance optimizations, and the capacity is attained by pure Gaussian signals with Gaussian displacement modulation, independently across channel uses. The proof combines geodesic convexity of entanglement for pure Gaussian states with a relative-entropy inequality for arbitrary states. We also obtain a closed formula for the entanglement cost of a family of four-mode bound entangled states.
\end{abstract}
\maketitle

\section{Introduction}

Gaussian states and operations are central to continuous-variable
quantum information, in which information is encoded in the quadratures
of light and other bosonic systems~\cite{BraunsteinVanLoock2005}.
In quantum optics, this framework encompasses coherent and squeezed
light, optical interferometers, and Gaussian channels modeling loss,
amplification, and thermal noise~\cite{Weedbrook2012}.
Gaussian states are completely specified by their first moments and
covariance matrices, while Gaussian channels are specified by their
action on these moments, giving a compact description even when many
modes are correlated~\cite{AdessoRagyLee2014,Serafini2017}. 

This framework brings
together a wealth of fundamental questions: on one hand, how much entanglement is required
to prepare a shared state or simulate a channel, and on the other hand, what rates of
classical communication can a channel support under an energy
constraint? Answering these questions would require in principle a formidable optimization over
arbitrary quantum protocols, including non-Gaussian states and
correlations across many copies or channel uses.

For a shared Gaussian state, a simple preparation starts with a pure Gaussian state and adds correlated random displacements. The least entanglement required by such an ensemble is the Gaussian entanglement of formation $\EG$: a minimum over pure covariance matrices lying below the target covariance~\cite{wolf}. The entanglement of formation $\EF$ instead minimizes the average entanglement over all pure-state decompositions~\cite{Bennett1996}, including possibly non-Gaussian ensembles, while the entanglement cost $\EC$ is the asymptotic entanglement consumed per copy under local operations and classical communication~\cite{Hayden2001,yamasaki}. The central questions are whether Gaussian ensembles are optimal for the entanglement of formation (\emph{Gaussian optimality}) and whether preparing many copies jointly can or cannot reduce the cost per copy (\emph{additivity}).

Gaussian optimality was established first for symmetric two-mode
states~\cite{giedke}. The general two-mode problem subsequently prompted
analytic proposals, bounds, and numerical
methods~\cite{AdessoIlluminati2005,marian,IvanSimon2008,AdessoDatta2010,akbari,tserkis2017,tserkis2019}.
Wilde established Gaussian optimality
and additivity for states obtained by sending one
mode of a two-mode squeezed vacuum through a quantum-limited attenuator
or amplifier~\cite{wilde2018}. More recently, Adesso established both
Gaussian optimality for every two-mode Gaussian state and additivity
on tensor products of such states, using a many-pair entropy inequality valid
for arbitrary correlated states~\cite{adesso2026}. Those results also
extend to bisymmetric multimode states, whose entanglement can be
concentrated onto a single pair of modes by local Gaussian
unitaries~\cite{SerafiniAdessoIlluminati2005}.

In this work, we solve the problem in full generality for arbitrary bipartitions comprising any finite number of modes. For every finite-energy Gaussian state on such a multimode bipartition, we prove
\begin{equation}\label{main:costeq}
 \EC(\rho_V)=\EF(\rho_V)=\EG(\rho_V)=G(V).
\end{equation}
Here $\rho_V$ denotes the centered bipartite Gaussian state with covariance $V$, and $G(V)$ is the attained minimum over pure Gaussian preparations described above. We also prove strong superadditivity: the entanglement of formation and entanglement cost across the full bipartition are at least the sum of their values for reduced states on disjoint groups of Alice's and Bob's modes. Both quantities are additive on tensor products.

The result also provides entanglement bounds beyond Gaussian states. Every finite-energy state with covariance $V$ has entanglement of formation and entanglement cost at least $G(V)$. Measured quadrature moments therefore give lower bounds on both quantities without any assumption of Gaussianity. We establish this fixed-covariance bound through a supporting entropy inequality, complementing the general Gaussian-extremality framework for strongly superadditive measures~\cite{extremality}. 

The theorem captured by Eq.~\eqref{main:costeq} for states has two closely connected applications to channels.
First, it determines the entanglement cost of every multimode
Gaussian channel in the sequential setting of Wilde~\cite{wilde2018},
where later inputs may depend on earlier outputs. Second, it determines
the classical capacity of every multimode Gaussian channel under a
mean energy constraint. This extends the recent single-mode result by Adesso, Lami, and Giovannetti~\cite{ALG2026}. For the capacity
problem, the connection to state entanglement is provided by the
identity between the minimum average output entropy at a fixed average
input state and the entanglement of formation of the corresponding
output--environment state in a dilation of the channel~\cite{MSW2004}.

A common bound on pure-state entanglement underlies all three results. Geodesic convexity supplies the bound within the Gaussian family, expressed through the expectation of a quadratic Hamiltonian. A relative-entropy inequality establishes it for arbitrary pure states. After averaging over any preparation ensemble, the bound depends only on the covariance matrix, connecting the geometry to optimal state preparation, channel simulation, and classical communication.

The paper is organized as follows. Section~\ref{sec:states} proves
Gaussian optimality, strong superadditivity, and covariance bounds on
the entanglement of arbitrary states. Sections~\ref{sec:channel-cost}
and~\ref{sec:capacity} determine the entanglement cost and classical
capacity of Gaussian channels. Section~\ref{sec:examples} develops
analytic state and channel examples and introduces the numerical
toolkit, while Sec.~\ref{sec:discussion} summarizes our results and their broader implications. The Appendices contain auxiliary
lemmas and technical parts of the proofs.

\section{Entanglement of formation and cost of multimode Gaussian states}\label{sec:states}

\subsection{Preliminaries}
We consider bosonic continuous-variable systems containing finitely many
modes. For a system of $k$ modes, we collect the canonical quadrature
operators in the vector $R=(x_1,p_1,\ldots,x_k,p_k)^{\mathsf T}$,
where $\mathsf T$ denotes matrix transposition and $\dagger$ denotes
the adjoint. Setting $\hbar=1$,
their commutation relations read
\begin{equation}\label{eq:omega}
 [R_j,R_l]=i(\Omega_k)_{jl},\qquad
 \Omega_k=\bigoplus_{j=1}^{k}
 \begin{pmatrix}0&1\\-1&0\end{pmatrix},
\end{equation}
where $\Omega_k$ is the standard symplectic form on the $2k$-dimensional
phase space; we omit its subscript when the number of modes is clear.

For a state $\rho$ with finite second moments, we write
$\langle O\rangle=\Tr(\rho O)$ and define its first-moment vector $d$
and covariance matrix $V$ by
\begin{equation}\label{eq:covariance-convention}
 d=\langle R\rangle,\qquad
 V_{jl}=\langle\{R_j-d_j,R_l-d_l\}\rangle,
\end{equation}
where $\{A,B\}=AB+BA$ denotes the anticommutator. This normalization
assigns to the vacuum state the identity covariance $\Id_{2k}$. Note that, relative to the convention in
Ref.~\cite{adesso2026}, here all quantum covariance matrices are doubled;
the quadrature operators and first moments are unchanged.

A real symmetric matrix $V$ is a physical covariance matrix if and only
if it satisfies the \emph{bona fide} condition~\cite{SimonMukundaDutta1994},
which expresses the uncertainty principle:
\begin{equation}\label{eq:bona-fide}
 V+i\Omega_k\geq0.
\end{equation}
A physical covariance matrix $\Gamma$ describes a pure Gaussian state
precisely when
\begin{equation}\label{eq:purity}
 \Gamma\Omega_k\Gamma=\Omega_k.
\end{equation}
Equivalently, $\det\Gamma=1$.

Every real symmetric positive-definite $2k\times2k$ matrix $V$
admits a {Williamson decomposition}~\cite{SimonMukundaDutta1994,AdessoRagyLee2014},
\begin{equation}\label{eq:williamson}
 V=\Xi D\Xi^{\mathsf T},\qquad D=\bigoplus_{j=1}^{k}\nu_j\Id_2,
\end{equation}
where $\Xi$ is real and symplectic, meaning
$\Xi\Omega_k\Xi^{\mathsf T}=\Omega_k$, and the numbers $\nu_j>0$ are the
symplectic eigenvalues of $V$, equivalently the positive
eigenvalues of $i\Omega_kV$. For a physical covariance matrix,
the uncertainty relation \eqref{eq:bona-fide} requires $\nu_j\geq1$. For a Gaussian
state, the normal modes with quadratures $\Xi^{-1}R$ are independent
thermal modes up to displacement, with mean occupations
$(\nu_j-1)/2$.

With natural logarithms, the von Neumann entropy is $S(\rho)=-\Tr\rho\log\rho$ and the Gaussian entropy is $S_G(V)=\sum_j g((\nu_j-1)/2)$, where
\begin{equation}\label{eq:g}
 g(u)=(u+1)\log(u+1)-u\log u,\qquad g(0)=0.
\end{equation}
Let $N=\sum_j a_j^\dagger a_j$, with $a_j=(x_j+ip_j)/\sqrt2$, be the total photon-number operator. Its expectation is $\Tr\rho N=\Tr V/4+\|d\|^2/2-k/2$; finite energy means that this expectation value is finite. For the finite number of modes considered here, this also guarantees
finite local entropies, which are continuous in trace norm on sets
with a common mean energy bound~\cite{EisertSimonPlenio2002,winter}.
Throughout the paper, entanglement measures and rates are expressed in nats; division by $\log2$ gives bits or ebits. 
The displacement operator $D(\xi)=e^{-i\xi^{\mathsf T}\Omega R}$ obeys $D(\xi)^\dagger R D(\xi)=R+\xi$. A classical random shift $\xi$ of ordinary covariance $K_\xi$ therefore adds $2K_\xi$ to $V$.

The Gaussian entanglement of formation quantifies the least average
entanglement required to prepare a bipartite Gaussian state from an
ensemble of pure Gaussian states~\cite{wolf}. For a fixed bipartition
$A:B$, it admits the covariance characterization
$\EG(\rho_V)=G(V)$, where
\begin{equation}\label{main:roof}
 G(V)=\min_{\substack{\Gamma\leq V\\\Gamma\text{ pure Gaussian}}}E(\Gamma),
 \qquad E(\Gamma)=S_G(\Gamma_A).
\end{equation}
Here the entanglement of a pure Gaussian state is the entropy of its
reduced state on $A$, with subscripts on states denoting reductions by
partial trace and those on covariance matrices denoting the
corresponding principal blocks.

The pure Gaussian covariance matrices satisfying $\Gamma\leq V$ form
a compact set, so the minimum is attained by a covariance $\Gamma_*$.
The corresponding pure Gaussian state provides an explicit
preparation of $\rho_V$: applying displacements drawn from a centered
Gaussian distribution with ordinary covariance $(V-\Gamma_*)/2$
produces the target state. Since these displacements are local
unitaries, every member of the ensemble has the same entanglement
$E(\Gamma_*)$, while a further fixed local displacement sets any
desired first moments without changing $G$.

For a general bipartite state, the entanglement of formation is the
least average pure-state entanglement over all ensemble
decompositions~\cite{Bennett1996,yamasaki},
\begin{equation}\label{main:ef}
 \EF(\rho)=\inf_{\mu:\int |\psi\rangle\langle\psi|\,d\mu=\rho}
       \int S(\psi_A)\,d\mu(\psi).
\end{equation}
Here $\psi_A=\Tr_B|\psi\rangle\langle\psi|$, and the infimum ranges
over probability measures on pure states whose average is $\rho$,
allowing both discrete and continuous
ensembles~\cite{HolevoShirokovWerner2005}. When the bipartition
needs to be explicit, we write $\EF^{A:B}(\rho)$ and $G^{A:B}(V)$.

The entanglement cost $\EC$ is the least asymptotic entanglement required
per copy to prepare a state by local operations and classical
communication (LOCC), allowing collective protocols with vanishing
trace-distance error. For the finite-energy states considered here,
it equals the regularized entanglement of
formation~\cite{Hayden2001,yamasaki},
\begin{equation}\label{main:coding}
 \EC(\rho)=\lim_{n\to\infty}\frac{\EF(\rho^{\otimes n})}{n}.
\end{equation}

\subsection{Main result}
\begin{theorem}[Optimality of multimode Gaussian entanglement]
\label{main:formation}
(a) For every finite-energy Gaussian state $\rho_V$ on an arbitrary
bipartition $A:B$,
\[
 \EC(\rho_V)=\EF(\rho_V)=\EG(\rho_V)=G(V).
\]

\noindent (b) For any disjoint mode groups $A_j\subseteq A$ and $B_j\subseteq B$,
the common value satisfies strong superadditivity,
\begin{equation}\label{main:super}
 G^{A:B}(V)\geq
 \sum_j G^{A_j:B_j}(V_{A_jB_j}),
\end{equation}
and it is additive on finite tensor products of bipartite Gaussian
states.

\noindent (c) Gaussian states also minimize both entanglement of formation and
entanglement cost at fixed covariance: every finite-energy state
$\rho$ with covariance $V$ satisfies
\begin{equation}\label{main:extremalityeq}
 \EF(\rho)\geq G(V),\qquad \EC(\rho)\geq G(V).
\end{equation}
\end{theorem}

\medskip

\begin{proof}[Proof of Theorem~\ref{main:formation}]
The following four subsections develop the proof, beginning with
strong superadditivity and additivity for the covariance minimum
$G$ in part~(b). Its convexity then provides a quadratic lower bound
on pure Gaussian entanglement, which we extend to arbitrary pure
states through a relative-entropy inequality. Averaging over
ensembles gives Gaussian optimality and the fixed-covariance bound
for entanglement of formation, while regularization gives the
corresponding statements for entanglement cost in parts~(a)
and~(c).

\subsection{Geometry of pure Gaussian preparations}\label{sec:gaussian-geometry}
To compare pure Gaussian preparations, we use paths that preserve both purity and the covariance constraint. For positive-definite matrices $X,Y$ and $0\leq t\leq1$, the weighted geometric mean is
\begin{equation}\label{main:mean}
 X\#_tY=X^{1/2}(X^{-1/2}YX^{-1/2})^tX^{1/2}.
\end{equation}
We abbreviate the midpoint as $X\#Y:=X\#_{1/2}Y$. The mean is covariant under congruences, jointly monotone in its arguments, and bounded above by the arithmetic mean~\cite{lami2017,lami2018}. Pure covariances are positive symplectic matrices: their positive powers and symplectic congruences preserve purity. Consequently $\Gamma_0\#_t\Gamma_1$ is pure whenever its endpoints are pure. Monotonicity also preserves the constraint $\Gamma\leq V$.

The Williamson decomposition~\eqref{eq:williamson} supplies
a pure Gaussian covariance $\Xi\Xi^{\mathsf T}\leq V$, since $\Xi$ is symplectic
and $D\geq\Id_{2k}$. If $L=\lambda_{\max}(V)$, every feasible pure covariance obeys
\begin{equation}\label{g:compact}
 L^{-1}\Id\leq\Gamma\leq L\Id,
 \qquad \Gamma^{-1}=\Omega\Gamma\Omega^{\mathsf T}.
\end{equation}
The purity equation and upper constraint are closed on this compact positive interval, proving attainment of $G$. The same argument gives the one-sided continuity
\begin{equation}\label{g:continuity}
 G(V+D)\longrightarrow G(V)
 \quad\text{as }D\geq0,\ \|D\|\longrightarrow0.
\end{equation}
Indeed, the minimizing pure covariances have convergent subsequences with feasible limits below $V$, while $G(V+D)\leq G(V)$ by monotonicity.

To prove convexity along these paths, we express $E$ through the relative spectrum of $\Gamma$ and a locally transformed copy. Let
\[
 \Pi=\Id_A\oplus(-\Id_B),\qquad
 R_\Gamma=\Gamma^{-1/2}\Pi\Gamma\Pi\Gamma^{-1/2}.
\]
The matrix $\Pi$ implements a local phase rotation by $\pi$ on Bob's modes, changing the signs of both quadratures. Local symplectic transformations commute with $\Pi$ and preserve the relative spectrum of $(\Gamma,\Pi\Gamma\Pi)$. These transformations bring any pure Gaussian state to Gaussian Schmidt form: a product of two-mode squeezed pairs and local vacua~\cite{botero}. A pair with squeezing $r$ contributes $e^{4r},e^{4r},e^{-4r},e^{-4r}$ to the relative spectrum, whereas each local pure mode contributes two unit eigenvalues. Therefore
\begin{align}\label{g:entropy}
 E(\Gamma)=\frac14\sum_i F\!\left(\log\lambda_i(R_\Gamma)\right),\quad 
 F(x)=g\!\left(\sinh^2(x/4)\right).
\end{align}
The scalar function $F$ is even, convex, and continuously differentiable, including at zero. Appendix~\ref{app:geometry} proves the required spectral comparison: the ordered log-eigenvalues of a relative matrix under geometric interpolation are majorized by the interpolated endpoint vectors. Convexity of $F$ then gives
\begin{equation}\label{main:puregeoconv}
 E(\Gamma_0\#_t\Gamma_1)
 \leq(1-t)E(\Gamma_0)+tE(\Gamma_1).
\end{equation}
Thus $E$ is convex along these geometric paths of pure Gaussian
covariances, a property we refer to as \emph{geodesic convexity}.

Strong superadditivity follows by using a local symmetry to remove correlations between groups without increasing entanglement. Group the coordinates as $(A_1B_1),(A_2B_2)$, write $V_i=V_{A_iB_i}$, and let $Q$ change the signs of all quadratures in the first group. This transformation is local across the total Alice--Bob cut. For any pure $\Gamma\leq V$, the midpoint
\[
 \overline\Gamma=\Gamma\#_{1/2}(Q\Gamma Q)
\]
is pure, has no greater entanglement, and is invariant under $Q$. It therefore splits into pure blocks $\overline\Gamma=\gamma_1\oplus\gamma_2$. The arithmetic--geometric inequality gives
\begin{align*}
 \overline\Gamma
 \leq\tfrac12(\Gamma+Q\Gamma Q)
 \leq\tfrac12(V+QVQ)=V_1\oplus V_2.
\end{align*}
Each block is feasible for its marginal, so
\[
 E(\Gamma)\geq E(\gamma_1)+E(\gamma_2)
 \geq G(V_1)+G(V_2).
\]
Minimizing proves Eq.~\eqref{main:super} for two groups, and
iteration handles any finite partition of the modes. If the
chosen groups do not exhaust the system, we include the remaining
modes as one further group and discard its nonnegative
contribution. For product covariances, product minimizers give
the matching upper bound:
\begin{equation}\label{main:add}
 G(V_1\oplus V_2)=G(V_1)+G(V_2).
\end{equation}

The same interpolation makes $G$ convex under ordinary mixing of covariance matrices. If $\Gamma_j$ minimizes $G(V_j)$, then
\[
 \Gamma_0\#_t\Gamma_1
 \leq V_0\#_tV_1\leq(1-t)V_0+tV_1.
\]
Using Eq.~\eqref{main:puregeoconv}, we obtain
\begin{equation}\label{main:covconv}
 G((1-t)V_0+tV_1)\leq(1-t)G(V_0)+tG(V_1).
\end{equation}

\subsection{A supporting quadratic}\label{sec:quadratic-support}
At a strictly physical covariance $V+i\Omega>0$, convexity of $G$ gives a supporting hyperplane: a global affine lower bound that is tight at $V$. Thus there is a real symmetric matrix $B_V$ such that
\[
 G(U)\geq G(V)+\Tr[B_V(U-V)].
\]
Adding a positive covariance can only enlarge the feasible set, so $B_V\leq0$. Set $W=-4B_V\geq0$. The support extends to boundary covariances by applying it to $(1-\epsilon)U+\epsilon V$ and using Eq.~\eqref{main:covconv}. For an optimal pure $\Gamma_*$, we have $G(\Gamma_*)=E(\Gamma_*)=G(V)$, whence
\[
 0\geq\tfrac14\Tr[W(V-\Gamma_*)]\geq0.
\]
It follows that
\begin{align}
 \Tr[W(V-\Gamma_*)]&=0,\label{main:complement}\\
 E(\Gamma)+\tfrac14\Tr(W\Gamma)&\geq
 E(\Gamma_*)+\tfrac14\Tr(W\Gamma_*)\label{main:gausssupport}
\end{align}
for every pure Gaussian $\Gamma$.

The physical quadratic $H_W=R^{\mathsf T}WR/2$ has centered expectation $\Tr(WV)/4$. A covariance subscript on an expectation denotes the centered Gaussian state with that covariance. Equation~\eqref{main:gausssupport} therefore says that $\Gamma_*$ minimizes the sum of entanglement and quadratic energy among pure Gaussian states. Both terms are geodesically convex: the entropy by Eq.~\eqref{main:puregeoconv}, and the positive quadratic by arithmetic--geometric order. Equation~\eqref{g:entropy} also shows that the functional is continuously differentiable on the pure manifold. Consequently, every stationary point is a global pure Gaussian minimum.

Gaussian optimality follows if this supporting inequality holds for every finite-energy pure state:
\begin{equation}\label{main:allstatesupport}
 S(\psi_A)+\langle H_W\rangle_\psi
 \geq E(\Gamma_*)+\langle H_W\rangle_{\Gamma_*}.
\end{equation}
Once established, this inequality can be averaged over any preparation ensemble. Complementarity~\eqref{main:complement} then cancels the quadratic term at the target covariance.

\subsection{From a Gaussian support to an unrestricted entropy bound}\label{sec:unrestricted-support}
The required entropy bound follows from a comparison of input and
output quantum relative entropies, defined by
$D(\rho\Vert\sigma)=\Tr\rho(\log\rho-\log\sigma)$.
For a centered Gaussian state $\sigma$ that is faithful, meaning
it has no zero eigenvalues, we write
$\sigma=e^{-H_\sigma}/\Tr e^{-H_\sigma}$ and choose the quadratic
Hamiltonian to have ground energy zero. Using the adjoint $\Phi^*$ of
a channel $\Phi$ defined by
$\Tr[\Phi(\rho)Z]=\Tr[\rho\Phi^*(Z)]$,
the following proposition provides the comparison, with its proof given
in Appendix~\ref{app:contraction}.

\begin{proposition}[Joint relative-entropy inequality]\label{main:contraction}
Let $\Phi$ be a Gaussian unitary followed by a partial trace. Let $\sigma$ and $\tau=\Phi(\sigma)$ be faithful centered Gaussian states. If $\lambda>0$ and the real coefficient matrix of
\begin{equation}\label{eq:strictB}
 B=\lambda H_\sigma-\Phi^*H_\tau
\end{equation}
in symmetric ordering is positive definite, then every finite-energy state obeys
\begin{equation}\label{main:contractioneq}
 D(\Phi(\rho)\Vert\tau)\leq\lambda D(\rho\Vert\sigma).
\end{equation}
\end{proposition}

The comparison becomes an entropy minimization principle through the Gibbs identities. For
\[
 F_\lambda(\rho)
 =S(\Phi(\rho))-\lambda S(\rho)+\Tr(B\rho),
\]
we have
\begin{equation}\label{main:variationalgap}
 F_\lambda(\rho)-F_\lambda(\sigma)
 =\lambda D(\rho\Vert\sigma)-D(\Phi(\rho)\Vert\tau).
\end{equation}
Proposition~\ref{main:contraction} makes the right-hand side nonnegative,
so $\sigma$ minimizes $F_\lambda$ over all finite-energy states.

\begin{figure*}[t]
\centering
\includegraphics[width=\textwidth]{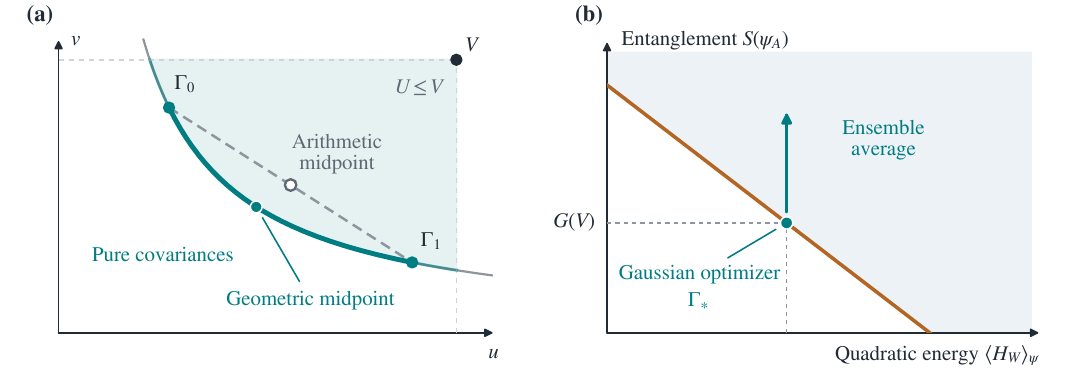}
\caption{Geometric structure of Gaussian optimality.
(a) \emph{Purity-preserving interpolation:} in the two-mode covariance slice
$U(u,v)=\Xi\diag(u,v,v,u)\Xi^{\mathsf T}$, with fixed symplectic $\Xi$, the shaded
region consists of physical covariances $U\leq V$. Its curved boundary
$uv=1$ consists of feasible pure covariances $\Gamma\leq V$, with the
geometric path from $\Gamma_0$ to $\Gamma_1$ highlighted.
The arithmetic and geometric midpoints remain below the target
covariance $V$, while the geometric midpoint also remains pure;
entanglement is convex along the geometric path, as in
Eq.~\eqref{main:puregeoconv}.
(b) \emph{The supporting entropy bound:} for a strictly physical
target covariance, the supporting inequality relates pure-state
entanglement $S(\psi_A)$ to quadratic energy
$\langle H_W\rangle_\psi$, with equality at the optimal Gaussian
seed $\Gamma_*$.
The relative-entropy argument based on
Proposition~\ref{main:contraction} establishes
Eq.~\eqref{main:allstatesupport} for arbitrary finite-energy pure
states, with the shaded half-plane indicating the bound.
For any ensemble of a centered state with covariance $V$,
complementarity fixes the average quadratic energy at
$\Tr(WV)/4=\Tr(W\Gamma_*)/4$. Averaging the entropy inequality
therefore gives the lower bound $G(V)$ on the average entanglement.}
\label{fig:proof-geometry}
\end{figure*}

The proof of Proposition~\ref{main:contraction} shows that the limiting
output of a regularized minimizing sequence for $F_\lambda$ is the
reference Gaussian state $\tau=\Phi(\sigma)$.
The strictly positive quadratic term bounds the energy of the sequence and supplies a strict remainder in the comparison of relative-entropy second variations. Gaussian covariance identities transfer this comparison between input and output; stationarity under Gaussian thermal relaxation then enforces the thermal balance equations characterizing $\tau$. At this limit, the output relative entropy in
Eq.~\eqref{main:variationalgap} vanishes, leaving the nonnegative
input term $\lambda D(\rho\Vert\sigma)$. The proof uses a weighted interpolation estimate with a strict remainder, established in Appendix~\ref{app:interpolation} from Ref.~\cite{interpolation}.

To apply this comparison to the supporting quadratic, we first separate the local pure factors and then approximate the remaining paired state by faithful Gaussian states. Write $\Gamma_*$ in Gaussian Schmidt form, with $n_{\mathrm p}$ entangled mode pairs and local pure modes. If $n_{\mathrm p}=0$, Gaussian stationarity makes the state with covariance $\Gamma_*$ a ground state of $H_W$, so Eq.~\eqref{main:allstatesupport} follows from the nonnegativity of entropy. For $n_{\mathrm p}>0$, completing a positive quadratic square gives
\begin{equation}\label{main:local-reduction}
 H_W=H+H_{\rm rem},\qquad
 H_{\rm rem}\geq0,\qquad
 \langle H_{\rm rem}\rangle_{\Gamma_*}=0.
\end{equation}
Here $H$ acts on the paired modes and has a positive-semidefinite real quadratic coefficient; its additive constant is fixed by the displayed operator identity. Appendix~\ref{app:support} gives this reduction in annihilation coordinates and verifies the stationarity of the paired seed.

The paired part of the optimal seed is a pure Gaussian state $\omega_c$ with entanglement $E_*:=S((\omega_c)_A)=E(\Gamma_*)$, since the discarded factors are local and pure. Choose annihilation operators $c_1,\ldots,c_{2n_{\mathrm p}}$, generally mixing Alice's and Bob's modes, for which $\omega_c$ is the joint vacuum, and define $N_c=\sum_{j=1}^{2n_{\mathrm p}}c_j^\dagger c_j$.
The stationary-point criterion above makes $\omega_c$ a global pure Gaussian minimizer of entropy plus $H$. Displaced pure Gaussian states obey the same bound because the quadratic coefficient of $H$ is positive semidefinite. Since every mixed Gaussian state admits a displacement ensemble of pure Gaussian states, concavity of the reduced entropy gives, for every Gaussian state $\rho$ on the paired modes,
\begin{equation}\label{s:Gaussianmixed}
 S(\rho_A)+\langle H\rangle_\rho\geq C_*,
 \qquad C_*=E_*+\langle H\rangle_{\omega_c}.
\end{equation}

Two small perturbations provide the strict positivity and faithfulness
required by Proposition~\ref{main:contraction}. For fixed $\eta>0$, adding
$\eta N_c$ makes the quadratic strictly positive and selects
$\omega_c$ uniquely, since equality in the preceding bound then
requires $\langle N_c\rangle=0$. We then introduce a small entropy
term and minimize over Gaussian states:
\begin{equation}\label{s:penalized}
 F_{\lambda,\eta}(\rho)
 =S(\rho_A)+\langle H+\eta N_c\rangle_\rho-\lambda S(\rho).
\end{equation}
The linear energy penalty dominates the entropy bound
\[
 S(\rho)\leq2n_{\mathrm p}\,g\!\left(\frac{\langle N_c\rangle}{2n_{\mathrm p}}\right),
\]
so a minimum exists in a bounded covariance set, uniformly for small $\lambda>0$ at fixed $\eta$. Its value is at most $C_*$, by evaluation at $\omega_c$. Equation~\eqref{s:Gaussianmixed} then implies that the minimum tends to $C_*$ and $\eta\langle N_c\rangle$ tends to zero as $\lambda\downarrow0$. Hence the minimizing covariances converge to that of $\omega_c$.

The entropy term makes the minimizer faithful. Its marginal on Alice's subsystem is already faithful for small $\lambda$, since every limiting Schmidt occupation is positive. If a global symplectic eigenvalue were $1$, increasing the corresponding thermal occupation from zero to $u$ would change the quadratic energy and the marginal entropy by $O(u)$, while $-\lambda g(u)=-\lambda u\log(1/u)+O(u)$ would decrease the functional. Thus the full minimizer $\sigma$ is faithful as well; positivity of the quadratic sets its first moments to zero.

At this interior Gaussian minimum, covariance differentiation gives
\begin{equation}\label{s:modular}
 H+\eta N_c=\lambda H_\sigma-\Phi^*H_\tau+\text{constant},
 \qquad \tau=\Phi(\sigma),
\end{equation}
where $\Phi$ traces out Bob's paired modes. Since the coefficient
on the left is strictly positive, Proposition~\ref{main:contraction} and
Eq.~\eqref{main:variationalgap} make $\sigma$ a minimum over all
finite-energy states. For a pure test state the global entropy
term vanishes, so letting first $\lambda\downarrow0$ at fixed
$\eta$, and then $\eta\downarrow0$, yields the paired-mode bound
\begin{equation}\label{s:corepure}
 S(\psi_A)+\langle H\rangle_\psi\geq C_*.
\end{equation}

Averaging this bound over pure-state ensembles gives
$\EF(\rho_{\rm p})+\Tr(H\rho_{\rm p})\geq C_*$
for every finite-energy state on the paired modes, since finite
mean photon number bounds all second moments and makes the
quadratic term integrable. 
For a pure state $|\psi\rangle$ of all modes, let $\rho_{\rm p}$ be its reduction to the paired modes. Local tracing of the unpaired modes gives $S(\psi_A)\geq\EF(\rho_{\rm p})$ by LOCC monotonicity. Restoring the nonnegative $H_{\rm rem}$ in Eq.~\eqref{main:local-reduction} proves Eq.~\eqref{main:allstatesupport}, including the local pure modes.

The geometric interpolation and the supporting entropy bound are
illustrated in Fig.~\ref{fig:proof-geometry}.

\subsection{Completion of the proof}\label{sec:state-proof-completion}
Since local displacements leave the entanglement measures unchanged,
we may take $\rho$ to be centered. For a strictly physical
covariance $V$, averaging Eq.~\eqref{main:allstatesupport} over
any pure-state ensemble replaces its quadratic term by
$\Tr(WV)/4$, which cancels by complementarity and gives
$\EF(\rho)\geq G(V)$.

For a boundary covariance, independent local Gaussian shifts of ordinary covariance $\epsilon\Id/2$ give a state $\rho_\epsilon$ with quantum covariance $V+\epsilon\Id$. LOCC monotonicity and Eq.~\eqref{g:continuity} yield
\[
 \EF(\rho)\geq\EF(\rho_\epsilon)
 \geq G(V+\epsilon\Id)\longrightarrow G(V).
\]
The entanglement of formation bound in part~(c) therefore holds
for every physical covariance, and the Gaussian displacement
ensemble attains it for $\rho_V$.

For Gaussian states, additivity~\eqref{main:add} removes the
regularization in Eq.~\eqref{main:coding}, proving part~(a).
The common value consequently inherits the strong superadditivity
and additivity established for $G$, proving part~(b).
For an arbitrary finite-energy state $\rho$ with covariance $V$,
applying the entanglement of formation bound to each tensor power
gives $\EF(\rho^{\otimes k})\geq G(V^{\oplus k})=kG(V)$;
dividing by $k$ and using Eq.~\eqref{main:coding} yields
$\EC(\rho)\geq G(V)$. This proves part~(c) and completes the
proof of Theorem~\ref{main:formation}.
\end{proof}

Interestingly, the lower bound in Eq.~\eqref{main:extremalityeq} is also attained
by a concrete family of non-Gaussian states: the Gaussian
displacement distribution in an optimal preparation can be
replaced by any non-Gaussian distribution with the same first
and second moments, preserving both the entanglement of formation
and the entanglement cost.

\begin{corollary}[Non-Gaussian states with exact entanglement cost]\label{cor:nongaussian-states}
Let $\Gamma_*$ attain $G(V)$, let $|\psi_{\Gamma_*}\rangle$ be its centered pure Gaussian state, and let $P$ be any centered probability distribution on real displacements with ordinary covariance $(V-\Gamma_*)/2$. Then
\begin{align}\label{eq:nongaussian-saturation}
 \rho_P&=\int D(\xi)|\psi_{\Gamma_*}\rangle\langle\psi_{\Gamma_*}|D(\xi)^\dagger\,dP(\xi),\nonumber\\
 \EF(\rho_P)&=\EC(\rho_P)=G(V).
\end{align}
\end{corollary}
\begin{proof}
The displayed ensemble has covariance $V$ and average entanglement $E(\Gamma_*)=G(V)$, because displacements are local unitaries. Thus $\EC(\rho_P)\leq\EF(\rho_P)\leq G(V)$, while Theorem~\ref{main:formation} gives the reverse bounds.
\end{proof}
When $V\neq\Gamma_*$, a finite-support displacement distribution gives a non-Gaussian state. The same covariance minimum therefore determines the exact preparation cost of an explicit family of Gaussian and non-Gaussian states.

\section{Entanglement cost of Gaussian channels}\label{sec:channel-cost}

In this Section we determine the entanglement cost of a Gaussian channel by
comparing the entanglement of the states it distributes with the
resources required for its teleportation
simulation~\cite{PLOB2017,WTB2017}.

A Gaussian channel $\Phi:A\to B$ from $m\geq1$ input modes to
$\ell\geq1$ output modes acts on covariance matrices
as~\cite{Caruso2008,Serafini2017}
\begin{equation}\label{eq:channel-covariance}
 V\longmapsto XVX^{\mathsf T}+Y,
 \qquad
 Y+i(\Omega_\ell-X\Omega_mX^{\mathsf T})\geq0,
\end{equation}
where $X\in\mathbb R^{2\ell\times2m}$ and
$Y=Y^{\mathsf T}\in\mathbb R^{2\ell\times2\ell}$, and the matrix
inequality is necessary and sufficient for complete positivity.
We take the first moments to transform as $d\mapsto Xd$, since
a fixed output displacement is a local unitary and can be
implemented at no additional entanglement cost.

We measure simulation errors using the trace distance between
states and the normalized diamond distance between channels,
\begin{equation}\label{eq:channel-distances}
 d_{\mathrm{tr}}(\rho,\sigma)
   =\frac12\|\rho-\sigma\|_1,
 \qquad
 d_\diamond(\Phi,\Psi)
   =\frac12\|\Phi-\Psi\|_\diamond,
\end{equation}
where $\|\cdot\|_1$ and $\|\cdot\|_\diamond$ denote the trace
norm and diamond norm, respectively; $d_\diamond$ is the
supremum of the output trace distance over all input states
and arbitrary reference systems.

Following Wilde~\cite{wilde2018}, we consider LOCC simulations
of $n$ sequential uses of $\Phi$, supplied with a maximally
entangled resource of Schmidt rank $M_n$. A test may retain
a quantum memory and choose later inputs using earlier
outputs; we denote by $\varepsilon_n$ the supremum, over all
such adaptive tests, of the trace distance between the final
states produced by the channel and its simulation. The
entanglement cost of the channel $\Phi$ is
\begin{equation}\label{eq:channel-cost-definition}
 \EC(\Phi)
 =\inf\left\{
    \limsup_{n\to\infty}\frac{\log M_n}{n}
    :\ \varepsilon_n\to0
   \right\},
\end{equation}
where the infimum runs over sequences of these simulations,
and rates are measured in nats per channel use.

The parallel entanglement cost $\EC^{\mathrm p}(\Phi)$ is defined
analogously for joint LOCC simulations of $\Phi^{\otimes n}$
whose diamond distance from the target channel tends to zero,
extending the finite-dimensional formulation of
Ref.~\cite{Berta2013}. In both settings, local operations may
have continuous classical outcomes, and input energy is
unrestricted.

We probe the channel with a product $\Psi_s^{RA}$ of $m$ two-mode
squeezed vacua, pairing each input mode in $A$ with a mode of an
$m$-mode reference $R$. Both marginals have covariance
$s\Id_{2m}$, with $s\geq1$, and applying $\Phi$ to $A$ gives
the joint reference--output state
\begin{equation}\label{eq:channel-probe}
 \omega_\Phi(s)
   =(\operatorname{id}_R\otimes\Phi)(\Psi_s^{RA}).
\end{equation}
With $Z_m=\bigoplus_{j=1}^m\operatorname{diag}(1,-1)$ and
$c_s=\sqrt{s^2-1}$, the covariance matrix of $\omega_\Phi(s)$ is
\begin{equation}\label{eq:channel-probe-covariance}
 V_\Phi(s)=
 \begin{pmatrix}
 s\Id_{2m}&c_sZ_mX^{\mathsf T}\\
 c_sXZ_m&sXX^{\mathsf T}+Y
 \end{pmatrix}.
\end{equation}
Both the probe $\Psi_s^{RA}$ and the output $\omega_\Phi(s)$
have finite energy for every finite $s$, and the state
entanglement measures in the following theorem are evaluated
across the reference--output bipartition $R:B$.

\begin{theorem}[Entanglement cost of multimode Gaussian channels]
\label{thm:channel-cost}
(a) For every multimode Gaussian channel $\Phi=(X,Y)$,
\begin{align}\label{eq:channel-cost}
 \EC(\Phi)=\EC^{\mathrm p}(\Phi)
 &=\sup_{s\geq1}\EF\bigl(\omega_\Phi(s)\bigr)\nonumber\\
 &=\lim_{s\to\infty}G\bigl(V_\Phi(s)\bigr).
\end{align}

\noindent (b) The common value is finite if and only if the noise
matrix $Y$ is positive definite; if $Y$ is singular, it is $+\infty$.

\noindent (c) The entanglement cost is additive on tensor products:
for any two multimode Gaussian channels,
\begin{equation}\label{eq:channel-cost-additivity}
 \EC(\Phi_1\otimes\Phi_2)=\EC(\Phi_1)+\EC(\Phi_2),
\end{equation}
including when either cost is infinite.
\end{theorem}

\begin{proof}[Proof of Theorem~\ref{thm:channel-cost}]
We prove parts~(a) and~(b) by obtaining a lower bound from Gaussian
probes and matching it by teleportation when $Y>0$. For singular $Y$,
a noiseless output quadrature forces the cost to diverge, while
part~(c) follows from additivity for Gaussian states.

\subsection{Probe bounds and teleportation}
A channel simulator must also reproduce its action on an entangled
probe. Fix $s<\infty$ while the number of simulated uses tends to
infinity. Alice prepares $(\Psi_s^{RA})^{\otimes n}$ locally and
feeds the input halves into a parallel simulation, obtaining an
approximation to $\omega_\Phi(s)^{\otimes n}$ with the same entanglement
consumption and no larger error. Theorem~\ref{main:formation}
therefore gives the lower bound
\begin{equation}\label{eq:channel-converse}
 G(V_\Phi(s))=\EC(\omega_\Phi(s))
 \leq\EC^{\mathrm p}(\Phi)\leq\EC(\Phi).
\end{equation}

The same probe outputs also serve as teleportation resources. Write
$\mathcal T_Q$ for the additive-noise channel that adds $Q\geq0$ to the
quantum covariance; its displacement law has ordinary covariance $Q/2$.
Braunstein--Kimble teleportation~\cite{BraunsteinKimble1998} with a resource state
$\Psi_s$ implements
$\mathcal T_{\epsilon_s\Id_{2m}}$, where
\begin{equation}\label{eq:channel-epsilon}
 \epsilon_s=2(s-c_s)\downarrow0.
\end{equation}
Since the channel maps an input displacement $\xi$ to an output
displacement $X\xi$, teleportation with a resource $\omega_\Phi(s)$
and correction $X\xi$ implements exactly~\cite{PLOB2017}
\begin{equation}\label{eq:channel-teleportation}
 \Phi_s=\Phi\circ\mathcal T_{\epsilon_s\Id_{2m}},
 \qquad (X,Y)_s=(X,Y+\epsilon_sXX^{\mathsf T}).
\end{equation}
Writing $A'$ for Alice's resource modes, the protocol measures the
commuting quadratures $R_A-Z_mR_{A'}$ and
displaces Bob's output by $X$ times the outcome. This description applies
to arbitrary inputs and their reference extensions.

Teleportation also establishes the existence of the limit in
Eq.~\eqref{eq:channel-cost}. Consider a fixed finite-energy Gaussian
probe $\psi^{RA}$, prepared locally by Alice. Since its simulated
output is obtained by LOCC from the shared resource $\omega_\Phi(s)$,
the output entanglement is at most $G(V_\Phi(s))$. Its covariance
differs from that of the actual channel output by
$0_R\oplus\epsilon_sXX^{\mathsf T}$, which tends to zero.
The covariance continuity in Eq.~\eqref{g:continuity} thus gives
\begin{equation}\label{eq:channel-fixed-probe}
 \EF\bigl((\operatorname{id}\otimes\Phi)(\psi)\bigr)
 \leq\liminf_{s\to\infty}G(V_\Phi(s)).
\end{equation}
The same bound holds after fixed local Gaussian processing of the output.
Taking $\psi=\Psi_t$ and then the supremum over $t$ yields
\begin{equation}\label{eq:channel-limit-exists}
 \begin{aligned}
 \sup_{t\geq1}G(V_\Phi(t))
 &\leq\liminf_{s\to\infty}G(V_\Phi(s))\\
 &\leq\limsup_{s\to\infty}G(V_\Phi(s))
 \leq\sup_{t\geq1}G(V_\Phi(t)).
 \end{aligned}
\end{equation}
The probe limit therefore exists and equals the supremum, with
$+\infty$ allowed.

\subsection{Achievability with positive noise}
When $Y>0$, the additional teleportation noise can be produced by a
small change of the environment in a fixed Gaussian dilation.
Choose a dilation with $k$ environmental modes in vacuum, whose
retained output quadratures are
\begin{equation}\label{eq:channel-dilation}
 R_B=XR_A+BR_E,
 \qquad Y=BB^{\mathsf T},
 \qquad B\Omega_kB^{\mathsf T}=\Delta,
\end{equation}
where $\Delta=\Omega_\ell-X\Omega_mX^{\mathsf T}$.
Such a dilation exists for arbitrary input and output mode counts,
with extra vacuum modes and discarded outputs as
needed~\cite{Caruso2008}.
When $Y>0$, the environmental coupling $B$ has the right inverse
\begin{equation}\label{eq:channel-right-inverse}
 L_B=B^{\mathsf T}Y^{-1},\qquad BL_B=\Id_{2\ell}.
\end{equation}
This right inverse lets us realize the teleportation noise by changing
only the environment. Keep the dilation unitary and replace the vacuum
state $\rho_{E,0}$ by a centered Gaussian environment $\rho_{E,s}$ of covariance
\begin{equation}\label{eq:channel-environment}
 V_E(s)=\Id_{2k}+\epsilon_sL_BXX^{\mathsf T}L_B^{\mathsf T}.
\end{equation}
The added noise is exactly that in Eq.~\eqref{eq:channel-teleportation}.
Both channels now use the same unitary with different environment
states, so trace-distance contractivity controls the simulation error
uniformly over all inputs and reference systems. The overlap of
$\rho_{E,s}$ with the vacuum yields
\begin{equation}\label{eq:channel-explicit-error}
 d_\diamond(\Phi_s,\Phi)
 \leq d_{\mathrm{tr}}(\rho_{E,s},\rho_{E,0})
 \leq\sqrt{\frac{\Tr(X^{\mathsf T}Y^{-1}X)}{2s}}.
\end{equation}
Appendix~\ref{app:channel-cost} gives the determinant estimate leading to
this bound. Environment-comparison proofs of uniform convergence were
developed in Refs.~\cite{wilde_uniform2018,pirandola_uniform2018}; the latter
also characterizes full-rank noise as the criterion for single-mode
Gaussian channels. The right inverse~\eqref{eq:channel-right-inverse}
gives the bound for every $Y>0$, including channels with unequal mode counts.

To prove that the cost is finite, it suffices to construct one finite-energy
resource that simulates $\Phi$ exactly. For teleportation with gain $X$,
the matrix selecting the resource contribution to the output noise is
\begin{equation}\label{eq:channel-error-matrix}
 M_{\rm tel}=\begin{pmatrix}-XZ_m&\Id_{2\ell}\end{pmatrix}.
\end{equation}
Both $M_{\rm tel}$ and $B$ have full row rank, and their rows have
the same pairwise symplectic products:
\begin{equation}\label{eq:channel-symplectic-gram}
 M_{\rm tel}\Omega_{m+\ell}M_{\rm tel}^{\mathsf T}
 =\Delta=B\Omega_kB^{\mathsf T}.
\end{equation}
The map matching corresponding rows extends to a symplectic
transformation after adding ancillary modes. Appendix~\ref{app:channel-completion}
uses this extension to construct a finite-energy Gaussian resource
with covariance $V_{\rm res}$ satisfying
\begin{equation}\label{eq:channel-exact-resource}
 M_{\rm tel}V_{\rm res}M_{\rm tel}^{\mathsf T}=Y.
\end{equation}
This simulator condition also appears in
Ref.~\cite[Proposition~8.8]{MishraWilde2026}, which proves the existence
of a finite-energy resource when $Y+i\Delta>0$. The symplectic
extension covers every $Y>0$, including cases where $Y+i\Delta$ is
singular. The characteristic-function calculation in
Appendix~\ref{app:channel-completion} verifies exact simulation
on arbitrary inputs and their reference extensions. Since the
resource has finite local entropy, it has
finite entanglement cost; preparing many copies and applying the
simulator therefore proves $\EC(\Phi)<\infty$. The lower bound
in Eq.~\eqref{eq:channel-converse} then makes the probe limit finite.

We attain the probe limit by letting the resource squeezing increase
with the number of simulated uses. The required pure Gaussian seeds
are prepared from maximally entangled states by entanglement
dilution~\cite{wilde2018}. The Gaussian Schmidt decomposition expresses
the reduced state of each seed on Alice's resource modes, up to a
local unitary, as a product of at most $m$ thermal states. For a thermal state with eigenvalues
$p_j$ and mean occupation $u>0$, its information variance is the
variance of $-\log p_j$ when $j$ is sampled with probability $p_j$.
Writing $\beta_{\rm th}=\log(1+u^{-1})$, this variance is
\begin{equation}\label{eq:channel-thermal-variance}
 u(u+1)\log^2(1+u^{-1})
 =\frac{\beta_{\rm th}^2}{4\sinh^2(\beta_{\rm th}/2)}\leq1.
\end{equation}
Vacuum factors contribute zero and variances add, so the information
variance of the seed's Schmidt probabilities is at most $m$.
Because this bound depends only on the number of modes, it also
applies when the seed changes with $n$ and its energy grows.
A cutoff of the Schmidt expansion with excess entanglement rate
$a_n>0$ consequently approximates $n$ copies with trace-distance
error at most $\sqrt{m/(na_n^2)}$; Lemma~\ref{lem:channel-variance}
gives the precise rank bound and proof.

Choose $s_n\to\infty$ so that
$d_\diamond(\Phi_{s_n},\Phi)\leq n^{-2}$, and let $\Gamma_n$ attain
$G(V_\Phi(s_n))$. Approximate $n$ copies of the corresponding pure
Gaussian seed using $a_n=n^{-1/4}$, and apply independent Gaussian
random displacements to the copies, with ordinary covariance
$[V_\Phi(s_n)-\Gamma_n]/2$ on each copy. The resulting state
approximates $\omega_\Phi(s_n)^{\otimes n}$ with no larger
trace-distance error.
Replacing channel calls one at a time in an adaptive test contributes
at most $n\,d_\diamond(\Phi_{s_n},\Phi)$~\cite{PLOB2017,pirandola_uniform2018}.
Contractivity bounds the additional error from resource preparation by
the trace distance of the joint resource. Hence
\begin{align}
 \varepsilon_n&\leq n^{-1}+\sqrt m\,n^{-1/4}\longrightarrow0,
 \label{eq:channel-total-error}\\
 \frac{\log M_n}{n}
 &\leq G(V_\Phi(s_n))+n^{-1/4}+\frac{\log2}{n}
 \longrightarrow\lim_{s\to\infty}G(V_\Phi(s)).
 \label{eq:channel-achievable-rate}
\end{align}
The right-hand side of Eq.~\eqref{eq:channel-achievable-rate}
converges to the probe limit, which is therefore achievable by
sequential simulation. Together with Eq.~\eqref{eq:channel-converse},
this proves parts~(a) and~(b) when $Y>0$.

\subsection{Noiseless quadratures and additivity}
If $Y$ is singular, at least one output quadrature receives no added noise.
We isolate a canonical input--output pair containing this quadrature and
show that its probe entanglement diverges. Choose a nonzero real
$w\in\ker Y$ and $z$ such that
$w^{\mathsf T}\Omega_\ell z=1$. Positivity of $Y+i\Delta$ implies
$\Delta w=0$, and therefore
\begin{equation}\label{eq:channel-noiseless-pair}
 (X^{\mathsf T}w)^{\mathsf T}\Omega_m(X^{\mathsf T}z)=1.
\end{equation}
Thus the output quadratures $w^{\mathsf T}R_B,z^{\mathsf T}R_B$
and their input counterparts
$w^{\mathsf T}XR_A,z^{\mathsf T}XR_A$ both form canonical pairs.
Completing them to symplectic coordinates makes the selected output
mode depend only on the selected input mode. Preparing the remaining
inputs in vacuum and discarding the remaining outputs therefore
gives an effective single-mode channel $\mathcal B_\nu$ with
\begin{equation}\label{eq:channel-b1}
 X_{\mathcal B_\nu}=\Id_2,\qquad
 Y_{\mathcal B_\nu}=\begin{pmatrix}0&0\\0&\nu\end{pmatrix},
 \qquad \nu=z^{\mathsf T}Yz\geq0.
\end{equation}
For a two-mode squeezed probe with marginal covariance $t\Id_2$, its
output marginal has symplectic eigenvalue $\sqrt{t(t+\nu)}$, while the
joint state has symplectic eigenvalues $1$ and $\sqrt{1+t\nu}$.
The coherent information of the joint state
$\omega_{\mathcal B_\nu}(t)$, defined as its output entropy minus
its joint entropy, is
\begin{equation}\label{eq:channel-b1-coherent-information}
 I_\nu(t)=
 g\!\left(\frac{\sqrt{t(t+\nu)}-1}{2}\right)
 -g\!\left(\frac{\sqrt{1+t\nu}-1}{2}\right).
\end{equation}
For $\nu=0$, this equals $g((t-1)/2)$; for fixed $\nu>0$, it grows as
$\frac12\log t+O(1)$.
Concavity of conditional entropy gives
$I_\nu(t)\leq\EF(\omega_{\mathcal B_\nu}(t))$, which equals
$\EC(\omega_{\mathcal B_\nu}(t))$ by Theorem~\ref{main:formation}.
Every simulation of $\Phi$ also simulates $\mathcal B_\nu$ by the
local Gaussian operations just described, so the probe lower bound
forces both channel costs to be infinite. The limit in
Eq.~\eqref{eq:channel-cost} diverges as well: for each fixed $t$,
apply Eq.~\eqref{eq:channel-fixed-probe} to the Gaussian input
preparation realizing this effective channel and its local output
processing, and then let $t\to\infty$. This proves parts~(a)
and~(b) for singular $Y$.

Finally, the probe covariance for $\Phi_1\otimes\Phi_2$ is
$V_{\Phi_1}(s)\oplus V_{\Phi_2}(s)$ up to local mode ordering. State
additivity and the established limits give
Eq.~\eqref{eq:channel-cost-additivity}, proving part~(c) and
completing the proof of Theorem~\ref{thm:channel-cost}.
\end{proof}

\subsection{Bounds from moments}
The probe argument also applies to non-Gaussian channels, because
Theorem~\ref{main:formation} bounds the entanglement cost of an
arbitrary output state using only its covariance matrix.
\begin{corollary}[Channel bounds from moments]\label{cor:channel-moments}
Let $\mathcal N:A\to B$ be a channel whose probe outputs
$\omega_{\mathcal N}(s)=(\operatorname{id}_R\otimes\mathcal N)(\Psi_s^{RA})$
have finite energy for every finite $s$, and denote their covariance
matrices by $V_{\mathcal N}(s)$. Then
\begin{equation}\label{eq:channel-moment-bound}
 \EC(\mathcal N)\geq\EC^{\mathrm p}(\mathcal N)
 \geq\sup_{s\geq1}G\bigl(V_{\mathcal N}(s)\bigr).
\end{equation}
If $\mathcal N$ has the same first- and second-moment action as a Gaussian
channel $\Phi=(X,Y)$ on every finite-energy input, the supremum
in Eq.~\eqref{eq:channel-moment-bound} equals $\EC(\Phi)$.
In particular, singular $Y$ gives
$\EC^{\mathrm p}(\mathcal N)=\EC(\mathcal N)=+\infty$.
\end{corollary}
\begin{proof}
For each fixed $s$, feeding locally prepared copies of $\Psi_s^{RA}$
into a parallel simulator gives
$\EC^{\mathrm p}(\mathcal N)\geq\EC(\omega_{\mathcal N}(s))$.
Theorem~\ref{main:formation} bounds the latter by
$G(V_{\mathcal N}(s))$ from below, and taking the supremum over $s$
proves Eq.~\eqref{eq:channel-moment-bound}.

If the first-moment actions agree on all finite-energy states,
polarization gives $\mathcal N^*(R_B)=XR_A$ as an identity of
quadratic forms, where $\mathcal N^*$ is the Heisenberg adjoint
of $\mathcal N$. This identity also fixes the cross moments
between the reference and output. Together with equality of the
second-moment actions, it gives
$V_{\mathcal N}(s)=V_\Phi(s)$, and
Theorem~\ref{thm:channel-cost} gives the remaining claims.
\end{proof}

Equation~\eqref{eq:channel-cost} expresses the entanglement
cost of a channel as the squeezing limit of a covariance minimum on $m+\ell$
modes. Finite-energy teleportation simulations for single-mode
phase-insensitive channels were developed in
Refs.~\cite{LiuzzoScorpo2017,tserkis2018} and used to bound private
communication in Refs.~\cite{Laurenza2018,Laurenza2019}.
Wilde determined the entanglement cost of pure-loss and
quantum-limited amplifier channels~\cite{wilde2018}; Gaussian
optimality and additivity for general two-mode states subsequently
extended these results to thermal-loss, thermal-amplifier, and
additive-noise channels~\cite{adesso2026}. The present work
extends this characterization to arbitrary multimode Gaussian
channels, including those with coupled modes and correlated noise.

\section{Classical capacity of Gaussian channels}\label{sec:capacity}
The state theorem also determines how much classical information can be
sent through a Gaussian channel. The connection is the output entropy:
a pure input in an isometric dilation produces a pure
output--environment state whose entanglement equals that entropy.
Minimizing the average output entropy at a fixed average input state
is therefore an entanglement of formation problem, to which the
covariance minimum applies.

Gaussian encodings attain the capacity of gauge-covariant channels,
which respect global phase rotations, under phase-invariant quadratic
energy constraints~\cite{GHG2015,GGCH2014}. For general single-mode
channels, Ji established Gaussian optimality of the single-use Holevo
information~\cite{Ji2026}, and Adesso, Lami, and Giovannetti established
additivity and the resulting classical capacity~\cite{ALG2026}.
The result in this Section allows arbitrary coupling between the modes within
each channel use.

We retain the channel notation of Eq.~\eqref{eq:channel-covariance},
with $m,\ell\geq1$, and constrain the input energy using the quadratic
Hamiltonian $H=\frac12R^{\mathsf T}hR+c$, where
$h=h^{\mathsf T}>0$; any linear term can be removed by an input
displacement. Its ground energy is
$E_0=c+\frac12\sum_{j=1}^m\nu_j(h)$, where $\nu_j(h)$ are the
symplectic eigenvalues of $h$, and its expectation in a state with
first moments $d$ and covariance $V$ is
$\frac14\Tr(hV)+\frac12d^{\mathsf T}hd+c$.

To communicate classical information through $\Phi$, an $n$-use code
encodes one of $K_n$ equiprobable messages into an input state
$\rho_x^{(n)}$, and the receiver performs a joint measurement on
the output of $\Phi^{\otimes n}$ to identify the message.
The rate is $\log K_n/n$ nats per channel use, and we denote the
average decoding error by $\varepsilon_n$.
The average input
$\bar\rho_n=K_n^{-1}\sum_{x=1}^{K_n}\rho_x^{(n)}$
must satisfy $\Tr(\bar\rho_nH_n)\leq nE$, where
$H_n=\sum_{j=1}^nH^{(j)}$.
The unassisted classical capacity without feedback is the largest
asymptotic rate achievable with vanishing error,
\begin{equation}\label{cap:capacity-definition}
 C(\Phi,H,E)
 =\sup\left\{
   \liminf_{n\to\infty}\frac{\log K_n}{n}:
   \varepsilon_n\to0
  \right\},
\end{equation}
where the supremum runs over sequences of codes obeying the
average-energy constraint.

For an input ensemble $\mathcal E=\{p_x,\rho_x\}$ with average
$\bar\rho_{\mathcal E}=\sum_xp_x\rho_x$, the output Holevo
information is~\cite{Holevo2004Constraints}
\begin{equation}\label{cap:holevo-information}
 \chi(\Phi,\mathcal E)
 =S\bigl(\Phi(\bar\rho_{\mathcal E})\bigr)
  -\sum_xp_xS\bigl(\Phi(\rho_x)\bigr).
\end{equation}
It is the entropy of the average output minus the average entropy
of the individual signal outputs, and bounds the classical
information recoverable by measuring the output.
Optimizing over ensembles on $n$ uses under the same average-energy
constraint gives
\begin{equation}\label{cap:holevo-optimization}
 \chi_n(E)
 =\sup_{\substack{\mathcal E_n\\
       \Tr(\bar\rho_{\mathcal E_n}H_n)\leq nE}}
       \chi(\Phi^{\otimes n},\mathcal E_n).
\end{equation}
The ensemble probabilities are arbitrary, and the signal states
may be non-Gaussian and entangled across uses. Continuous ensembles
are also included, with sums replaced by integrals~\cite{HolevoShirokov2006}.

A common input displacement induces a common output displacement,
so centering the average input preserves its Holevo information and
removes the nonnegative contribution of its first moments to the
energy. For a centered Gaussian average with covariance $V$, a pure
Gaussian seed with covariance $\gamma\leq V$ can be modulated by
centered Gaussian displacements with ordinary covariance
$(V-\gamma)/2$. We write $\mathcal V_H(E)$ for the allowed average
covariances and $f_\Phi(V)$ for the largest Holevo information
obtained from these displacement ensembles at fixed $V$:
\begin{align}
 \mathcal V_H(E)
 &=\left\{V:V+i\Om_m\geq0,\quad
              \tfrac14\Tr(hV)+c\leq E\right\},
 \label{cap:energyset}\\
 \mu_\Phi(V)
 &=\min_{\substack{\gamma\leq V\\
                   \gamma>0,\ \gamma\Om_m\gamma=\Om_m}}
       S_G(X\gamma X^{\mathsf T}+Y),
 \label{cap:mu}\\
 f_\Phi(V)
 &=S_G(XVX^{\mathsf T}+Y)-\mu_\Phi(V).
 \label{cap:objective}
\end{align}
Here $\mu_\Phi(V)$ is the least output entropy of a compatible pure
Gaussian seed, while the first term in $f_\Phi(V)$ is the entropy of
the Gaussian average output.
For the usual constraint on the mean total photon number, the input
Hamiltonian is
$H=\sum_{j=1}^m a_j^\dagger a_j
  =\frac12R^{\mathsf T}R-\frac m2$.
Thus $h=\Id_{2m}$ and $c=-m/2$, and $E$ is the allowed mean total
photon number per channel use. For the centered average inputs
considered above, the constraint becomes $\Tr V\leq4E+2m$.

\begin{theorem}[Classical capacity of multimode Gaussian channels]
\label{cap:theorem}
(a) For every multimode Gaussian channel $\Phi$, quadratic Hamiltonian
$H$ as above, and finite $E\geq E_0$,
\begin{align}\label{cap:formula}
 C(\Phi,H,E)
 &=\max_{V\in\mathcal V_H(E)}f_\Phi(V),\nonumber\\
 \chi_n(E)
 &=nC(\Phi,H,E)\qquad(n\geq1).
\end{align}

\noindent (b) The covariance maximum and the seed minimum are attained.
An optimal ensemble consists of Gaussian-distributed displacements
of a pure Gaussian seed, which may be entangled across the input
modes within one use. Independent copies of this ensemble attain
$\chi_n(E)$, and the capacity is achievable with product signal
states across uses and collective decoding.

\noindent (c) The capacity is concave and nondecreasing in $E$, with
$C(\Phi,H,E_0)=0$.
\end{theorem}

\begin{proof}[Proof of Theorem~\ref{cap:theorem}]
We first express the minimum average output entropy as entanglement
of formation. The fixed-covariance bound and strong superadditivity
then control arbitrary encodings across multiple uses, while Gaussian
modulation attains the resulting Holevo bound.

\subsection{Output entropy as entanglement of formation}

Choose a Gaussian isometry $T:A\to BF$ realizing
$\Phi(\rho)=\Tr_F(T\rho T^\dagger)$, obtained by adjoining $q$
vacuum modes and applying a Gaussian unitary $U$ with symplectic
matrix $\Xi$~\cite{Caruso2008}:
\begin{equation}\label{cap:dilation}
 T|\psi\rangle
 =U\bigl(|\psi\rangle\otimes|0\rangle^{\otimes q}\bigr),
 \qquad
 \Sigma(V)=\Xi(V\oplus\Id_{2q})\Xi^{\mathsf T}.
\end{equation}
The output $B$ contains $\ell$ modes and has covariance
$XVX^{\mathsf T}+Y$, while $F$ contains the discarded modes,
with an idle vacuum mode included if necessary.
Such vacuum dilations also cover unequal input and output mode
counts, as detailed in Appendix~\ref{app:capacity}.

Refining mixed signals into pure states preserves the average input
and cannot increase the average output entropy, by concavity.
The isometry maps pure ensembles of $\rho$ to pure ensembles of
$T\rho T^\dagger$. Conversely, the vectors in any pure ensemble of
$T\rho T^\dagger$ lie in the range of $T$ almost surely and can be
pulled back by $T^\dagger$. Writing $\psi$ for a pure-state projector,
each dilated signal has entanglement $S(\Phi(\psi))$, which gives
the Matsumoto--Shimono--Winter identity~\cite{MSW2004},
\begin{equation}\label{cap:msw}
 \widehat S_\Phi(\rho)
 :=\inf_{\rho=\int\psi\,d\pi(\psi)}
       \int S(\Phi(\psi))\,d\pi(\psi)
 =\EF^{B:F}(T\rho T^\dagger).
\end{equation}
For finite-energy inputs, the output and environment entropies are
finite, and the ensemble correspondence applies to continuous
ensembles as well.

Every pure input covariance $\gamma\leq V$ gives a pure covariance
$\Gamma\leq\Sigma(V)$ through
\begin{equation}\label{cap:seedbijection}
 \Gamma=\Xi(\gamma\oplus\Id_{2q})\Xi^{\mathsf T}.
\end{equation}
Conversely, undoing the dilation gives
$\Gamma'=\Xi^{-1}\Gamma\Xi^{-\mathsf T}
 \leq V\oplus\Id_{2q}$.
The centered Gaussian state with covariance $\Gamma'$ has an
ancilla marginal with covariance at most $\Id_{2q}$, hence
nonpositive mean photon number. Positivity of the number operator
forces this marginal to be the vacuum, which factors from the
remaining state. Thus
$\Gamma'=\gamma\oplus\Id_{2q}$ with $\gamma$ pure and
$\gamma\leq V$, establishing a bijection between the input seeds
and the pure covariances below $\Sigma(V)$.
Minimizing their entanglement therefore gives
\begin{equation}\label{cap:seedroof}
 G^{B:F}(\Sigma(V))
 =\min_{\gamma\leq V\;\text{pure}}
       S_G(X\gamma X^{\mathsf T}+Y)
 =\mu_\Phi(V).
\end{equation}

For an arbitrary input state $\rho$ with covariance $V_\rho$,
combining Eqs.~\eqref{cap:msw} and~\eqref{cap:seedroof} with
Theorem~\ref{main:formation} yields
\begin{equation}\label{cap:fixedcovroof}
 \widehat S_\Phi(\rho)\geq\mu_\Phi(V_\rho),
 \qquad
 \widehat S_\Phi(\rho_V)=\mu_\Phi(V).
\end{equation}
Thus $\mu_\Phi(V)$ bounds the average output entropy of every
ensemble whose average input has covariance $V$, and Gaussian
displacement modulation attains this bound for the Gaussian
average $\rho_V$.

\subsection{Additivity and optimal encoding}

The identity $\mu_\Phi(V)=G^{B:F}(\Sigma(V))$ makes $\mu_\Phi$
convex, by Eq.~\eqref{main:covconv} and the affine dependence of
$\Sigma(V)$ on $V$. Gaussian entropy is concave in the covariance:
a mixture of centered Gaussian states has at least their average
entropy, and the Gaussian state with the mixture's covariance has
at least the entropy of that mixture. Consequently,
$S_G(XVX^{\mathsf T}+Y)$ is concave in $V$, and so is
$f_\Phi(V)$.

For an arbitrary average input $\bar\rho$ on $n$ channel uses,
let $V_j$ be its $j$th marginal covariance and let $\Sigma_n$
be the covariance after the dilation $T^{\otimes n}$.
Applying the fixed-covariance bound and Gaussian strong
superadditivity from Theorem~\ref{main:formation} gives
\begin{equation}\label{cap:blockroof}
 \widehat S_{\Phi^{\otimes n}}(\bar\rho)
 \geq G^{B^n:F^n}(\Sigma_n)
 \geq\sum_{j=1}^n\mu_\Phi(V_j).
\end{equation}
Here the block of $\Sigma_n$ on $B_jF_j$ is $\Sigma(V_j)$,
including when $\bar\rho$ is correlated and non-Gaussian.
Entropy subadditivity and Gaussian maximal entropy at fixed
covariance also bound the entropy of the average output by
$\sum_jS_G(XV_jX^{\mathsf T}+Y)$.
The Holevo information of any allowed ensemble therefore satisfies
\begin{align}
 \chi
 &\leq\sum_{j=1}^n f_\Phi(V_j)
 \leq n f_\Phi\!\left(\frac1n\sum_{j=1}^nV_j\right)
 \nonumber\\
 &\leq n\max_{V\in\mathcal V_H(E)}f_\Phi(V).
 \label{cap:blockchi}
\end{align}
The second inequality is Jensen's inequality.
For the last, the average covariance
$n^{-1}\sum_jV_j$ is physical and belongs to $\mathcal V_H(E)$,
since the first moments make a nonnegative contribution to the
total input energy.

Both covariance optimizations are attained on the joint domain
\begin{equation}\label{cap:compactdomain}
 \left\{(V,\gamma):
 V\in\mathcal V_H(E),\quad
 \gamma\leq V,\quad
 \gamma\text{ pure}\right\}.
\end{equation}
This set is compact: the energy constraint and $h>0$ bound $V$
and $\gamma$ above, purity bounds the eigenvalues of $\gamma$
away from zero, and the constraints are closed.
Gaussian entropy is continuous there, so
$S_G(XVX^{\mathsf T}+Y)-S_G(X\gamma X^{\mathsf T}+Y)$
attains its maximum. The same argument at fixed $V$ proves
attainment of the seed minimum.

Let $(V_*,\gamma_*)$ be an optimizing pair.
Displacing the pure Gaussian state $\rho_{\gamma_*}$ according
to a centered Gaussian law with ordinary covariance
$(V_*-\gamma_*)/2$ produces the average state $\rho_{V_*}$.
Every signal output has entropy
$S_G(X\gamma_*X^{\mathsf T}+Y)$, so the ensemble has Holevo
information $f_\Phi(V_*)$.
Independent copies saturate Eq.~\eqref{cap:blockchi} for every $n$.

Concavity of $f_\Phi$ and the affine energy constraint make the
optimal value concave in $E$, while inclusion of the feasible
sets makes it nondecreasing. Being finite and concave, it is
continuous for $E>E_0$.
At $E=E_0$, the only allowed average input is the unique pure
ground state of $H$, so every signal is that state and both
the Holevo optimum and the capacity vanish.

To identify the Holevo optimum with operational capacity,
observe that $h>0$ makes the input energy constraint bound
the total input photon number. The channel's affine action on
moments then bounds output photon number and hence output
entropy on each allowed energy set.
The constrained coding theorem, together with the Holevo bound
for the converse, identifies $C(\Phi,H,E)$ with
$\lim_{n\to\infty}\chi_n(E)/n$~\cite[Proposition~3]{Holevo2004Constraints}.

For $E>E_0$, choose an optimal Gaussian ensemble at a slightly
smaller energy. Truncating and discretizing its displacement
modulation gives finite signal alphabets within the budget,
with Holevo information approaching that of the continuous
ensemble; Appendix~\ref{app:capacity} supplies the energy and
approximation estimates. Continuity of the optimal value allows
the smaller energy to approach $E$, proving achievability with
product Gaussian signal states and collective decoding.
This establishes parts~(a) and~(b), while the energy dependence
proved above gives part~(c), completing the proof of
Theorem~\ref{cap:theorem}.
\end{proof}

\section{Examples and applications}\label{sec:examples}
We now apply the state and channel results from the previous Sections to two relevant examples. 

The first example uses Theorem~\ref{main:formation} to calculate the exact
entanglement cost of a four-mode Gaussian family containing bound
entangled states, quantifying how much entanglement is needed to
prepare states from which no entanglement can be distilled back. The optimal
preparation also yields non-Gaussian states with the same entanglement cost.

The second example applies Theorems~\ref{thm:channel-cost}
and~\ref{cap:theorem} to a two-mode optical channel with unequal
losses and a correlated environment, determining exactly the entanglement
cost for simulating such channel and its classical capacity above a finite
input-energy threshold.

We further introduce a numerical package~\cite{GaoAdessoGEoFToolkit}
for evaluating the entanglement $G(V)$ of general
multimode covariance matrices $V$. The package provides upper and lower bounds, whose gap
quantifies the remaining optimization error on the exact value.

\subsection{Entanglement cost of four-mode bound entangled states}
\label{ex:bound}
Bound entangled states are entangled states with zero distillable
entanglement: by LOCC one cannot
extract maximally entangled pairs from many copies of such states at a nonzero
asymptotic rate~\cite{Horodecki1998Bound,Horodecki2009Entanglement}.
In the Gaussian realm, bound entangled states with a positive
partial transpose (PPT) can exist when both parties hold at least two
modes~\cite{wernerwolf2001}, as also demonstrated experimentally in optical systems~\cite{DiGuglielmo2011}.

Here we consider a four-mode family in which every reduction to one
mode of Alice and one mode of Bob is separable. The exact
entanglement cost therefore quantifies the resources needed to
prepare entanglement shared by the two multimode parties, even
though no individual pair of modes is entangled.

Alice and Bob each hold two modes. In the reordered coordinates
$(x_{A_1},x_{A_2},x_{B_1},x_{B_2},
p_{A_1},p_{A_2},p_{B_1},p_{B_2})$,
consider the covariance matrices
\begin{gather}
 V_{a,c}=X_{a,c}\oplus P_{a,c},\qquad
 X_{a,c}=
 \begin{pmatrix}
  a\Id_2&cZ\\
  cZ&a\Id_2
 \end{pmatrix},\nonumber\\
 P_{a,c}=
 \begin{pmatrix}
  a\Id_2&cM\\
  cM&a\Id_2
 \end{pmatrix},
 \label{ex:wwcov}
\end{gather}
where $Z=\diag(1,-1)$,
$M=\left(\begin{smallmatrix}0&1\\1&0\end{smallmatrix}\right)$,
and $c\geq0$. Each mode has marginal covariance matrix $a\Id_2$, corresponding
to position and momentum ordinary variances $a/2$, while $c$ controls
the strength of the correlations between the parties.
Position correlations connect $A_1$ with $B_1$ and $A_2$ with
$B_2$, whereas momentum correlations connect $A_1$ with $B_2$
and $A_2$ with $B_1$.

Positive definiteness of $V_{a,c}$ requires $a>c$, and the
anticommutation relation $ZM+MZ=0$ gives the squared symplectic
eigenvalues of the four-mode states as
\[
 \nu_\pm^2=a^2\pm c\sqrt{2a^2-c^2},
\]
each with multiplicity two. The uncertainty condition
$\nu_-\geq1$ therefore amounts to the physical region
\begin{equation}\label{ex:physical-region}
 a\geq a_0(c):=\sqrt{1+c^2+\sqrt2c}.
\end{equation}
Partial transposition on Bob's modes replaces $cM$ by $-cM$
without changing the symplectic spectrum, so every physical
state in the family is PPT.

Separability criteria and the effect of noise on Gaussian bound
entanglement have been studied in
Refs.~\cite{chen2023,shchukin2026,singh2025}.
We now use Theorem~\ref{main:formation} to determine the
entanglement of formation and entanglement cost throughout the
physical region by evaluating $G(V_{a,c})$. This calculation
identifies the bound entangled members of the family and gives
the exact entanglement required to prepare them.

\begin{figure*}[t]
\centering
\includegraphics[width=\textwidth]{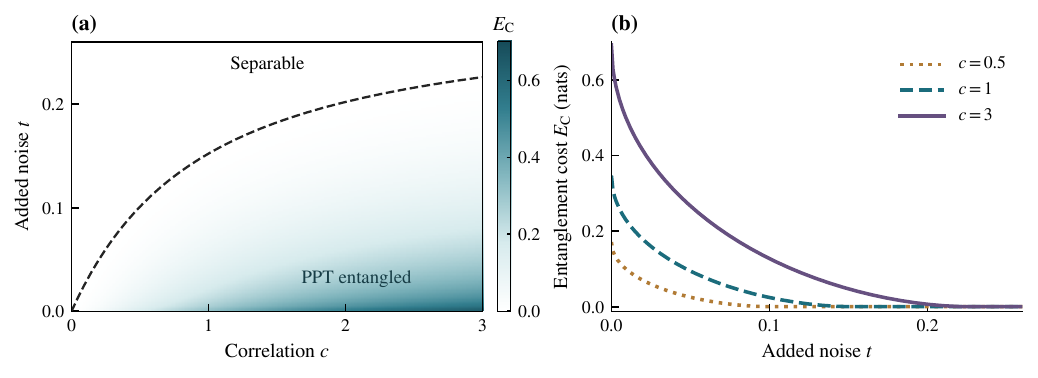}
\caption{Entanglement cost of the four-mode Gaussian
family~\eqref{ex:wwcov}. Independent Gaussian displacement noise gives
$a=a_0(c)+t$, with ordinary noise covariance $t\Id_8/2$.
(a) The gradient shows $\EC=\EF=G$ in nats, and the dashed line
$t=c+1-a_0(c)$ marks the separability threshold.
(b) The exact cost decreases to zero along three noise curves.
Every state shown is PPT and has zero distillable entanglement and
logarithmic negativity; below the threshold it nevertheless requires
entanglement to prepare, while all shared two-mode reductions remain
separable.}
\label{fig:state-example}
\end{figure*}

\paragraph*{Optimal preparation.}
An optimal pure Gaussian seed can be chosen from the family
\begin{gather}
 \Gamma(u)=Q(u)\oplus Q(u)^{-1},\nonumber\\
 Q(u)=\begin{pmatrix}
 \cosh u\,\Id_2&\frac{\sinh u}{\sqrt2}(Z-M)\\
 \frac{\sinh u}{\sqrt2}(Z-M)&\cosh u\,\Id_2
 \end{pmatrix}.
 \label{ex:wwseed}
\end{gather}
The marginal on either party has two symplectic eigenvalues equal
to $\cosh u$. Thus, up to local Gaussian unitaries, the seed consists
of two equally squeezed pairs, each with squeezing $u/2$, and its
entanglement is $2g(\sinh^2(u/2))$. Adding Gaussian displacements of
ordinary covariance $[V_{a,c}-\Gamma(u)]/2$ prepares the target whenever
$\Gamma(u)\leq V_{a,c}$. Since the seed entanglement increases with
$u\geq0$, the task is to find the smallest admissible squeezing. The solution is
\begin{align}\label{ex:wwu}
 u_*&=\left[\operatorname{atanh}\frac{c}{\sqrt2a}
 -\operatorname{arcosh}\frac{a^2-c^2+1}{\sqrt{4a^2-2c^2}}\right]_+,\\
 G(V_{a,c})&=2g\!\left(\sinh^2\frac{u_*}{2}\right),
 \qquad [x]_+:=\max\{x,0\}.\label{ex:wwcost}
\end{align}
Theorem~\ref{main:formation} identifies Eq.~\eqref{ex:wwcost}
with both $\EF$ and $\EC$ of the Gaussian target;
Fig.~\ref{fig:state-example}(a) shows the resulting
entanglement cost across the family.

The reduction to Eq.~\eqref{ex:wwseed} is exact because target
symmetries can be imposed by geometric averaging while preserving
purity, feasibility, and the entanglement bound. To see this, define
\begin{gather*}
 J=\begin{pmatrix}0&1\\-1&0\end{pmatrix},\quad
 \mathsf R_0=J\oplus(-J),\quad
 \mathsf W_0=\begin{pmatrix}0&\Id_2\\\Id_2&0\end{pmatrix},\\
 \mathsf U=(Z+M)/\sqrt2,\qquad K=\mathsf U\oplus\mathsf U.
\end{gather*}
The target is invariant under congruence by
\begin{gather*}
 \Theta=\Id_4\oplus(-\Id_4),\qquad \mathsf R=\mathsf R_0\oplus\mathsf R_0,\\
 \mathsf W=\mathsf W_0\oplus\mathsf W_0,\qquad
 \mathsf F=\begin{pmatrix}0&K\\-K&0\end{pmatrix}.
\end{gather*}
Here $\Theta$ represents global complex conjugation in phase space,
$\mathsf R$ and $\mathsf F$ are local symplectic transformations, and
$\mathsf W$ interchanges the parties. All preserve pure-state
entanglement, and their congruence actions commute and square to the
identity. The successive geometric midpoints
$\Gamma\mapsto\Gamma\#(L\Gamma L^{\mathsf T})$, for
$L=\Theta,\mathsf R,\mathsf W,\mathsf F$, therefore impose all four
symmetries without increasing the objective, by
Eq.~\eqref{main:puregeoconv}.

Invariance under $\Theta$ gives $\Gamma=Q\oplus Q^{-1}$, and the
$\mathsf R$ and $\mathsf W$ symmetries restrict $Q$ to
\[
 Q=\begin{pmatrix}\alpha\Id_2&vZ+wM\\vZ+wM&\alpha\Id_2\end{pmatrix}.
\]
The remaining condition $Q=KQ^{-1}K$ forces
$\alpha^2-v^2-w^2=1$ and $w=-v$, giving Eq.~\eqref{ex:wwseed}.
Replacing $u$ by $|u|$ leaves the entropy unchanged and improves
feasibility for $c\geq0$, so it suffices to use $u\geq0$.

The position and momentum constraints are equivalent by the
$\mathsf F$ symmetry and reduce to the scalar inequality
\begin{equation}\label{ex:scalar-feasibility}
 a-\cosh u\geq\sqrt{c^2-\sqrt2c\sinh u+\sinh^2u}.
\end{equation}
Writing $\beta=\operatorname{atanh}(c/(\sqrt2a))$, its square becomes
\[
 \sqrt{4a^2-2c^2}\cosh(u-\beta)\leq a^2-c^2+1.
\]
Physicality ensures
$(a^2-c^2+1)/\sqrt{4a^2-2c^2}\geq1$, so this inequality specifies an
interval of allowed $u$ centered at $\beta$. Its smallest nonnegative
point is $u_*$ in Eq.~\eqref{ex:wwu}, which obeys $u_*\leq\beta$ and $\cosh u_*\leq\cosh\beta\leq a$, so it
also satisfies the unsquared constraint~\eqref{ex:scalar-feasibility}.

\paragraph*{Irreversibility and displacement noise.}
The entanglement cost is positive precisely for
\[
 a_0(c)\leq a<c+1.
\]
For $a\geq c+1$, the product covariance $\Id_8$ is feasible and the
state is separable. In the entangled region, PPT implies zero
distillable entanglement and zero logarithmic negativity. Every
shared two-mode marginal has correlations in at most one quadrature;
it is PPT and hence separable. Equation~\eqref{ex:wwcost} therefore
quantifies an irreversible preparation cost carried exclusively by the joint
four-mode correlations.

Independent Gaussian displacement noise increases $a$ and leaves $c$
unchanged: ordinary displacement covariance $t\Id_8/2$ gives
$V_{a,c}+t\Id_8=V_{a+t,c}$. An initially entangled member becomes
separable at $t=c+1-a$. Figure~\ref{fig:state-example}(b) follows these
noise curves from the uncertainty boundary, setting $a=a_0(c)+t$.
The boundary value also has a simple expression. Parameterizing it by
\[
 a=\frac{\sqrt{1+\kappa^2}}{1-\kappa},\qquad
 c=\frac{\sqrt2\kappa}{1-\kappa},\qquad 0\leq\kappa<1,
\]
gives $u_*=\operatorname{arsinh}\kappa$ and
\[
 G(V_{a,c})=2g\!\left(\frac{\sqrt{1+\kappa^2}-1}{2}\right).
\]

\paragraph*{Non-Gaussian states with the same cost.}
The entanglement cost calculated in Eq.~\eqref{ex:wwcost} also
provides a lower bound on $\EF$ and $\EC$ for every non-Gaussian
state with covariance $V_{a,c}$, by
Theorem~\ref{main:formation}. To construct states attaining this
bound, we retain the optimal pure Gaussian seed $\Gamma(u_*)$
from Eq.~\eqref{ex:wwseed} and replace the Gaussian displacement
distribution by a discrete one with the same first and second
moments.

In the entangled region $a_0(c)\leq a<c+1$, equality in
Eq.~\eqref{ex:scalar-feasibility} implies that
$V_{a,c}-\Gamma(u_*)$ has rank four. We can therefore choose a
real $8\times4$ matrix $L_*$ and define the random displacement
$\xi$ by
\begin{equation}\label{ex:binary-modulation}
 L_*L_*^{\mathsf T}=\frac{V_{a,c}-\Gamma(u_*)}{2},
 \qquad \xi=L_*z,
\end{equation}
where the four components of $z$ independently take the values
$+1$ and $-1$ with equal probability. Averaging over these
sixteen displacements produces a non-Gaussian mixture of
sixteen equally weighted pure Gaussian states. Since the
displacement has zero mean and ordinary covariance
$L_*L_*^{\mathsf T}$, the mixture has covariance $V_{a,c}$.
Corollary~\ref{cor:nongaussian-states} therefore gives
$\EF=\EC=G(V_{a,c})$ for the resulting non-Gaussian state.

These finite mixtures are distillable. Indeed, the mixture
has rank at most sixteen, whereas each displaced seed has
infinite Schmidt rank for $u_*>0$. Local projection onto
seventeen matching Schmidt vectors of one component succeeds
with nonzero probability and produces a state of rank at most
sixteen whose two marginals each have rank seventeen.
The rank criterion for distillability~\cite{Horodecki2003Rank}
then applies to the projected state, establishing distillability
of the original mixture. Thus the Gaussian and non-Gaussian
preparations have the same covariance and entanglement cost,
but the Gaussian state is bound entangled while the
non-Gaussian mixture is distillable and has a negative partial
transpose.

\subsection{Entanglement cost and classical capacity of a correlated two-mode channel}\label{ex:channel}
Our second example compares the entanglement needed to simulate a
noisy optical device with the classical information that it can
transmit. Each input mode is mixed at a beam splitter with one of two
environmental modes, which are jointly prepared in a two-mode
squeezed vacuum with squeezing $r\geq0$; the beam splitters
have unequal transmissivities $0<\eta_1,\eta_2<1$.
The resulting channel $\Phi_r$ retains the two signal outputs and discards the
two environmental outputs.
When $\eta_1\ne\eta_2$ and $r>0$, the correlated noise and unequal
losses prevent a separation into independent one-mode channels. Nevertheless, symmetry reduces the simulation
cost and the minimum output entropy to one-dimensional optimizations,
allowing us to obtain an explicit capacity formula above a finite energy threshold.

Using $Z=\diag(1,-1)$, we write the pure covariance matrix of a two-mode squeezed vacuum state with
squeezing $t\in\mathbb R$ as
\begin{equation}\label{ex:gamma-t}
 \gamma_t=\begin{pmatrix}
 \cosh(2t)\Id_2&\sinh(2t)Z\\
 \sinh(2t)Z&\cosh(2t)\Id_2
 \end{pmatrix}.
\end{equation}
The environmental state has covariance $\gamma_r$, so the
channel $\Phi_r$ is described by the covariance map
$V\mapsto XVX^{\mathsf T}+Y$ of
Eq.~\eqref{eq:channel-covariance}, with
\begin{gather}\label{ex:loss-channel}
 X=\bigoplus_{j=1}^2\sqrt{\eta_j}\Id_2,\qquad
 L=\bigoplus_{j=1}^2\sqrt{1-\eta_j}\Id_2,\\
 Y=L\gamma_rL^{\mathsf T}.\nonumber
\end{gather}
Here $X$ describes attenuation of the input quadratures,
while $L$ gives the contribution of the environmental
quadratures to the signal outputs. The noise matrix $Y$
therefore incorporates the correlations between the
environmental modes.

For $\eta_1\ne\eta_2$ and $r>0$, the channel $\Phi_r$ cannot
be transformed into a product of one-mode channels by Gaussian
input and output unitaries. To establish this, consider the
matrices $X\Omega_2X^{\mathsf T}\Omega_2^{-1}$ and
$Y\Omega_2^{-1}$. Both undergo the same similarity transformation
associated with the output unitary, so whether they commute
is unchanged by Gaussian input and output unitaries.
For a product of one-mode channels, both matrices are block
diagonal, and the first is proportional to the identity on
each one-mode block; they must therefore commute.
For the channel in Eq.~\eqref{ex:loss-channel}, however, the
upper-right $2\times2$ block of their commutator is
\[
 -(\eta_1-\eta_2)\sqrt{(1-\eta_1)(1-\eta_2)}\sinh(2r)
 \begin{pmatrix}0&1\\1&0\end{pmatrix}\ne0,
\]
which rules out such a product representation.

For equal losses, a balanced mixing of the input and output
modes separates the channel into two independent one-mode
channels. This reduction and the advantages of entangled
inputs were studied in Ref.~\cite{Lupo2009Correlated}.
With unequal losses, the nonzero commutator shows that no
Gaussian input and output unitaries can make both the
attenuation and the noise act independently on two modes,
making the problem genuinely multimode.

\paragraph*{A common symmetry reduction.}
Capacity depends on keeping the entropy of the individual signal
outputs low. We first minimize this entropy over all finite-energy
inputs; the same symmetry argument will then reduce the probe
optimization that determines the simulation cost. Put $J=\left(\begin{smallmatrix}0&1\\-1&0\end{smallmatrix}\right)$, $K=J\oplus(-J)$, and $\Theta=Z\oplus Z$. The channel and its beam-splitter dilation are covariant under the opposite phase rotation $K$ and under global transposition $\Theta$. For a pure Gaussian input $\gamma$, the dilation gives a pure output--environment state. Its entanglement equals the output entropy, so Eq.~\eqref{main:puregeoconv} makes that entropy convex along geometric means of $\gamma$.

The midpoint $\gamma\#(K\gamma K^{\mathsf T})$ has no larger output entropy and is fixed by $K$, since $K^2=-\Id_4$. Taking its midpoint with its $\Theta$ transform also imposes $\Theta$ invariance, while retaining $K$ invariance because $\Theta K\Theta=-K$. Global transposition preserves pure-state entanglement. A covariance fixed by both actions has local blocks $a\Id_2,b\Id_2$ and correlation block $cZ$; purity forces $a=b$ and $a^2-c^2=1$. The optimum thus belongs to the family $\gamma_t$.

Define
\begin{equation}\label{ex:output-entropy}
 s_r(t)=S_G(X\gamma_tX^{\mathsf T}+Y).
\end{equation}
Since $\gamma_{t_0}\#_u\gamma_{t_1}=\gamma_{(1-u)t_0+ut_1}$, this is an ordinary convex function of $t$. Its minimum is attained: the output covariance is bounded below by $Y>0$ and has an eigenvalue tending to infinity as $|t|\to\infty$, so $S_G(V)\geq\tfrac12\log\det V$ gives diverging entropy. For every finite-energy input $\rho$, Eq.~\eqref{cap:fixedcovroof} gives
\[
 S(\Phi_r(\rho))\geq\widehat S_{\Phi_r}(\rho)
 \geq\mu_{\Phi_r}(V_\rho)\geq\min_t s_r(t).
\]
A minimizing $\gamma_t$ attains the last value, which is therefore the minimum output entropy over all finite-energy inputs.

\begin{figure*}[t]
\centering
\includegraphics[width=\textwidth]{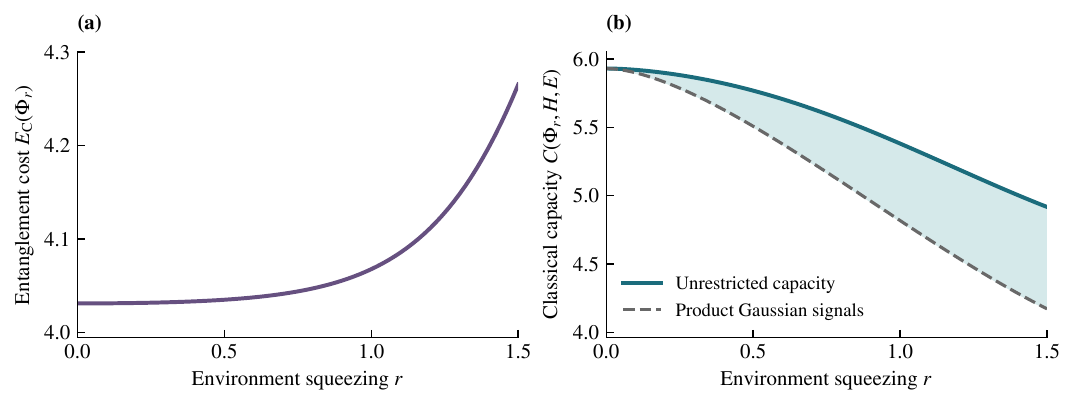}
\caption{Resources and communication rates of the correlated
two-mode attenuator~\eqref{ex:loss-channel}, with $\eta_1=0.55$ and
$\eta_2=0.8$, as the environmental squeezing $r$ increases.
(a) Entanglement cost of simulating the channel, from
Eq.~\eqref{ex:channel-cost-scalar}.
(b) Classical capacity at total input photon budget $E=20$, compared
with the optimal rate for product Gaussian signals across the
physical input ports. The modulation conditions for both formulas
hold throughout the plotted range. Both encoding classes allow
collective decoding, and the shaded gap is the benefit of entangling
the input ports within each use. Rates are in nats per use of the
two-mode device. At $r=0$, the channel is a product of two pure-loss
channels.}
\label{fig:channel-example}
\end{figure*}

\paragraph*{Exact entanglement cost.}
For simulation, the optimized entropy is that of one party's share
of a pure teleportation resource. This leads to a second scalar
minimization. Define
\begin{equation}\label{ex:resource-marginal}
 W_r(t)=L^{-1}(\gamma_t+X\gamma_rX^{\mathsf T})L^{-1}.
\end{equation}
Then
\begin{equation}\label{ex:channel-cost-scalar}
 \EC(\Phi_r)=\min_{t\in\mathbb R}S_G(W_r(t)).
\end{equation}
For the upper bound, we construct a finite-energy resource that implements the channel exactly. Let $Z_2=Z\oplus Z$, $A_\eta=L^{-1}$,
$B_\eta=L^{-1}X$, and
\[
 \Xi_\eta=\begin{pmatrix}
 A_\eta&B_\eta Z_2\\B_\eta Z_2&A_\eta
 \end{pmatrix}.
\]
The identity $A_\eta^2-B_\eta^2=\Id_4$ makes $\Xi_\eta$ symplectic. The pure resource covariance matrix $\Xi_\eta(\gamma_t\oplus\gamma_r)\Xi_\eta^{\mathsf T}$ has Alice's marginal $W_r(t)$. Moreover, the teleportation noise matrix $M_{\rm tel}=(-XZ_2\ \ \Id_4)$ obeys
\[
 M_{\rm tel}\Xi_\eta=(0\ \ L),
\]
so this resource implements $\Phi_r$ exactly. Its entanglement is convex in $t$ by Eq.~\eqref{main:puregeoconv}; a uniform positive lower eigenvalue of $W_r(t)$ and a diverging largest eigenvalue make the minimum attained.

For the matching converse, Gaussian conditioning relates the probe
entanglement in Theorem~\ref{thm:channel-cost} to a pure Gaussian
measurement on its purifying environment. The same two symmetries
restrict this measurement to the family $\gamma_t$. As the probe
squeezing grows, the conditional entropy tends to $S_G(W_r(t))$;
bounded-entropy minimizers remain in a compact interval of $t$, so
the limit gives the reverse inequality in
Eq.~\eqref{ex:channel-cost-scalar}. Appendix~\ref{app:correlated-loss}
provides the conditioning formula and the limit argument. The resulting entanglement cost is illustrated in
Fig.~\ref{fig:channel-example}(a).

\paragraph*{Capacity and entanglement between the input ports.}
The capacity calculation balances two entropy requirements: each
signal should produce as little output entropy as possible, while
the average output should have as much entropy as the energy budget
permits. A signal seed minimizing $s_r(t)$ meets the first requirement.
At sufficiently large input energy, correlated classical
displacements also make the average output a product of thermal
states with the optimal occupations, so one ensemble attains both
extrema.

Let $t_*$ minimize $s_r(t)$ and impose total input photon budget $E>0$ for $H=N_1+N_2$, where $N_j=a_j^\dagger a_j$. Put $n_{\rm e}=\sinh^2r$. Every output obeys
\[
 \sum_j\frac{\langle N_j\rangle}{\eta_j}
 \leq E+n_{\rm e}\sum_j\frac{1-\eta_j}{\eta_j}.
\]
The Gibbs maximum-entropy principle determines the optimal average output occupations from the unique $\lambda>0$ satisfying
\begin{equation}\label{ex:waterfill}
 \bar n_j=\frac1{e^{\lambda/\eta_j}-1},\qquad
 \sum_j\frac{\bar n_j}{\eta_j}
 =E+n_{\rm e}\sum_j\frac{1-\eta_j}{\eta_j}.
\end{equation}
Its covariance is $\bigoplus_j(2\bar n_j+1)\Id_2$, with inverse-image input covariance
\[
 V_*=X^{-1}\left[\bigoplus_j(2\bar n_j+1)\Id_2-Y\right]X^{-\mathsf T}.
\]
The condition $V_*\geq\gamma_{t_*}$ is precisely the requirement
that this average can be prepared by displacing the optimal seed.
When it holds, modulation with ordinary covariance
$(V_*-\gamma_{t_*})/2$ attains both entropy extrema, giving
\begin{equation}\label{ex:capacity-scalar}
 C(\Phi_r,H,E)=\sum_jg(\bar n_j)-s_r(t_*).
\end{equation}
The condition holds above a finite energy threshold: the
off-diagonal block of $V_*$ is independent of $E$, whereas its
diagonal blocks diverge as $E\to\infty$. At lower energies,
Theorem~\ref{cap:theorem} still determines the capacity through
the constrained covariance optimization.

We can now quantify the benefit of entangling the two physical
input ports. As a benchmark, restrict each Gaussian signal to be a
product across the ports, while allowing classical correlations in
the modulation, correlations across uses within either port, and
collective decoding. For a pure one-mode covariance $A$, $JAJ^{\mathsf T}=A^{-1}$, so the first symmetry midpoint of a product seed is vacuum. Its minimum output entropy is therefore $s_r(0)$. The same argument includes block codes: a signal has the form $\rho_{1^n}\otimes\rho_{2^n}$, where either Gaussian port state may correlate its $n$ uses. After refining to pure Gaussian factors, the simultaneous opposite rotations send each port covariance $A$ to $\Omega_n A\Omega_n^{\mathsf T}=A^{-1}$. The midpoint of the full product covariance and its transform is $\Id_{4n}$, giving output entropy at least $ns_r(0)$. Entropy concavity extends this bound to mixed factors and classical mixtures. Thus, when $V_*\geq \Id_4$, the optimal rate for this class is
\begin{equation}\label{ex:product-capacity}
 C_{\rm prod,G}=\sum_jg(\bar n_j)-s_r(0).
\end{equation}
Correlated modulation and time sharing are included in the same average-output entropy bound.

When both modulation conditions hold, the advantage is strictly positive for $r>0$. The symplectic eigenvalues of a covariance matrix with blocks $a\Id_2,b\Id_2,cZ$ are $[\sqrt{(a+b)^2-4c^2}\pm(a-b)]/2$. Writing the blocks of $X\gamma_tX^{\mathsf T}+Y$ as $a(t)\Id_2$, $b(t)\Id_2$, and $c(t)Z$, one has $a'(0)=b'(0)=0$, $c'(0)=2\sqrt{\eta_1\eta_2}>0$, and $c(0)>0$. Both symplectic eigenvalues exceed one and have negative derivatives at $t=0$. Hence $s_r'(0)<0$ and
\[
 C(\Phi_r,H,E)-C_{\rm prod,G}=s_r(0)-s_r(t_*)>0.
\]
The gap is an advantage from entangling modes within one use of
the device; independent uses of the optimal ensemble still attain
the unrestricted capacity. 
Figure~\ref{fig:channel-example} compares this communication
advantage with the entanglement cost of simulating the same
channel.

\subsection{Numerical evaluation with upper and lower bounds}\label{ex:computation}
The analytic examples presented earlier in this Section exploit special symmetries, whereas the evaluation of the quantities in Eq.~\eqref{main:costeq} for arbitrary
multimode covariance matrices generally requires numerical optimization. Here we provide a
standalone Python toolkit~\cite{GaoAdessoGEoFToolkit} that takes a physical
covariance matrix and a mode bipartition, searches for an optimal pure
Gaussian preparation, and evaluates upper and lower bounds on $G(V)$.
For a Gaussian state, these bounds apply to both its entanglement of
formation and its entanglement cost. For any non-Gaussian state with
the same covariance, the lower endpoint remains a lower bound on
both quantities.

The covariance formulation replaces an optimization over ensembles
by a finite matrix problem~\cite{wolf}. Existing approaches include
optimization on pure Gaussian manifolds~\cite{windt2021} and conic
upper bounds~\cite{gaoklco2026}. Here geodesic convexity also controls
the error of a candidate solution. A feasible pure seed
$\Gamma_0\leq V$ specifies a preparation and gives the upper bound
$G(V)\leq E(\Gamma_0)$, while its first variation bounds how much the
entanglement can decrease elsewhere in the feasible set.

For $k$ total modes, define the normalized target covariance $M_0$
and a radius $d_V$ that bounds the distance of every feasible pure
seed from the candidate:
\begin{equation}\label{num:radius}
 M_0=\Gamma_0^{-1/2}V\Gamma_0^{-1/2},\qquad
 d_V=\sqrt{2k}\log\lambda_{\max}(M_0).
\end{equation}
Let $K_0$ represent the first variation along pure Gaussian paths,
\[
 \left.\frac{d}{dt}E(\Gamma_0^{1/2}e^{tQ}\Gamma_0^{1/2})\right|_{t=0}
 =\Tr(K_0Q),
\]
where $Q=Q^{\mathsf T}$ and $Q\Omega+\Omega Q=0$. The orthogonal
projection onto this tangent space is
$\mathcal T_\Omega(A)=(A+\Omega A\Omega)/2$; Appendix~\ref{app:numerical}
gives the analytic derivative. A positive semidefinite matrix
$\Lambda$ acts as a multiplier for the covariance constraint in
these normalized coordinates. It gives the error bound
\begin{equation}\label{num:certificate}
 \varepsilon(\Lambda)=\Tr[\Lambda(M_0-\Id)]
 +d_V\|\mathcal T_\Omega(K_0+\Lambda)\|_F,
\end{equation}
where $\|\cdot\|_F$ is the Frobenius norm, and hence the interval
\begin{equation}\label{ex:numerical-bracket}
 \max\{0,E(\Gamma_0)-\varepsilon(\Lambda)\}
 \leq G(V)\leq E(\Gamma_0).
\end{equation}
The first term of $\varepsilon$ measures slack in the covariance
constraint, while the second measures the remaining first-order
variation. Their sum controls the possible improvement over the
candidate throughout the feasible set.

Indeed, any feasible pure $\Gamma$ gives a positive symplectic matrix
$C=\Gamma_0^{-1/2}\Gamma\Gamma_0^{-1/2}\leq M_0$. Its ordinary
eigenvalues occur in reciprocal pairs, so
$\|\log C\|_F\leq d_V$. Apply the first-order geodesic-convexity
inequality to $E(\Gamma)+\Tr(W\Gamma)/4$, with
$W=4\Gamma_0^{-1/2}\Lambda\Gamma_0^{-1/2}$. Covariance order bounds
the quadratic term, and Cauchy--Schwarz bounds the tangent term,
giving Eq.~\eqref{ex:numerical-bracket}.

For a fixed candidate, minimizing $\varepsilon(\Lambda)$ improves
the lower endpoint through a semidefinite program. If $v$ stacks
the columns of $\mathcal T_\Omega(K_0+\Lambda)$, introduce a scalar
$t$ and impose
\[
 \Lambda\geq0,\qquad
 \begin{pmatrix}t&v^{\mathsf T}\\v&t\Id\end{pmatrix}\geq0.
\]
Minimizing $\Tr[\Lambda(M_0-\Id)]+d_Vt$ gives the best bound of
this form, while any feasible weight already gives a valid lower
bound. Thus the pure-seed search is a nonlinear matrix optimization,
followed by a convex optimization that bounds how far its
entanglement lies above the global minimum.

The implementation uses Williamson coordinates to separate mixed
normal modes from pure ones. With the mixed modes ordered first,
it writes a candidate as $\Xi(e^Q\oplus\Id)\Xi^{\mathsf T}$, where
$Q$ is symmetric and anticommutes with the corresponding symplectic
form. This parameterization enforces purity, while the remaining
normal modes are fixed to vacuum by the covariance constraint.
The search then imposes covariance order and evaluates
Eq.~\eqref{ex:numerical-bracket}. The package evaluates these bounds in double precision and
returns the candidate seed, both endpoints, and their gap,
together with purity and covariance-order residuals.
The same covariance
minimum can be used in the channel formulas of
Secs.~\ref{sec:channel-cost} and~\ref{sec:capacity}.

\section{Conclusions and outlook}\label{sec:discussion}

In this paper we have resolved the Gaussian optimality problem for
entanglement of formation on arbitrary multimode bipartitions,
listed as Problem~29 in the Open Quantum Problems
collection~\cite{KruegerWerner2005}. Theorem~\ref{main:formation} proves that Gaussian
entanglement of formation, unrestricted entanglement of formation,
and entanglement cost coincide for every finite-energy Gaussian
state, extending the two-mode and bisymmetric results of
Ref.~\cite{adesso2026} to general bipartitions. The supporting entropy
inequality underpinning Theorem~\ref{main:formation}  also enables us to determine the entanglement cost of simulating multimode
Gaussian channels (Theorem~\ref{thm:channel-cost}) and their
energy-constrained classical capacity
(Theorem~\ref{cap:theorem}). These channel results extend earlier
single-mode solutions~\cite{wilde2018,adesso2026,ALG2026} to correlated
multimode devices, including those that cannot be reduced to
independent single-mode channels by Gaussian input and output
unitaries.

The additivity established here contrasts with the general quantum
setting, where entanglement of formation can be nonadditive and
entangled inputs across channel uses can increase classical
communication rates~\cite{Shor2004,Hastings2009}. Here the asymptotic
quantities are determined by covariance optimizations for a single
state or channel use, together with a squeezing limit for channel
simulation. This reduction holds even though the underlying
operational problems allow non-Gaussian states and arbitrary
correlations across repetitions. The Gaussian description therefore
determines the unrestricted preparation costs and communication
rates, rather than merely the performance of Gaussian protocols.

The proofs reveal a useful and insightful geometry of pure Gaussian preparations.
Their covariances can be interpolated along paths that preserve
purity and the target covariance constraint, while the entanglement
entropy is convex along those paths. This allows the symmetries used in the state and channel examples
of Secs.~\ref{ex:bound} and~\ref{ex:channel} --- including the exact determination of the entanglement cost of a family of four-mode bound entangled states with all separable two-mode marginals --- to reduce the optimization over pure Gaussian preparations. Supporting quadratic observables connect this
geometry to entropy inequalities for arbitrary states and provide
global lower bounds on the Gaussian entanglement of formation. Comparing these bounds with
the entanglement of a feasible preparation quantifies the remaining
optimization error, as implemented in the numerical toolkit of
Sec.~\ref{ex:computation}. The same structure offers a route to
further exactly solvable entanglement quantification in relevant multimode families and to quantitative
resource estimates when no symmetry reduction is available.

Our results establish a general operational instance of Gaussian
extremality~\cite{extremality} for entanglement
preparation: among all states with the same second moments, the
Gaussian state minimizes both entanglement of formation and
entanglement cost. For non-Gaussian sources, Eq.~\eqref{main:extremalityeq} therefore gives
lower bounds on both entanglement quantifiers from measured second moments alone.
This makes it possible to estimate entanglement experimentally without full state tomography, extending
the two-mode applications of Ref.~\cite{adesso2026} to arbitrary
multimode bipartitions. The bounds apply in particular to the non-Gaussian
optical states produced by conditional measurements,
including photon addition and subtraction~\cite{Walschaers2021}.
For larger systems, bounds obtained from disjoint groups of
modes can be added to lower-bound the entanglement across
the full bipartition, reducing the number of correlations
that must be measured and the size of the covariance
optimizations.

Corollary~\ref{cor:nongaussian-states} also identifies
non-Gaussian families for which the covariance bound is
exact. Replacing the Gaussian displacement distribution
of an optimal preparation by a non-Gaussian distribution
with the same first and second moments produces states
with exactly known entanglement of formation and
entanglement cost.

For quantum communication, our multimode results address Gaussian
devices whose loss and noise cannot be separated into
independent one-mode channels. Theorem~\ref{cap:theorem}
reduces their classical capacity to a covariance optimization
for one use of the full device. The unequal-loss example in
Sec.~\ref{ex:channel} gives an advantage for an entangled
Gaussian seed over Gaussian signals that are product across
the physical ports, while independent uses of the optimal
ensemble attain the full capacity with collective decoding.
The optimization therefore identifies how squeezing,
correlations among input modes, and classical modulation
should be matched to the device's noise and energy budget.

This description is increasingly accessible experimentally:
recent work has reconstructed the full transformation and
noise matrices of 16-mode Gaussian optical processes,
including processes combining mode-dependent loss with
parametric interactions~\cite{Gwak2026}. Such
characterizations provide precisely the matrices entering
our capacity and simulation cost formulas. For non-Gaussian
devices, Corollary~\ref{cor:channel-moments} still gives
lower bounds on the entanglement cost from the covariances
of finite-energy probe outputs, extending the resource
assessment beyond devices described by a Gaussian model.

A central next step is to approach these capacities with
practical collective receivers, building on progress in
joint optical detection~\cite{TakeokaGuha2014,Cui2025GreenMachine}.
Quantifying achievable rates at finite code length would
connect the optimal encodings to realistic decoding
resources, while characterizing the tradeoff among
entanglement, squeezing, and simulation accuracy would
extend existing finite-resource teleportation methods to
coupled multimode devices~\cite{LiuzzoScorpo2017,tserkis2018}.

By determining optimal encodings and exact channel
simulation costs, the results in this paper establish a foundation for
designing communication protocols that optimally exploit correlations
among modes. The challenge now is to translate this control
of multimode quantum resources into practical devices that approach
their fundamental transmission limits.

\paragraph*{Note added.} Parts of this work build on results first reported in preprints
by one of the authors~\cite{gao_geometry,gao_optimality}. Toward completion of this manuscript, we became aware of He's preprint~\cite{He2026frontier}, which also gives a general multimode classical capacity formula as the zero-assistance case of an energy--entanglement tradeoff. We are also aware of related results in preparation by Shao-Hua Hu and Ludovico Lami.

\begin{acknowledgments}
Artificial intelligence tools based on OpenAI and Anthropic models assisted with mathematical development, literature searches, numerical methods, analytic checks, and manuscript preparation. The authors are responsible for the results and the final manuscript.
GA acknowledges financial support from the Engineering and Physical Sciences Research Council (EPSRC Grant No.~UKRI6536).

\end{acknowledgments}
\section*{Data availability}
The data that support the findings of this study are available from the authors upon reasonable request.
The numerical toolkit, including usage instructions and example
calculations, is provided as accompanying
software~\cite{GaoAdessoGEoFToolkit}.

\appendix
\section{Spectral properties of the covariance minimum}\label{app:geometry}
\subsection{Relative spectra under geometric means}
The geodesic convexity in Sec.~\ref{sec:gaussian-geometry} follows from a comparison of relative spectra. Let $A_0,A_1,B_0,B_1$ be positive-definite matrices, and for $0\leq t\leq1$ set $A_t=A_0\#_tA_1$ and $B_t=B_0\#_tB_1$. Let $x(t)$ be the decreasing vector of log-eigenvalues of $A_t^{-1/2}B_tA_t^{-1/2}$. Then
\begin{equation}\label{g:major}
 x(t)\prec(1-t)x(0)+tx(1).
\end{equation}
Here $x\prec y$ denotes vector majorization: the sum of the largest $j$ entries of $x$ is at most that for $y$ for every $j$, with equality for the full sum.

To bound the largest $k$ relative eigenvalues at once, apply the $k$th exterior-power representation. Its eigenvalues are products of $k$ original eigenvalues, and it commutes with multiplication, inversion, positive powers, and hence the geometric mean. The endpoint inequalities
\[
 \wedge^k B_j\leq
 \exp\!\left(\sum_{i=1}^k x_i(j)\right)\wedge^k A_j
\]
interpolate by joint monotonicity and the homogeneity $(aX)\#_t(bY)=a^{1-t}b^t\,X\#_tY$. They give every partial-sum inequality in Eq.~\eqref{g:major}; the identity $\det(X\#_tY)=(\det X)^{1-t}(\det Y)^t$ gives equality for the full sum. Karamata's inequality and convexity therefore imply, for every convex scalar function $F$,
\begin{equation}\label{g:traceconv}
 \sum_iF(x_i(t))\leq(1-t)\sum_iF(x_i(0))+t\sum_iF(x_i(1)).
\end{equation}

For the entropy function in Eq.~\eqref{g:entropy}, convexity follows from the scalar derivatives. Put $f_{\rm sq}(r)=g(\sinh^2r)$ for $r>0$. Then
\begin{align*}
 f_{\rm sq}'(r)&=2\sinh(2r)\log\coth r,\\
 f_{\rm sq}''(r)&=4\cosh(2r)\log\coth r-4>0.
\end{align*}
The last inequality is equivalent to $(z+z^{-1})\operatorname{atanh}z>1$ for $z=e^{-2r}\in(0,1)$, which holds because $\operatorname{atanh}z>z$. Also $f_{\rm sq}'(0+)=0$, so its even extension is continuously differentiable. Congruence by $\Pi$ commutes with geometric means, so $\Pi\Gamma_t\Pi=(\Pi\Gamma_0\Pi)\#_t(\Pi\Gamma_1\Pi)$. Equation~\eqref{g:traceconv}, applied to $(A_j,B_j)=(\Gamma_j,\Pi\Gamma_j\Pi)$, therefore gives Eq.~\eqref{main:puregeoconv}.

\subsection{Derivative of pure Gaussian entanglement}\label{app:numerical}
The spectral formula~\eqref{g:entropy} gives the analytic derivative used in Sec.~\ref{ex:computation}. Put $A_0=\Gamma_0^{-1/2}$, $R_{\Gamma_0}=A_0\Pi\Gamma_0\Pi A_0$, and $f(x)=F(\log x)$. Differentiating the trace of a function of the generalized eigenvalues of $(\Pi\Gamma\Pi,\Gamma)$ gives a full-matrix gradient extending the pure-state differential:
\begin{align}\label{num:gradient}
 D_E&=\tfrac14[\Pi A_0f'(R_{\Gamma_0})A_0\Pi-A_0R_{\Gamma_0}f'(R_{\Gamma_0})A_0],\\
 K_0&=\Gamma_0^{1/2}D_E\Gamma_0^{1/2}.\nonumber
\end{align}
Here $f'(x)=F'(\log x)/x$, with
\[
 F'(x)=\tfrac12\sinh(x/2)\log\coth(|x|/4),\qquad F'(0)=0.
\]
The value $F'(0)=0$ includes local pure modes. The tangent projection $\mathcal T_\Omega(K_0)$ determines the variation on pure covariances, independently of the chosen extension to the surrounding matrix space.

\section{Quadratic reduction of local pure modes}\label{app:support}
The supporting quadratic may couple entangled Schmidt pairs to local pure modes. Completing its square separates these modes into a nonnegative remainder and leaves a stationary quadratic support on the paired modes. This gives the decomposition~\eqref{main:local-reduction} used in Sec.~\ref{sec:unrestricted-support}, including its additive constant.
Put the optimal pure covariance in Gaussian Schmidt form, with $n_{\mathrm p}$ entangled pairs. Let $a_{Aj},a_{Bj}$ be the annihilation operators in these coordinates, and set $q_j=\tanh r_j\in(0,1)$ for the pair squeezing $r_j$, $j=1,\ldots,n_{\mathrm p}$. Define
\begin{align}
 c_{Aj}&=(a_{Aj}-q_ja_{Bj}^\dagger)/\sqrt{1-q_j^2},\nonumber\\
 c_{Bj}&=(a_{Bj}-q_ja_{Aj}^\dagger)/\sqrt{1-q_j^2},\label{s:annihilators}\\
 \alpha_j&=-2\log q_j,\qquad n_j=q_j^2/(1-q_j^2).\nonumber
\end{align}
Collect the paired annihilators in a vector $c_{\mathrm p}$, with all Alice labels before all Bob labels, and the local vacuum annihilators in $c_{\mathrm u}$; write $c=(c_{\mathrm p}^{\mathsf T},c_{\mathrm u}^{\mathsf T})^{\mathsf T}$. The normalized full Gaussian seed $|\psi_{\Gamma_*}\rangle$ is characterized by $c_j|\psi_{\Gamma_*}\rangle=0$, and its reduction to the paired modes is the state $\omega_c$ defined in Sec.~\ref{sec:unrestricted-support}. Thus $E_*=S((\omega_c)_A)=\sum_{j=1}^{n_{\mathrm p}}g(n_j)$. Let $H_A=\sum_j\alpha_ja_{Aj}^\dagger a_{Aj}$ be the thermal Hamiltonian of the paired Alice marginal, with zero coefficient on her unpaired modes; the sum is zero if $n_{\mathrm p}=0$.

The first variation of $E$ at the minimizer equals that of $\langle H_A\rangle$. For positive marginal occupations this is the thermal entropy derivative. For a vacuum occupation, any analytic two-sided covariance path has occupation $O(t^2)$: the eigenvalue branches of $i\Gamma_A(t)^{1/2}\Omega_A\Gamma_A(t)^{1/2}$ are analytic, and uncertainty for both signs of $t$ forces their first derivatives at $1$ to vanish. Their entropy contribution is $O(t^2|\log|t||)=o(t)$. The vacuum modes therefore contribute zero to the first variation.

Stationarity at $\Gamma_*$ of the functional on the left-hand side of Eq.~\eqref{main:gausssupport} under all squeezing variations eliminates terms with two creation or two annihilation operators. In the $c$ coordinates,
\begin{equation}\label{s:parent}
 H_W+H_A=c^\dagger\mathsf Kc+C_0,
 \quad \mathsf K\geq0,\quad
 C_0=\frac{\Tr\mathsf K-\sum_j\alpha_j}{2}.
\end{equation}
The sum $H_W+H_A$ is bounded below, so evaluation on coherent states forces $\mathsf K\geq0$; the constant follows from symmetric ordering. In particular $\langle H_W\rangle_{\psi_{\Gamma_*}}=C_0-\sum_j\alpha_jn_j$.

If $n_{\mathrm p}=0$, then $H_A=0$, so $|\psi_{\Gamma_*}\rangle$ is a ground state of $H_W$, as used in Sec.~\ref{sec:unrestricted-support}. Henceforth assume $n_{\mathrm p}>0$.

Partition $\mathsf K$ according to the paired and unpaired components of $c$. Completing the square for this positive block matrix gives
\begin{align}
 c^\dagger\mathsf Kc
 &=c_{\mathrm p}^\dagger\mathsf K_{\mathrm{eff}}c_{\mathrm p}
   +(c_{\mathrm u}+\mathsf K_{\mathrm{uu}}^{+}\mathsf K_{\mathrm{up}}c_{\mathrm p})^\dagger
     \mathsf K_{\mathrm{uu}}(c_{\mathrm u}+\mathsf K_{\mathrm{uu}}^{+}\mathsf K_{\mathrm{up}}c_{\mathrm p}),\nonumber\\
 \mathsf K_{\mathrm{eff}}
 &=\mathsf K_{\mathrm{pp}}-\mathsf K_{\mathrm{pu}}\mathsf K_{\mathrm{uu}}^{+}\mathsf K_{\mathrm{up}}\geq0.
 \label{s:schur}
\end{align}
Here $+$ denotes the Moore--Penrose inverse; the necessary range condition follows from $\mathsf K\geq0$. The operator $c_{\mathrm p}^\dagger\mathsf K_{\mathrm{eff}}c_{\mathrm p}-H_A$ acts only on the paired modes and has a positive-semidefinite real quadratic coefficient. Indeed, in the classical quadratic forms, minimizing that of $H_W\geq0$ over the quadratures of the unpaired modes sets the square in Eq.~\eqref{s:schur} to zero. The coefficient of $c_{\mathrm p}^\dagger\mathsf K_{\mathrm{eff}}c_{\mathrm p}-H_A$ is therefore the Schur complement of $W$ on the paired quadratures. Expressed in creation and annihilation operators, this positivity condition is
\begin{equation}\label{s:cone}
 \begin{pmatrix}\mathsf K_{\mathrm{eff}}-D&-P\\-P&\mathsf K_{\mathrm{eff}}^{\mathsf T}-D\end{pmatrix}\geq0,
\end{equation}
where $Q=\diag(q_j)$, $D_\alpha=\diag(\alpha_j/(1-q_j^2))$,
$D=\diag(D_\alpha,QD_\alpha Q)>0$, and
$P=\left(\begin{smallmatrix}0&D_\alpha Q\\D_\alpha Q&0\end{smallmatrix}\right)$.
In particular $\mathsf K_{\mathrm{eff}}\geq D>0$.

Define the reduced quadratic on the paired modes by
\[
 H=c_{\rm p}^\dagger\mathsf K_{\rm eff}c_{\rm p}-H_A+C_0,
\]
and let $H_{\rm rem}$ be the nonnegative square in Eq.~\eqref{s:schur}. Equations~\eqref{s:parent} and~\eqref{s:schur} give $H_W=H+H_{\rm rem}$. Every annihilator in that square annihilates $|\psi_{\Gamma_*}\rangle$, so the remainder has zero expectation at the Gaussian seed. Equation~\eqref{s:cone} proves that $H$ has a positive-semidefinite real coefficient.

Finally, the first variation of the paired marginal entropy is that of $H_A$. The first variation of $\langle c_{\rm p}^\dagger\mathsf K_{\rm eff}c_{\rm p}\rangle$ at its vacuum is zero. Hence the paired seed $\omega_c$ is stationary for entropy plus $H$, and the geodesic-convex criterion in Sec.~\ref{sec:quadratic-support} makes it a global pure Gaussian minimum. This completes the reduction.

\section{Relative-entropy inequality for joint Gaussian references}
\label{app:contraction}

\begin{proof}[Proof of Proposition~\ref{main:contraction}]
We argue by contradiction, ruling out a negative
variational gap. Thermal replacement and a quartic number penalty
make the minimization regular enough to use its stationarity equations.
A weighted comparison then controls the departure of the output from
thermal equilibrium and identifies its limit as the reference state.
Subsection~\ref{app:contraction-reference} rewrites the hypotheses in thermal coordinates and
introduces the balance defects. Subsection~\ref{app:contraction-variational} regularizes a putative
counterexample and derives the Gibbs equation of its minimizer.
Subsections~\ref{app:contraction-hessian} and~\ref{app:contraction-comparisons} turn second-order optimality and the strict
hypothesis into a quantitative bound on the defects. Subsection~\ref{a:tail}
controls the error from thermal replacement, and Subsection~\ref{app:contraction-limit} evaluates
stationarity along the thermal semigroup to reach the contradiction.

The reference states, Gaussian unitaries and thermal coordinates are
fixed throughout this appendix. Constants may depend on their thermal
occupations; unless stated otherwise, they are uniform in the quartic
penalty introduced below. We use
\(N=\sum_{i=1}^n a_i^\dagger a_i\) and \(K=\Id+N\), where \(n\) is the input mode count. A fixed
Gaussian change of coordinates preserves finite energy.

\subsection{Reference identities and weak thermal balance}\label{app:contraction-reference}

We first express the covariance relation \(\tau=\Phi(\sigma)\) and the
strict quadratic hypothesis in thermal coordinates. The faithful input
and output reference Hamiltonians are
\[
 \begin{aligned}
 H_\sigma&=\sum_i h_i a_i^\dagger a_i,
 &H_\tau&=\sum_j h_{0j}b_j^\dagger b_j,\\
 h_i&=\log\frac{n_{\sigma,i}+1}{n_{\sigma,i}},
 &h_{0j}&=\log\frac{n_{\tau,j}+1}{n_{\tau,j}},
 \end{aligned}
\]
where \(n_{\sigma,i}=\Tr\sigma a_i^\dagger a_i>0\) and
\(n_{\tau,j}=\Tr\tau b_j^\dagger b_j>0\) are thermal occupations.
These Hamiltonians have ground energy zero.
In particular, their scalar ordering terms are included in
\(B=\lambda H_\sigma-\Phi^*H_\tau\).
Write the pullbacks of the retained annihilators as \(\Phi^*(b)=\mathsf U a+\mathsf V a^\dagger\), and set
\[
 \mathsf C=\begin{pmatrix}\mathsf U&\mathsf V\\\overline{\mathsf V}&\overline{\mathsf U}\end{pmatrix},
 \quad \mathsf H=\operatorname{diag}(h_i,h_i),\quad
 \mathsf H_0=\operatorname{diag}(h_{0j},h_{0j}).
\]
The canonical relations are \(\mathsf U\mathsf U^\dagger-\mathsf V\mathsf V^\dagger=\Id\) and
\(\mathsf U\mathsf V^{\mathsf T}=\mathsf V\mathsf U^{\mathsf T}\).  With
\(D_+=\operatorname{diag}(n_{\sigma,i},n_{\sigma,i}+1)\),
\(D_-=\operatorname{diag}(n_{\sigma,i}+1,n_{\sigma,i})\), and the analogous output
matrices \(T_\pm\), the unsymmetrized second-moment identity reads
\begin{equation}\label{a:cov}
                 \mathsf C D_\pm\mathsf C^\dagger=T_\pm.
\end{equation}
These matrices contain \(\langle a_i^\dagger a_i\rangle\) and
\(\langle a_i a_i^\dagger\rangle\), so their entries are unchanged by
our doubled-covariance convention. The identity includes the vanishing
anomalous moments \(\langle b_jb_k\rangle_\tau\). The hypothesis
of Proposition~\ref{main:contraction} is equivalent to
\(\lambda\mathsf H-\mathsf C^\dagger\mathsf H_0\mathsf C>0\). Comparing the nonzero singular values
of \(\mathsf H_0^{1/2}\mathsf C\mathsf H^{-1/2}\) and its adjoint gives some
\(\delta\in(0,\lambda)\) such that
\begin{equation}\label{a:gain}
              \mathsf C\mathsf H^{-1}\mathsf C^\dagger
                         \leq(\lambda-\delta)\mathsf H_0^{-1}.
\end{equation}

To express thermal balance for an arbitrary finite-energy joint state
\(\rho\), let \(\mu_i=\Tr\rho a_i\) and \(a_i'=a_i-\mu_i\Id\).
Its centered balance defects are the complex-linear functionals on
bounded observables defined by
\begin{align}\label{a:defect}
 q_i^+(Z)&=(n_{\sigma,i}+1)\Tr\rho(a_i')^\dagger Z
                   -n_{\sigma,i}\Tr\rho Z(a_i')^\dagger,\\
 q_i^-(Z)&=(n_{\sigma,i}+1)\Tr\rho Za_i'-n_{\sigma,i}\Tr\rho a_i'Z.
 \notag
\end{align}
These weak traces are well defined: \(a_i\rho^{1/2}\) and
\(a_i^\dagger\rho^{1/2}\) are Hilbert--Schmidt, so multiplication
by \(\rho^{1/2}\) makes each required state--ladder product trace
class.  Thus the functionals act on every bounded \(Z\).
The functionals vanish on \(\Id\), and
\(q_i^-(Z)=\overline{q_i^+(Z^\dagger)}\).

Put \(\rho_{\mathrm{out}}=\Phi(\rho)\), \(\mu_{\mathrm{out},j}=\Tr\rho_{\mathrm{out}} b_j\), and define its functionals using the
reference occupations \(n_{\tau,j}\) and these actual means. Then
\begin{equation}\label{a:defectmap}
 \begin{pmatrix}q_{\mathrm{out},+}\\q_{\mathrm{out},-}\end{pmatrix}
       =\overline{\mathsf C}\begin{pmatrix}q_+\\q_-\end{pmatrix}\Phi^*,
\end{equation}
where \(q\Phi^*\) denotes the functional \(Z\mapsto q(\Phi^*Z)\).
For example, expanding the plus functional as
\(\Tr\rho(a_i')^\dagger Z+n_{\sigma,i}\Tr\rho[a_i^\dagger,Z]\)
and using
\[
 [a_i^\dagger,\Phi^*Z]
 =\sum_j\bigl(\mathsf U_{ji}\Phi^*[b_j^\dagger,Z]
       -\overline{\mathsf V}_{ji}\Phi^*[b_j,Z]\bigr)
\]
gives~\eqref{a:defectmap}; \eqref{a:cov} supplies the thermal
coefficient and cancels the anomalous coefficient.  These are weak
commutator identities.  They hold first on number-state cutoffs;
Hilbert--Schmidt factorization passes every state--ladder product to
the trace-norm limit, establishing the identity for arbitrary
finite-energy joint states.

\subsection{A regularized variational problem}\label{app:contraction-variational}

We regularize a hypothetical violation in two stages: thermal replacement
bounds the output logarithm, and a quartic number penalty gives the
minimizer the moments needed in the variational argument. Suppose a
finite-energy state strictly violates the contraction inequality.
Let \(P_L\) project onto total photon number at most \(L\), and set
\(\rho_L=P_L\rho P_L/\Tr(P_L\rho)\). These normalized compressions converge in the sandwiched trace
norm \(\|K^{1/2}(\rho_L-\rho)K^{1/2}\|_1\).  Thus their quadratic
expectations converge.  Input and output entropies converge by
oscillator energy-constrained continuity~\cite{winter}.  Consequently a
counterexample supported on finitely many number states suffices.

Choose a small \(\theta>0\), put \(w=1-\theta\), and mix the channel
output with its thermal reference: \(\rho_0=w\Phi\rho+\theta\tau\).
For \(\zeta>0\), minimize over states with finite fourth moment
\begin{equation}\label{a:objective}
 \begin{split}
 J_\zeta(\rho)={}&S(\rho_0)-S(\tau)
       -w\lambda[S(\rho)-S(\sigma)]\\
 &+w\Tr B(\rho-\sigma)+\zeta\Tr\rho K^4.
 \end{split}
\end{equation}
Continuity at this fixed counterexample permits \(\theta\) to be chosen
small enough that \(\inf J_\zeta\leq-\beta<0\) for all sufficiently
small \(\zeta\).  Additional restrictions on \(\theta\) below
depend only on the fixed reference data and preserve this property.

Strict positivity of the real quadratic coefficient gives
\(B\geq cN-C\Id\).  The input entropy satisfies
\(S(\rho)\leq n g(\Tr\rho N/n)=O(\log(1+\Tr\rho N))\).
Since \(S(\rho_0)\geq0\), a fixed sublevel of \(J_\zeta\) has a
uniform energy bound, independent of \(\zeta\); for fixed
\(\zeta\) it also has a fourth-moment bound.  Compactness of
oscillator energy sublevels, entropy continuity on those sublevels,
and lower semicontinuity of positive energy forms give a minimizer. For fixed \(\zeta\), write \(\rho=\rho_\zeta\) for such a minimizer.

Let \(\mathcal H_0=-\log\rho_0\).  Operator monotonicity of the logarithm
and \(\rho_0\geq\theta\tau\) imply the form bound
\begin{equation}\label{a:outlog}
              0\leq \mathcal H_0\leq-\log\tau-\log\theta\leq C_\theta K_{\rm out},
\end{equation}
where \(K_{\rm out}=\Id+\sum_jb_j^\dagger b_j\) is the output number weight. Pullback of the positive
form, defined by increasing bounded spectral truncations, obeys
\(0\leq\Phi^*\mathcal H_0\leq C_\theta K\). For a competitor \(\widetilde\rho\), set
\(\widetilde\rho_0=w\Phi(\widetilde\rho)+\theta\tau\), and replace its output entropy
in~\eqref{a:objective} by \(\Tr\widetilde\rho_0\mathcal H_0\).
Relative-entropy positivity makes this a majorant, with equality at
the minimizer.  Hence that minimizer also minimizes the resulting
input Gibbs functional~\cite{bertasutterwalter}.

\begin{lemma}[Quartic Gibbs regularity]\label{a:regularity}
Let \(K=\Id+N\), \(c>0\), and let \(T\geq0\) be bounded.
The closed form
\[
 A=cK^4+K^{1/2}TK^{1/2},\qquad
 \mathcal D(A^{1/2})=\mathcal D(K^2),
\]
has compact resolvent and \(0<\Tr e^{-A}<\infty\).
Its Gibbs state is faithful and has finite seventh number moment.
It is the unique minimizer of \(\Tr\omega A-S(\omega)\) among
states with finite fourth moment.
\end{lemma}
\begin{proof}[Proof of Lemma~\ref{a:regularity}]
The form is \(\|F^{1/2}K^2\psi\|^2\), where
\(F=c\Id+K^{-3/2}TK^{-3/2}\) is bounded and boundedly invertible.
Thus it is closed and its inverse is the compact operator
\(K^{-2}F^{-1}K^{-2}\).  The min--max principle applied to
\(A\geq cK^4\) bounds its eigenvalue counting function by that
of \(cK^4\), which grows polynomially.  For an eigenvector
\(A\psi=u\psi\), the number-coordinate form equation gives
\[
 cK^{7/2}\psi=uK^{-1/2}\psi-TK^{1/2}\psi.
\]
The right side is square summable because
\(\|K^2\psi\|^2\leq u/c\).  It follows that
\(\psi\in\mathcal D(K^{7/2})\), with its graph norm bounded by
a polynomial in \(u\).  Summation against \(e^{-u}\) proves the
trace and seventh-moment assertions.  Finally,
\(\Tr\omega A-S(\omega)=D(\omega\Vert e^{-A}/\Tr e^{-A})
-\log\Tr e^{-A}\); all terms are finite on the stated domain.
\end{proof}

After a scalar shift, the Gibbs Hamiltonian of the preceding majorant
has exactly the form in Lemma~\ref{a:regularity}.  Therefore the
minimizer is faithful and satisfies the exact form identity
\begin{equation}\label{a:gibbs}
 w\lambda \mathcal H_\rho=w\Phi^*\mathcal H_0+wB+\zeta K^4+c_\zeta\Id,
 \qquad \mathcal H_\rho=-\log\rho,
\end{equation}
together with \(\Tr\rho K^7<\infty\) and
\(0\leq \mathcal H_\rho\leq C_\zeta K^4\).
These moment bounds hold at each fixed \(\zeta\); only the first
moment is uniform as \(\zeta\downarrow0\).

The output and the replaced output also have finite seventh moments.
Let \(\mathcal U\) be the Gaussian unitary in \(\Phi\), and let
\(K_{\rm tot}\) denote one plus total photon number before discarding modes.
Conjugating a number power by this unitary expands
it into finitely many ladder monomials.  Splitting each monomial into
two factors and using the number-shift bounds proves
\(\mathcal U^\dagger K_{\rm tot}^k\mathcal U\leq C_kK_{\rm tot}^k\) as forms for every
integer \(k\geq1\).  Partial trace then transfers the moment bound.
The same shift estimates show that
\(K_{\rm tot}^s\mathcal UK_{\rm tot}^{-s}\) is bounded for integer \(s\),
which will transfer sandwiched trace convergence below.

\subsection{The Hessian bound on a common centered core}\label{app:contraction-hessian}

The second variation at a minimizer is nonnegative. We express this
condition as a norm bound for output observables pulled back to the input.
For a faithful state \(\omega=\operatorname{diag}(p_l)\), define the
logarithmic mean and its associated weighted norm by
\[
 \begin{gathered}
 \ell(p,q)=\frac{p-q}{\log p-\log q},\qquad \ell(p,p)=p,\\
 \|Z\|_{\ell,\omega}^2=\sum_{l,k}\ell(p_l,p_k)|Z_{lk}|^2.
 \end{gathered}
\]
We use the centered core consisting of operators
\(Z-\Tr(\omega Z)\Id\), where \(Z\) is supported on finitely many
eigenvectors of \(\omega\). The subtraction leaves a constant diagonal
tail outside that finite block.

The Gibbs identity gives, for finite-fourth-moment competitors,
\[
 J_\zeta(\widetilde\rho)-J_\zeta(\rho)
 =w\lambda D(\widetilde\rho\Vert\rho)
                  -D(\widetilde\rho_0\Vert\rho_0)\geq0.
\]
We claim that every bounded output test obeys
\begin{equation}\label{a:hessian}
 w\|\Phi^*Z-\Tr\rho_{\mathrm{out}} Z\Id\|_{\ell,\rho}^2
                       \leq\lambda\|Z\|_{\ell,\rho_0}^2.
\end{equation}
We derive this bound first on finite blocks and then extend it by
duality~\cite{lesniewskiruskai}. Let \(X=X^\dagger\) be trace zero and
supported on finitely many input eigenvectors.  Both signs of
\(\rho+tX\) are admissible for small \(t\), and
\[
 D(\rho+tX\Vert\rho)
   =\tfrac12t^2 Q_\rho(X)+o(t^2),\qquad
 Q_\rho(X)=\sum_{l,k}\frac{|X_{lk}|^2}{\ell(p_l,p_k)}.
\]
Let \(Z\) be a Hermitian output-core test, centered so that
\(\Tr\rho_0Z=0\). The output of \(\widetilde\rho=\rho+tX\) is
\(\widetilde\rho_0=\rho_0+wt\Phi(X)\). The Gibbs variational inequality
\(D(\widetilde\rho_0\Vert\rho_0)\geq\Tr\widetilde\rho_0Y-\log\Tr e^{\log\rho_0+Y}\),
with \(Y=sZ\), and the finite-block expansion
\(\log\Tr e^{\log\rho_0+sZ}=s^2\|Z\|_{\ell,\rho_0}^2/2+o(s^2)\), combined
with the preceding display, give
\[
 \tfrac12w\lambda t^2Q_\rho(X)+o(t^2)
 \geq swt\Tr(X\Phi^*Z)-\tfrac12s^2\|Z\|_{\ell,\rho_0}^2+o(s^2).
\]
Choosing \(s=wt\Tr(X\Phi^*Z)/\|Z\|_{\ell,\rho_0}^2\) and letting
\(t\to0\) yields
\[
 w|\Tr X\Phi^*Z|^2
       \leq\lambda Q_\rho(X)\|Z\|_{\ell,\rho_0}^2.
\]
Finite trace-zero eigenblocks are dense in the trace-zero
\(Q_\rho\)-space: compress the weighted coordinate sum and correct
the residual trace at one fixed diagonal entry with \(p_l>0\).
Trace is continuous there by Cauchy--Schwarz and \(\sum_lp_l=1\).
Taking the supremum over \(X\) therefore gives the centered
logarithmic-mean norm on the left of~\eqref{a:hessian}.
Real and imaginary parts give complex tests.

For later extensions use the arithmetic norm
\begin{equation}\label{a:arithmetic}
 \|Z\|_{\mathrm a,\omega}^2
   =\sum_{l,k}(p_l+p_k)|Z_{lk}|^2
   =\Tr\omega(ZZ^\dagger+Z^\dagger Z).
\end{equation}
For every bounded \(Z\), its centered eigenblock compressions
converge to \(Z-\Tr(\omega Z)\Id\) in this norm.  The Schwarz inequalities for \(\Phi^*\), followed by
the variance identity for \(\rho_0=w\rho_{\mathrm{out}}+\theta\tau\), give the
same convergence for their actual centered pullbacks.  Since
\(\ell(p,q)\leq(p+q)/2\), this proves~\eqref{a:hessian} for every
bounded \(Z\), and identifies its continuous extension with the
physical pullback.  Each additional metric below is bounded above
by a fixed multiple of the arithmetic metric. The same core is
dense in each centered weighted space: truncate the weighted
coordinate sum, then subtract the residual mean. This correction
converges because every diagonal weight is a fixed positive
multiple of \(p_l\).
Thus all endpoint comparisons use the same operator on the same core.

\subsection{Four comparisons and a strict dual remainder}\label{app:contraction-comparisons}

We combine the strict matrix bound~\eqref{a:gain} with the Hessian
bound~\eqref{a:hessian}. These give four weighted norm comparisons,
whose interpolation yields the norms of the thermal balance defects.
The strict gap will remain as a quantitative bound on those defects.

Associate plus and minus components with each mode. For an output vector
\(Z\), centered componentwise against \(\rho_0\), define
\[
                  LZ=\sqrt w\,\mathsf C^\dagger
                      (\Phi^*Z-\Tr\rho_{\mathrm{out}} Z\Id).
\]
Every image component is centered against \(\rho\).
For each input or output mode, write \(h\) for its reference inverse temperature and \(\bar n=(e^h-1)^{-1}\) for its occupation. For a coordinate \((l,k)\) in the eigenbasis of \(\rho\) or \(\rho_0\), set
\[
 x=\tfrac12\log(p_l/p_k),\qquad m=1/h,
 \qquad y=\mp h/2
\]
on the plus/minus mode, respectively, and use base measure
\(\ell(p_l,p_k)\).  Let \(f(x)=x/\sinh x\), with \(f(0)=1\).
The first pair of four weights
\begin{equation}\label{a:four}
                  mf(x)e^{\pm x},\qquad mf(y)e^{\pm y}
\end{equation}
becomes \(p_l/h,p_k/h\) after multiplication by the base measure.
These two metrics compare under \(L\) with constant \(\lambda-\delta\).
Indeed, put \(Y_\beta=\Phi^*Z_\beta-\Tr(\rho_{\mathrm{out}}Z_\beta)\Id\) for the
output components \(\beta\). The Gram matrix
\(G_{\beta\beta'}=\Tr\rho\,Y_\beta Y_{\beta'}^\dagger\) is positive
semidefinite, so~\eqref{a:gain} gives
\[
 \sum_\alpha h_\alpha^{-1}\Tr\rho\,(LZ)_\alpha(LZ)_\alpha^\dagger
 \leq(\lambda-\delta)w\sum_\beta h_{0\beta}^{-1}G_{\beta\beta}.
\]
With \(\operatorname{Var}_\omega(Z)=\Tr\omega ZZ^\dagger-|\Tr\omega Z|^2\),
the Kadison--Schwarz inequality gives
\(G_{\beta\beta}\leq\operatorname{Var}_{\rho_{\mathrm{out}}}(Z_\beta)\), and
concavity of the variance under \(\rho_0=w\rho_{\mathrm{out}}+\theta\tau\) gives
\(w\operatorname{Var}_{\rho_{\mathrm{out}}}(Z_\beta)\leq\operatorname{Var}_{\rho_0}(Z_\beta)\).
The same argument applies with \(ZZ^\dagger\) replaced by \(Z^\dagger Z\).
The second pair becomes \(\bar n\ell,(\bar n+1)\ell\), with the order
exchanged on the minus sign. Since \(\mathsf C D_\pm\mathsf C^\dagger=T_\pm\)
is diagonal,
\[
 \sum_\alpha(D_\pm)_\alpha\|(LZ)_\alpha\|_{\ell,\rho}^2
 =w\sum_\beta(T_\pm)_\beta\|Y_\beta\|_{\ell,\rho}^2,
\]
and~\eqref{a:hessian}, applied to each output component, gives
comparison constant \(\lambda\) for each of these two metrics. These
comparisons use the Hessian of the full joint input state.

To combine these four comparisons, we apply Lemma~\ref{i:four}.
Its test-space hypothesis holds because each source-core test has finitely
many values of \((m,x,y)\), and a function of this tuple multiplies
all diagonal entries, including the identity tail, by the same
constant within each mode/sign.  It therefore preserves the core and
its centering.  The resulting metrics, including their base measures,
are
\begin{equation}\label{a:metric}
 \begin{split}
 M_+(p_l,p_k)&=\ell(\bar n p_l,(\bar n+1)p_k),\\
 M_-(p_l,p_k)&=\ell((\bar n+1)p_l,\bar n p_k).
 \end{split}
\end{equation}
The subscripts \(M_{i,\pm}\) and \(M_{0j,\pm}\) below use the input occupations \(n_{\sigma,i}\) and output occupations \(n_{\tau,j}\), respectively. Input and output metric weights use the eigenvalues of \(\rho\) and \(\rho_0\), respectively. Vector norms sum over modes, signs, and state-eigenbasis coordinates.
The lemma also retains the strict source remainder
\begin{equation}\label{a:gap}
 \|LZ\|_{M,\rho}^2
       \leq\lambda\|Z\|_{M,\rho_0}^2-\|Z\|_\Delta^2,
 \qquad 0<\Delta\leq\delta M,
\end{equation}
where \(\Delta\) is the output weight in~\eqref{i:gap}, multiplied
by \(\ell(p_l,p_k)\).
For each fixed reference temperature, its explicit formula gives
\begin{equation}\label{a:arithmeticgap}
 \Delta\geq\frac{c(p_l+p_k)}{1+\log^2(p_l/p_k)},
 \qquad \frac{\Delta}{M^2}\geq\frac{c'}{p_l+p_k}.
\end{equation}
To obtain both bounds, set \(b(x)=x\coth x\), with \(b(0)=1\), and use the
minimizer \(\kappa\) in~\eqref{i:minimizer}.  With \(y\) fixed,
the auxiliary weight \(E\) in Eq.~\eqref{i:E} satisfies \(E/m=O(|x|^{-1})\), and
\[
 \begin{array}{ll}
 \kappa=1-[b(y)+y]/x+O(x^{-2}),&x\longrightarrow+\infty,\\
 \kappa=-1+[b(y)-y]/|x|+O(x^{-2}),&x\longrightarrow-\infty.
 \end{array}
\]
Since \(b(y)\pm y>0\), substitution gives the claimed positive
endpoint bounds; continuity supplies the bounds on compact intervals.
Finitely many modes allow common constants.  The ratio \(\Delta/M\)
may approach zero.

We next justify applying the dual inequality to the weak defects.
Write \(\omega=\sum_kp_k|\psi_k\rangle\langle\psi_k|\), \(\mathcal H_\omega=-\log\omega\), and let \(a'=a-\Tr(\omega a)\Id\) be the centered ladder operator under consideration.
Let \(K_\omega\) be one plus total photon number on the system of \(\omega\). For a state satisfying \(\Tr\omega K_\omega^7<\infty\) and
\(\mathcal H_\omega\leq CK_\omega^4\), a plus-defect has coordinate coefficients
of modulus
\( |[(\bar n+1)p_k-\bar n p_l]a'_{lk}|\).  Hence
\[
 \frac{|q_{lk}|^2}{M_{+,lk}}
 \leq C_{\bar n}(p_l+p_k)(1-\log p_l-\log p_k)|a'_{lk}|^2.
\]
The crossed logarithmic sum, for example, is bounded by
\[
 \sum_{l,k}p_k(-\log p_l)|a'_{lk}|^2
 =\sum_kp_k\langle a'\psi_k,\mathcal H_\omega a'\psi_k\rangle
 \leq C\Tr\omega K_\omega^5.
\]
For the same-index sum, use
\[
 ( -\log p_k)\|a'\psi_k\|^2
 \leq C\langle K_\omega^4\rangle_{\psi_k}\langle K_\omega\rangle_{\psi_k}
 \leq C\langle K_\omega^5\rangle_{\psi_k}.
\]
The last inequality is the nonnegative covariance of the increasing
functions \(u^4\) and \(u\) in the spectral distribution of \(K_\omega\).
Interchanging indices gives the creation-operator terms.  Thus all
defects have finite \(M^{-1}\) norm.  They also have finite arithmetic
dual norm, already from finite energy, because
\( |q_{lk}|^2/(p_l+p_k)\leq C_{\bar n}(p_l+p_k)|a'_{lk}|^2\).

Consequently their pairings with a bounded physical pullback agree
with their completed \(M\)-pairings: approximate that pullback by
centered eigenblocks in the arithmetic norm, which also gives
\(M\)-convergence, and use both dual continuities.  This identifies
the functional transpose \(L' q\), defined by
\((L'q)(Z)=q(LZ)\), with the actual weak functional in
\eqref{a:defectmap}.  Moreover, \(M_\pm(p,p)=p/h\), so the identity is
orthogonal to centered tests.  Since \(q(\Id)=0\), restriction to
the centered space does not alter the displayed dual norms.

Put
\[
 \begin{gathered}
 Q_{\rm in}=\sum_i\|q_i^+\|_{M_{i,+}^{-1}}^2,\qquad
 Q_{\rm main}=\sum_j\|q_{{\rm main},j}^+\|_{M_{0j,+}^{-1}}^2,\\
 q_{\rm main}=\sqrt w\,L'q=wq_{\mathrm{out}}.
 \end{gathered}
\]
Transpose of the eigenvalue indices makes the minus and plus norms
equal.  Applying the dual part of Lemma~\ref{i:four} to both signs
therefore yields
\begin{equation}\label{a:D}
 Q_{\rm gap}:=w\lambda Q_{\rm in}-Q_{\rm main}
   \geq\frac1{2\lambda}\sum_{\alpha\text{ in both signs}}
             |q_{{\rm main},\alpha}|^2
                         \frac{\Delta_\alpha}{M_\alpha^2}\geq0.
\end{equation}
In particular, for every bounded componentwise output test,
\begin{equation}\label{a:weaktest}
 |q_{{\rm main},j}^{\pm}(Z)|^2
       \leq C Q_{\rm gap}\,\Tr\rho_0(ZZ^\dagger+Z^\dagger Z)
       \leq 2CQ_{\rm gap}\|Z\|^2.
\end{equation}
The constants are uniform in the minimizing state and in \(\zeta\).

\subsection{The replacement error and its logarithmic tail}
\label{a:tail}

We now relate the balance defects of \(\rho_0\), which enters the
variational functional, to those of the actual channel output \(\rho_{\mathrm{out}}\).
Let \(q_0\) denote the functionals of \(\rho_0\) with reference occupations \(n_{\tau,j}\).
Centering the mixture gives the exact identity
\begin{equation}\label{a:tailidentity}
 \begin{split}
 q_0&=q_{\rm main}+q_{\rm tail},\\
 q_{{\rm tail},j}^+(Z)
   &=\theta w\overline\mu_{\mathrm{out},j}\Tr[(\rho_{\mathrm{out}}-\tau)Z].
 \end{split}
\end{equation}
The minus expression has the conjugate mean.  In the \(\rho_0\)
eigenbasis, the absolute values of these coefficients are
\(\theta|\mu_{\mathrm{out},j}|\,| (\rho_0-\tau)_{lk}|\).
By the first inequality in~\eqref{a:arithmeticgap},
\begin{equation}\label{a:tailinitial}
 \|q_{\rm tail}\|_{\Delta^{-1}}^2
 \leq C\theta^2\|\mu_{\mathrm{out}}\|^2
 \sum_{l,k}\frac{|(\rho_0-\tau)_{lk}|^2
              [1+\log^2(p_l/p_k)]}{p_l+p_k}.
\end{equation}
Order gives \(\tau\rho_0^{-1}\tau\leq\tau/\theta\) as a form.
It bounds the arithmetic sum by \(O(\theta^{-1})\).
The logarithmic sum contains only the off-diagonal entries of \(\tau\);
using \((u-v)^2\leq2u^2+2v^2\) bounds it by
\(4\Tr\tau \mathcal H_0^2/\theta\).  We control this logarithmic second
moment separately as follows.

Let \((\tau_i,u_i)\) and \((p_k,v_k)\) be spectral data of \(\tau\) and \(\rho_0\), respectively, and use the probability distribution
\(\tau_i|\langle u_i,v_k\rangle|^2\), as in
Ref.~\cite{nussbaum}.  For
\(Y=\log\tau_i-\log p_k\) and \(0<s<1\), inverse order and
Schatten H\"older give
\begin{equation}\label{a:petz}
 \mathbb E e^{sY}=\Tr\tau^{1+s}\rho_0^{-s}\leq\theta^{-s},
 \qquad
 \mathbb E e^{-sY}=\Tr\tau^{1-s}\rho_0^s\leq1.
\end{equation}
For the negative power first add a positive scalar to the operators
and then take the monotone limit; all spectral sums are nonnegative.
Since \(Y^2\leq C_s(e^{sY}+e^{-sY})\),
\begin{equation}\label{a:logmoment}
 \Tr\tau \mathcal H_0^2
 \leq2\Tr\tau(\log\tau)^2+2C_s(\theta^{-s}+1).
\end{equation}
The reference has finite logarithmic second moment.  Taking \(s=1/2\)
in~\eqref{a:tailinitial}--\eqref{a:logmoment} proves the uniform bound
\begin{equation}\label{a:tailbound}
             \|q_{\rm tail}\|_{\Delta^{-1}}^2
                       \leq C\theta^{1/2}\|\mu_{\mathrm{out}}\|^2.
\end{equation}
Here \(C\) is independent of both \(0<\theta<1/2\) and \(\zeta\).

\subsection{Stationarity and the thermal limit}\label{app:contraction-limit}

We evaluate stationarity along thermal Ornstein--Uhlenbeck evolution
with invariant state \(\sigma\). Its generator is
\[
 \begin{gathered}
 \mathcal L_\sigma\rho
  =\sum_i\bigl[(n_{\sigma,i}+1)\mathcal D[a_i]\rho
                         +n_{\sigma,i}\mathcal D[a_i^\dagger]\rho\bigr],\\
 \mathcal D[a]\rho=a\rho a^\dagger-\tfrac12\{a^\dagger a,\rho\}.
 \end{gathered}
\]
The output generator \(\mathcal L_\tau\) is defined analogously.
Their linear drifts agree,
and their diffusion covariances transform by~\eqref{a:cov}; hence
\begin{equation}\label{a:OU}
                    \Phi\mathcal L_\sigma\rho
                         =\mathcal L_\tau\Phi\rho.
\end{equation}
For a finite number cutoff, this follows by applying the annihilator
transformation to the characteristic function
\(\widehat\chi_\rho(z)=\Tr\rho\exp[\sum_i(z_i a_i^\dagger-\bar z_i a_i)]\),
where \(z\) is a vector of complex amplitudes:
\[
 \widehat\chi_{\mathcal L_\sigma\rho}(z)
 =-\tfrac12(z\cdot\partial_z+\overline z\cdot\partial_{\overline z})
           \widehat\chi_\rho(z)
  -\sum_i(n_{\sigma,i}+\tfrac12)|z_i|^2\widehat\chi_\rho(z).
\]
The seventh-moment bound extends this identity from number cutoffs
to the minimizing state. With
\(A_\rho=K^3\rho K^3\), every term of
\(K^2\mathcal L_\sigma\rho K^2\) is a bounded-factor product
around the trace-class operator \(A_\rho\).  For example,
\[
 K^2a\rho a^\dagger K^2
 =(K^2aK^{-3})A_\rho(K^{-3}a^\dagger K^2).
\]
The anticommutator terms satisfy the same bound.  Number compressions
give \(K^3P_L\rho P_LK^3=P_LA_\rho P_L\to A_\rho\) in trace norm.
The fixed-unitary bounds established above and partial trace transfer
this weighted convergence to the output. Equality of the Weyl
characteristic functions then gives
\eqref{a:OU} for the minimizing state, with trace-zero generator
expressions.

The bounded form \(K^{-2}\mathcal H_\rho K^{-2}\) defines its logarithmic
pairing with \(K^2\mathcal L_\sigma\rho K^2\). The same input pairing
applies to \(\Phi^*\mathcal H_0\), \(B\), and \(K^4\), while the
output pairing with \(\mathcal H_0\) uses \(K_{\rm out}\). To pull back
the output logarithm, take the positive and negative parts of the
sandwiched input generator and conjugate them by \(K^{-2}\). Each resulting positive
operator has finite fourth moment, so the positive form pullback
identity applies.  Subtraction gives
\begin{equation}\label{a:logslicing}
 \Tr\mathcal L_\tau\rho_0\mathcal H_0
             =w\Tr\mathcal L_\sigma\rho\,\Phi^*\mathcal H_0.
\end{equation}
This identity uses the weighted trace pairing defined above.

The input logarithmic pairing admits the identity
\begin{equation}\label{a:production}
 \Tr\mathcal L_\sigma\rho \mathcal H_\rho
      =Q_{\rm in}-\sum_i h_i(E_i-|\mu_i|^2-n_{\sigma,i}),
 \qquad E_i=\Tr\rho a_i^\dagger a_i.
\end{equation}
To derive Eq.~\eqref{a:production}, work in the eigenbasis of the
joint state and put
\(k_{lk}=(n_{\sigma,i}+1)p_k-n_{\sigma,i}p_l\) and
\(v_{lk}=\log(p_l/p_k)/h_i\).  The defect norm divided by \(h_i\)
has summands \((1-v_{lk})k_{lk}|a'_{i,lk}|^2\), whereas the
logarithmic pairing divided by \(h_i\) has summands
\(-v_{lk}k_{lk}|a'_{i,lk}|^2\).
Their difference sums to \(E_i-|\mu_i|^2-n_{\sigma,i}\) by the canonical
commutation relation.  Centering changes only diagonal ladder
entries, where \(v_{ll}=0\).  The logarithmic sums are absolutely
summable by the estimates above.  For the anticommutator expansion,
every state eigenvector belongs to \(\mathcal D(K^{7/2})\), so
\(a_i^\dagger a_i\psi\in\mathcal D(K^2)\); finite eigenstate
compressions converge in the \(K^3\)-sandwiched trace norm.  Thus
the same weighted pairing justifies the displayed spectral calculation.

Pair~\eqref{a:gibbs} with \(\mathcal L_\sigma\rho\), and use
\eqref{a:OU}, \eqref{a:logslicing}, and~\eqref{a:production}.
The centered energy terms cancel because
\(B=\lambda H_\sigma-\Phi^*H_\tau\).  The result is
\begin{equation}\label{a:stationarity}
 \begin{split}
 0&=Q_0-w\lambda Q_{\rm in}-M_{\rm mean}+\zeta P_4,\\
 M_{\rm mean}
   &=w\left[\lambda\sum_i h_i|\mu_i|^2
                    -w\sum_jh_{0j}|\mu_{\mathrm{out},j}|^2\right],
 \end{split}
\end{equation}
where \(Q_0=\sum_j\|q_{0j}^+\|_{M_{0j,+}^{-1}}^2\) and
\(P_4=\Tr\mathcal L_\sigma\rho K^4\).
Strict quadratic positivity implies
\(M_{\rm mean}\geq wc\|\mu\|^2\), while
\(\|\mu_{\mathrm{out}}\|\leq C\|\mu\|\).
Also \(P_4\leq C\), uniformly in the state: the number-coordinate
formula for \(\mathcal L_\sigma^*K^4\) has leading term
\(-4N^4\) and remainder bounded above by \(C(1+N)^3\).

Expand \(Q_0\) by~\eqref{a:tailidentity}.
The strict dual bound~\eqref{a:D}, Cauchy--Schwarz with weights
\(\Delta/M^2\) and \(\Delta^{-1}\), and~\eqref{a:tailbound}
give
\[
 |\text{cross term}|\leq C\theta^{1/4}\sqrt{Q_{\rm gap}}\,\|\mu\|,
 \qquad
 \|q_{\rm tail}\|_{M^{-1}}^2
                       \leq C\theta^{1/2}\|\mu\|^2.
\]
The second bound uses \(\Delta\leq\lambda M\).
The constants in these bounds are independent of \(0<\theta<1/2\)
and \(\zeta\).
Choose \(\theta<1/2\) sufficiently small once, then apply Young's
inequality in~\eqref{a:stationarity}.  This yields
\begin{equation}\label{a:absorption}
                 \tfrac12Q_{\rm gap}+\tfrac12wc\|\mu\|^2\leq C\zeta.
\end{equation}

With the references, strict gap and replacement parameter \(\theta\)
fixed, the uniform energy bound supplies a sequence
\(\rho_\zeta\to\widehat\rho\) in trace norm as
\(\zeta\downarrow0\), with \(\Phi\rho_\zeta\to\widehat\rho_{\mathrm{out}}:=\Phi(\widehat\rho)\).
Equation~\eqref{a:absorption} gives vanishing means.
Equation~\eqref{a:weaktest}, applied to every fixed finite-number
output matrix, gives \(q_{{\rm main},j}^\pm(Z)\to0\).
For such a test both ladder products are bounded finite-rank
operators; hence their weak traces pass to the trace-norm limit.
Using \(q_{\rm main}=wq_{\mathrm{out}}\) and the vanishing means, the
limit satisfies the thermal balance relation on all number-state
matrix elements:
\[
           (n_{\tau,j}+1)b_j\widehat\rho_{\mathrm{out}}
                     =n_{\tau,j}\widehat\rho_{\mathrm{out}} b_j.
\]
Let \(\boldsymbol n,\boldsymbol m\) be output number multiindices and \(e_j\) the \(j\)th unit vector. With \(s_j=n_{\tau,j}/(n_{\tau,j}+1)\), the balance relation reads
\(\sqrt{n_j+1}\,(\widehat\rho_{\mathrm{out}})_{\boldsymbol n+e_j,\boldsymbol m}
 =s_j\sqrt{m_j}\,(\widehat\rho_{\mathrm{out}})_{\boldsymbol n,\boldsymbol m-e_j}\).
Iteration reaches a zero boundary element whenever \(n_j>m_j\);
self-adjointness handles \(n_j<m_j\).  Thus the state is diagonal,
and the diagonal recurrence and trace one give
\(\widehat\rho_{\mathrm{out}}=\tau\).

Let \(J_0\) denote the objective in Eq.~\eqref{a:objective} with \(\zeta=0\). Entropy continuity on the common energy sublevel and lower
semicontinuity of \(B+C\Id\) give
\(J_0(\widehat\rho)\leq\liminf J_\zeta(\rho_\zeta)\leq-\beta\).
But its output is \(\tau\), so the exact Gibbs-reference identities
give \(J_0(\widehat\rho)=w\lambda D(\widehat\rho\Vert\sigma)\geq0\).
This contradiction completes the proof of Proposition~\ref{main:contraction}.
\end{proof}

\section{Four-weight interpolation with a strict remainder}
\label{app:interpolation}

The following lemma strengthens the diagonal-form interpolation theorem
of Ref.~\cite[Theorem 3.1]{interpolation} by retaining a quantitative gap
from one pair of endpoint bounds. After duality, this gap controls the
thermal balance defects in Appendix~\ref{app:contraction}.

\begin{lemma}[Strict four-weight comparison]\label{i:four}
Put \(f(x)=x/\sinh x\), with \(f(0)=1\). At each source and target coordinate fix \(m>0\) and real \(x,y\),
with any fixed positive coordinate measure.  In coordinate formulas
for dual norms, that measure is absorbed into the weights.
Let \(\mathcal T\) be a source test space such that each element
is supported on coordinates at which \((m,x,y)\) takes finitely
many values, and suppose \(\mathcal T\) is invariant under
multiplication by functions of this tuple.
The subscripts ``out'' and ``in'' label the source and target,
respectively, as in the observable pullback in Appendix~\ref{app:contraction}.
We say that a linear map \(L\) from \(\mathcal T\) into the target coordinate space compares a positive diagonal weight \(\omega\) with constant \(\Lambda\) if \(\|LZ\|_{\omega,\rm in}^2\leq\Lambda\|Z\|_{\omega,\rm out}^2\) for every \(Z\in\mathcal T\).
Suppose \(L\) compares \(mf(x)e^{\pm x}\) with constant
\(\lambda-\delta\), and \(mf(y)e^{\pm y}\) with constant
\(\lambda\), where \(0<\delta<\lambda\).
Then, for \(M=mf(x)f(y)/f(x+y)\),
\[
        \|LZ\|_{M,\rm in}^2
        \leq\lambda\|Z\|_{M,\rm out}^2-\|Z\|_\Delta^2,
\]
with a strictly positive source weight \(\Delta\leq\delta M\)
given in~\eqref{i:gap}.  The continuous extension between the
completed form spaces has the dual estimate~\eqref{a:dualgap}.
\end{lemma}

\begin{proof}
Positive sums, pointwise limits and parallel sums preserve a common
comparison constant. The parallel sum \(X:Y=XY/(X+Y)\) satisfies
\[
 (X:Y)|z|^2=\min_{u+v=z}(X|u|^2+Y|v|^2),\qquad
 (u,v)=\frac{(Y,X)}{X+Y}z.
\]
The source minimizing split belongs to \(\mathcal T\); on the
target one uses only the lower bound by the minimum.
For pointwise limits, a source test has finitely many parameter
values, and Fatou's lemma applies to target coordinate sums.
These arguments work with unbounded diagonal forms.
Positive quadrature sums and parallel sums give the two logarithmic-mean
operations
\[
 \begin{aligned}
 \ell(X,Y)^{-1}&=\int_0^1\frac{du}{(1-u)X+uY},\\
 \frac{XY}{\ell(X,Y)}&=\int_0^1\frac{du}{u/X+(1-u)/Y}.
 \end{aligned}
\]
Applied to the opposite-exponent pairs, they give comparisons for
\(m,ms,mt\), where \(s=f(x)^2\), \(t=f(y)^2\).

The remaining scalar ingredient is the Stieltjes representation proved
in Ref.~\cite[Lemma 3.2]{interpolation}. The function defined by
\(\mathcal B(f(x)^2)=b(x)=x\coth x\), with \(b(0)=1\), extends across the positive
axis and satisfies
\begin{equation}\label{i:stieltjes}
 -\mathcal B'(s)=c+\int_{[0,\infty)}\frac{d\nu(r)}{s+r}>0,
 \quad c\geq0,\quad \int\frac{d\nu(r)}{1+r}<\infty.
\end{equation}
This representation supplies the positive combinations needed below.
Define
\begin{equation}\label{i:E}
 \frac1E=\frac2m\int_0^1[-\mathcal B'(us+(1-u)t)]\,du,
 \qquad
 E=\frac{m(t-s)}{2[\mathcal B(s)-\mathcal B(t)]},
\end{equation}
with \(E=-m/[2\mathcal B'(s)]\) at \(s=t\).
The Stieltjes formula constructs \(E\) from \(m,ms,mt\) by
positive combinations and parallel sums: each reciprocal integrand
has denominator \(u(ms)+(1-u)(mt)+rm\), together with a possible
multiple of \(1/m\).  Truncate the positive measure and use positive
quadrature sums, then the preceding limit argument.  Thus \(E\) also
compares with constant \(\lambda\).

For scalars \(\xi,\phi\), put
\begin{equation}\label{i:quadratic}
 \begin{split}
 Q(\xi,\phi)={}&\tfrac14\sum_\pm mf(x)e^{\pm x}|\xi\mp\phi|^2\\
 &+\tfrac14\sum_\pm mf(y)e^{\pm y}|\xi\pm\phi|^2+E|\phi|^2.
 \end{split}
\end{equation}
With \(a=b(x)+b(y)\), \(e=E/m\), direct minimization gives
\begin{equation}\label{i:minimizer}
 \begin{gathered}
 \phi=\kappa\xi,\qquad \kappa=\frac{x-y}{a+2e},\\
 \min_\phi Q=\frac m2\left[a-\frac{(x-y)^2}{a+2e}\right]|\xi|^2
             =M|\xi|^2.
 \end{gathered}
\end{equation}
The last equality follows from \(b(x)^2=x^2+f(x)^2\) and the
addition formula for \(\sinh\). For \(x^2\neq y^2\), the same identity and~\eqref{i:E} give
\(a+2e=(x^2-y^2)/[b(x)-b(y)]\). Consequently, by continuity,
\(\kappa=[b(x)-b(y)]/(x+y)\) whenever \(x+y\neq0\).  At \(x=y\) the minimum is
\(mb(x)|\xi|^2\); at \(x=-y\) it is \(mf(x)^2|\xi|^2\).
The zero cases follow by continuity.

Use the source minimizing split \(\phi=\kappa Z\), which is
admissible by tuple invariance.  Apply the first pair of comparisons
with constant \(\lambda-\delta\), and the remaining comparisons
with constant \(\lambda\).  On the target minimize freely.
The retained source contribution is
\begin{equation}\label{i:gap}
 \Delta=\frac{\delta mf(x)}4
       \left[e^x(1-\kappa)^2+e^{-x}(1+\kappa)^2\right].
\end{equation}
It is strictly positive, and \(\Delta\leq\delta M\) because it
is \(\delta\) times the first positive contribution to the
minimum in~\eqref{i:quadratic}.  The minimizing multiplier is
applied only on the source space.

To obtain the dual estimate, let \(\mathcal A=M_{\rm in}^{1/2}
LM_{\rm out}^{-1/2}\) on the completed spaces, and
\(\mathcal R=M_{\rm out}^{-1/2}\Delta M_{\rm out}^{-1/2}\).
The interpolated bound reads \(\mathcal A^\dagger\mathcal A\leq\lambda\Id-\mathcal R\),
where \(0\leq\mathcal R\leq\delta\Id\). Tuple invariance ensures that
the multiplier preserves the closure of \(\mathcal T\).
For an input functional \(q\), set
\(v=M_{\rm in}^{-1/2}q\), \(u=\mathcal A^\dagger v\).
Expansion of \(\lambda\|v-\mathcal Au/\lambda\|^2\geq0\)
then gives
\begin{equation}\label{a:dualgap}
 \lambda\|q\|_{M_{\rm in}^{-1}}^2
       -\|L'q\|_{M_{\rm out}^{-1}}^2
 \geq\frac1\lambda\sum_\alpha
           |(L'q)_\alpha|^2\frac{\Delta_\alpha}{M_\alpha^2}.
\end{equation}
Here coefficients are identified by their Riesz representation;
\(L'q\) is the functional \(Z\mapsto q(LZ)\).
The estimates extend to the completed spaces by boundedness of
\(\mathcal A\) and \(\mathcal R\).
\end{proof}

\section{Gaussian channel simulation estimates}\label{app:channel-cost}
This appendix proves the environment comparison, resource-preparation
estimate, and symplectic extension used in Section~\ref{sec:channel-cost},
together with the probe reduction for the correlated-loss example in
Sec.~\ref{ex:channel}.

\subsection{Environment comparison}
The vacuum environment $\rho_{E,0}$ and the Gaussian state $\rho_{E,s}$ in
Eq.~\eqref{eq:channel-environment} pass through the same dilation unitary.
Contractivity and the overlap with the vacuum give
\begin{align}\label{eq:channel-uniform-bound}
 d_\diamond(\Phi_s,\Phi)
 &\leq d_{\mathrm{tr}}(\rho_{E,s},\rho_{E,0})\nonumber\\
 &\leq\sqrt{1-
 \det\!\left(\Id_{2k}+\tfrac{\epsilon_s}{2}L_BXX^{\mathsf T}L_B^{\mathsf T}\right)^{-1/2}}.
\end{align}
Since $L_B^{\mathsf T}L_B=Y^{-1}$, Sylvester's determinant identity reduces
the determinant to $\det(\Id_{2m}+\epsilon_sX^{\mathsf T}Y^{-1}X/2)$.
Using $\log\det(\Id+A)\leq\Tr A$, $1-e^{-u}\leq u$, and
$\epsilon_s\leq2/s$ proves Eq.~\eqref{eq:channel-explicit-error}.

\subsection{Uniform Gaussian dilution}
The following dilution bound depends on the number of modes and is
uniform in the resource energy.
\begin{lemma}[Uniform Schmidt information variance]
\label{lem:channel-variance}
For a pure Gaussian state $\varphi^{AB}$ with at most $m$ modes on Alice's
side, the reduced state satisfies
\begin{equation}\label{eq:channel-variance}
 \Tr\varphi_A\bigl[-\log\varphi_A-S(\varphi_A)\bigr]^2\leq m.
\end{equation}
Here $\log\varphi_A$ is restricted to the support of $\varphi_A$.
For any sequence of such states $\varphi_n$ and any $a_n>0$, $n$ copies
of $\varphi_n$ can consequently be approximated by LOCC from a maximally
entangled state of Schmidt rank
\begin{equation}\label{eq:channel-dilution-rank}
 M_n=\left\lceil\exp\{n[S((\varphi_n)_A)+a_n]\}\right\rceil
\end{equation}
with trace-distance error at most $\sqrt{m/(na_n^2)}$, whenever that bound
is less than one.
\end{lemma}
\begin{proof}
The Gaussian Schmidt decomposition expresses $\varphi_A$, up to a
unitary, as a product of at most $m$ thermal states. Their information
variances are bounded by one in Eq.~\eqref{eq:channel-thermal-variance};
vacuum factors contribute zero. Adding variances proves
Eq.~\eqref{eq:channel-variance}.

For each $n$, let $P_n$ project onto eigenvalues of
$(\varphi_n)_A^{\otimes n}$ at least
$\exp\{-n[S((\varphi_n)_A)+a_n]\}$. Chebyshev's inequality gives
\begin{equation}\label{eq:channel-typical-weight}
 \begin{gathered}
 p_n:=\Tr P_n(\varphi_n)_A^{\otimes n}
 \geq1-\frac{m}{na_n^2},\\
 \operatorname{rank}P_n\leq
 \exp\{n[S((\varphi_n)_A)+a_n]\}.
 \end{gathered}
\end{equation}
The normalized projection of $\varphi_n^{\otimes n}$ has Schmidt rank at
most $M_n$ and can be prepared by finite-dimensional teleportation. Its
trace distance from $\varphi_n^{\otimes n}$ is $\sqrt{1-p_n}$.
\end{proof}

\subsection{Exact simulation with finite-energy resources}
\label{app:channel-completion}
For $Y>0$, the full-row-rank matrices $M_{\rm tel}$ and $B$ in
Eq.~\eqref{eq:channel-symplectic-gram} have the same symplectic Gram
matrix. Pad them with zero columns to a common finite symplectic space.
The correspondence between their rows is an isometry of the restricted
alternating forms. By Witt's extension theorem for symplectic spaces,
it extends to a symplectic automorphism of the whole space. To construct the extension, choose
canonical pairs in each row space, complete the nullspace of its
restricted form by conjugate partners, and then complete to symplectic
bases. Mapping the two bases gives a symplectic matrix $\Xi$ such that
\begin{equation}\label{eq:channel-completion}
 \widetilde M_{\rm tel} \Xi=\widetilde B.
\end{equation}
The pure covariance $\Xi\Xi^{\mathsf T}$ has a marginal $V_{\rm res}$ on the
original $m+\ell$ resource modes satisfying
$M_{\rm tel}V_{\rm res}M_{\rm tel}^{\mathsf T}=BB^{\mathsf T}=Y$.

Writing $\chi_\rho(z)=\Tr\rho e^{iz^{\mathsf T}R}$ for the
characteristic function, teleportation with gain $X$ and the centered
Gaussian resource $\rho_{\rm res}$ of covariance $V_{\rm res}$ gives,
for an arbitrary input and every $z\in\mathbb R^{2\ell}$,
\begin{align*}
 \chi_{\rm out}(z)
 &=\chi_{\rm in}(X^{\mathsf T}z)\,
   \chi_{\rho_{\rm res}}(-Z_mX^{\mathsf T}z,z)\\
 &=\chi_{\rm in}(X^{\mathsf T}z)e^{-z^{\mathsf T}Yz/4}.
\end{align*}
The resulting map agrees with $\Phi$ on all input characteristic
functions, establishing exact simulation, including for inputs
entangled with a reference.

\subsection{Optimal probe conditioning for the correlated-loss channel}
\label{app:correlated-loss}
We establish the lower bound in Eq.~\eqref{ex:channel-cost-scalar}
for the channel of Sec.~\ref{ex:channel}. The pure resource constructed
there gives the matching upper bound, so it remains to evaluate the
large-squeezing limit of the reference--output probes in
Theorem~\ref{thm:channel-cost}. Pure covariances below a Gaussian probe covariance can be parameterized by rank-one Gaussian measurements on its purification~\cite{AdessoDatta2010,wolf}. The inverse conditioning relation makes this correspondence explicit. A pure covariance below the probe covariance coincides with it on its pure normal modes, by the vacuum-factorization argument after Eq.~\eqref{cap:seedbijection}, so only the mixed normal modes matter. In mixed Schmidt normal modes, write $D=\bigoplus_j\nu_j\Id_2$, $\nu_j>1$, and $C=\bigoplus_j\sqrt{\nu_j^2-1}Z$. A pure covariance $\gamma<D$ is obtained by measuring the purifying modes with covariance
\[
 M=C^{\mathsf T}(D-\gamma)^{-1}C-D.
\]
Substitution of $\gamma\Omega\gamma=\Omega$ shows that $M$ is positive and pure. Boundary seeds follow by limits.

In the present dilation of the reference--output probe, the environment covariance $V_{E,r}(s)=sL^2+X\gamma_rX^{\mathsf T}$ satisfies $V_{E,r}(s)+i\Omega_2\geq(s-1)L^2>0$ for $s>1$. Both symplectic eigenvalues of the environment covariance therefore exceed one, so these two modes provide a minimal Gaussian purification of the probe output, related to the purification used above by a symplectic transformation of the environment. The reference--environment cross block $-c_sZ_2L$ is invertible, so the conditional covariance determines the measurement covariance uniquely. The probe symmetries are $(-K)_R\oplus K_B$ and $\Theta_R\oplus\Theta_B$, with actions $K_E$ and $\Theta_E$ on the environment. Taking the two geometric midpoints imposes these symmetries on a minimizing pure probe seed, and uniqueness transfers them to $M$. A two-mode covariance invariant under $K$ and $\Theta$ has local blocks $a\Id_2$, $b\Id_2$ and correlation block $cZ$, and purity forces $a=b$ and $a^2-c^2=1$. Consequently $M=\gamma_t$ for some $t\in\mathbb R$. For boundary seeds, geometric interpolation toward the invariant seed obtained by measuring vacuum makes the covariance difference strictly positive on the mixed Schmidt modes. Applying the inverse relation and then taking the interpolation parameter to zero preserves the same reduction.

Gaussian conditioning now gives the reference marginal
\begin{align}\label{ex:conditional-reference}
 A_s(t)&=s\Id_4-(s^2-1)Z_2L
 [sL^2+X\gamma_rX^{\mathsf T}+\gamma_t]^{-1}LZ_2\nonumber\\
 &=Z_2f_s(W_r(t))Z_2,\qquad
 f_s(W)=(sW+\Id_4)(s\Id_4+W)^{-1}.
\end{align}
Thus $G(V_{\Phi_r}(s))=\inf_tS_G(f_s(W_r(t)))$. The remaining step is to identify its limit at large squeezing with the scalar minimum in Eq.~\eqref{ex:channel-cost-scalar}. For fixed $t$, $f_s(W_r(t))\to W_r(t)$ as $s\to\infty$. To pass the limit through the infimum, observe that
\[
 W_r(t)\geq L^{-1}X\gamma_rX^{\mathsf T}L^{-1}>0
\]
uniformly in $t$, while its largest eigenvalue diverges with $|t|$. For $s\geq1$ and $w>0$,
\[
 f_s(w)\geq\min\{1,w\},\qquad
 f_s(w)\geq\tfrac12\min\{s,w\}.
\]
Therefore $\det f_s(W_r(t))\to\infty$ along every joint limit $s\to\infty$, $|t|\to\infty$. Bounded-entropy minimizing sequences stay in a compact interval of $t$, on which convergence is uniform. The channel cost theorem then gives Eq.~\eqref{ex:channel-cost-scalar}.

\section{Dilations, ensembles, and constrained coding}\label{app:capacity}
We give the Gaussian dilation and ensemble details used in
Section~\ref{sec:capacity}, followed by the approximation that turns
continuous Gaussian modulation into codes with a finite signal alphabet.

\subsection{Rectangular dilations and continuous ensembles}
For equal input and output mode counts, every Gaussian channel has a
Gaussian unitary dilation with a finite vacuum environment, including
rank-deficient gains and noise~\cite[Corollary~1]{Caruso2008}.
If $\ell>m$, extend the input by $\ell-m$ modes discarded by the channel,
replacing $X$ by $(X\ \ 0)$. Prepare these modes in vacuum and count them
as part of the environment. If $\ell<m$, append $m-\ell$ vacuum outputs, replacing $X$ by
$\left(\begin{smallmatrix}X\\0\end{smallmatrix}\right)$ and $Y$ by
$Y\oplus\Id_{2(m-\ell)}$, and include them among the discarded modes. In both constructions the
extended complete-positivity condition contains the original condition
as a block. This gives Eq.~\eqref{cap:dilation}; an idle vacuum mode may
also be added to $F$.

For the ensemble correspondence in Eq.~\eqref{cap:msw}, let $Q$ project
onto the orthogonal complement of the range of $T$. Since
$\Tr(QT\rho T^\dagger)=0$ and every ensemble expectation of $Q$ is
nonnegative, almost every vector of a pure ensemble lies in the range
of $T$.

\subsection{Energy bounds and finite-alphabet coding}
A confining quadratic Hamiltonian bounds the total input photon
number. Since a Gaussian channel acts affinely on first and second
moments, its output photon number is bounded on every input
energy-constrained set, and the thermal maximum-entropy bound gives
\begin{equation}\label{cap:boundedentropy}
 \sup_{\Tr\rho H\leq E}S(\Phi(\rho))<\infty.
\end{equation}
The same holds for the pure dilation and every fixed number of uses.
This is the hypothesis of the constrained coding theorem, applied to the
nonnegative Hamiltonian $H-E_0\Id$~\cite[Proposition~3]{Holevo2004Constraints},
which yields
\begin{equation}\label{cap:regularization}
 C(\Phi,H,E)=\lim_{n\to\infty}\frac{\chi_n(E)}n.
\end{equation}
Its direct construction obeys the mean energy bound on each codeword;
the converse also holds for energy averaged over messages.

A Gaussian displacement ensemble with mean energy strictly below $E$
can be approximated by finite ensembles of pure Gaussian signals:
truncate the displacement tail and discretize the remaining compact
region within its support, which may be a proper subspace.
The averages converge in trace norm with
uniformly bounded output energy, so their entropies converge~\cite{winter}.
Every signal output has the same entropy, and sufficiently accurate
approximations respect the energy constraint. Coding with this finite
alphabet uses product displaced copies of the seed and collective
measurements. For $E>E_0$, continuity of the finite concave capacity
expression allows the energy of these ensembles to approach $E$, giving
the claimed rate.

\bibliography{references}

@article{wolf,
  author        = {M. M. Wolf and G. Giedke and O. Kr\"uger and R. F. Werner and J. I. Cirac},
  title         = {{Gaussian entanglement of formation}},
  journal       = {Phys. Rev. A},
  volume        = {69},
  pages         = {052320},
  year          = {2004},
  doi           = {10.1103/PhysRevA.69.052320}
}

@article{yamasaki,
  author        = {H. Yamasaki and K. Kuroiwa and P. Hayden and L. Lami},
  title         = {{Entanglement cost for infinite-dimensional physical systems}},
  journal       = {Commun. Math. Phys.},
  volume        = {406},
  pages         = {277},
  year          = {2025},
  doi           = {10.1007/s00220-025-05431-1}
}

@article{giedke,
  author        = {G. Giedke and M. M. Wolf and O. Kr\"uger and R. F. Werner and J. I. Cirac},
  title         = {{Entanglement of formation for symmetric Gaussian states}},
  journal       = {Phys. Rev. Lett.},
  volume        = {91},
  pages         = {107901},
  year          = {2003},
  doi           = {10.1103/PhysRevLett.91.107901}
}

@article{marian,
  author        = {P. Marian and T. A. Marian},
  title         = {{Entanglement of formation for an arbitrary two-mode Gaussian state}},
  journal       = {Phys. Rev. Lett.},
  volume        = {101},
  pages         = {220403},
  year          = {2008},
  doi           = {10.1103/PhysRevLett.101.220403}
}

@article{akbari,
  author        = {Y. Akbari-Kourbolagh and H. Alijanzadeh-Boura},
  title         = {{On the entanglement of formation of two-mode Gaussian states: a compact form}},
  journal       = {Quantum Inf. Process.},
  volume        = {14},
  pages         = {4179--4199},
  year          = {2015},
  doi           = {10.1007/s11128-015-1119-5}
}

@article{tserkis2017,
  author        = {S. Tserkis and T. C. Ralph},
  title         = {{Quantifying entanglement in two-mode Gaussian states}},
  journal       = {Phys. Rev. A},
  volume        = {96},
  pages         = {062338},
  year          = {2017},
  doi           = {10.1103/PhysRevA.96.062338}
}

@article{tserkis2019,
  author        = {S. Tserkis and S. Onoe and T. C. Ralph},
  title         = {{Quantifying entanglement of formation for two-mode Gaussian states: analytical expressions for upper and lower bounds and numerical estimation of its exact value}},
  journal       = {Phys. Rev. A},
  volume        = {99},
  pages         = {052337},
  year          = {2019},
  doi           = {10.1103/PhysRevA.99.052337}
}

@article{extremality,
  author        = {M. M. Wolf and G. Giedke and J. I. Cirac},
  title         = {{Extremality of Gaussian quantum states}},
  journal       = {Phys. Rev. Lett.},
  volume        = {96},
  pages         = {080502},
  year          = {2006},
  doi           = {10.1103/PhysRevLett.96.080502}
}

@article{wilde2018,
  author        = {M. M. Wilde},
  title         = {{Entanglement cost and quantum channel simulation}},
  journal       = {Phys. Rev. A},
  volume        = {98},
  pages         = {042338},
  year          = {2018},
  doi           = {10.1103/PhysRevA.98.042338}
}

@article{lami2017,
  author        = {L. Lami and C. Hirche and G. Adesso and A. Winter},
  title         = {{From log-determinant inequalities to Gaussian entanglement via recoverability theory}},
  journal       = {IEEE Trans. Inf. Theory},
  volume        = {63},
  pages         = {7553--7568},
  year          = {2017},
  doi           = {10.1109/TIT.2017.2737546}
}

@article{lami2018,
  author        = {L. Lami and A. Serafini and G. Adesso},
  title         = {{Gaussian entanglement revisited}},
  journal       = {New J. Phys.},
  volume        = {20},
  pages         = {023030},
  year          = {2018},
  doi           = {10.1088/1367-2630/aaa654}
}

@article{botero,
  author        = {A. Botero and B. Reznik},
  title         = {{Modewise entanglement of Gaussian states}},
  journal       = {Phys. Rev. A},
  volume        = {67},
  pages         = {052311},
  year          = {2003},
  doi           = {10.1103/PhysRevA.67.052311}
}

@article{winter,
  author        = {A. Winter},
  title         = {{Tight uniform continuity bounds for quantum entropies: conditional entropy, relative entropy distance and energy constraints}},
  journal       = {Commun. Math. Phys.},
  volume        = {347},
  pages         = {291--313},
  year          = {2016},
  doi           = {10.1007/s00220-016-2609-8}
}

@article{bertasutterwalter,
  author        = {M. Berta and D. Sutter and M. Walter},
  title         = {{Quantum Brascamp--Lieb dualities}},
  journal       = {Commun. Math. Phys.},
  volume        = {401},
  pages         = {1807--1830},
  year          = {2023},
  doi           = {10.1007/s00220-023-04678-w}
}

@article{lesniewskiruskai,
  author        = {A. Lesniewski and M. B. Ruskai},
  title         = {{Monotone Riemannian metrics and relative entropy on noncommutative probability spaces}},
  journal       = {J. Math. Phys.},
  volume        = {40},
  pages         = {5702--5724},
  year          = {1999},
  doi           = {10.1063/1.533053}
}

@article{nussbaum,
  author        = {M. Nussbaum and A. Szko{\l}a},
  title         = {{The Chernoff lower bound for symmetric quantum hypothesis testing}},
  journal       = {Ann. Statist.},
  volume        = {37},
  pages         = {1040--1057},
  year          = {2009},
  doi           = {10.1214/08-AOS593}
}

@article{wernerwolf2001,
  author        = {R. F. Werner and M. M. Wolf},
  title         = {{Bound entangled Gaussian states}},
  journal       = {Phys. Rev. Lett.},
  volume        = {86},
  pages         = {3658--3661},
  year          = {2001},
  doi           = {10.1103/PhysRevLett.86.3658}
}

@article{chen2023,
  author        = {X.-y. Chen and M. Miao and R. Yin and J. Yuan},
  title         = {{Bosonic Gaussian channel and Gaussian witness entanglement criterion of continuous variables}},
  journal       = {Phys. Rev. Research},
  volume        = {5},
  pages         = {033066},
  year          = {2023},
  doi           = {10.1103/PhysRevResearch.5.033066}
}

@article{shchukin2026,
  author        = {E. Shchukin and P. van Loock},
  title         = {{Revisiting Gaussian genuine entanglement witnesses with modern software}},
  journal       = {Phys. Rev. Research},
  volume        = {8},
  pages         = {023124},
  year          = {2026},
  doi           = {10.1103/drmr-qf2q}
}

@article{singh2025,
  author        = {G. Singh and S. Roy and {Arvind}},
  title         = {{Noisy dynamics of Gaussian entanglement: A transient bound entangled phase before separability}},
  journal       = {Phys. Rev. A},
  volume        = {112},
  pages         = {052424},
  year          = {2025},
  doi           = {10.1103/vpq9-691b}
}

@article{windt2021,
  author        = {B. Windt and A. Jahn and J. Eisert and L. Hackl},
  title         = {{Local optimization on pure Gaussian state manifolds}},
  journal       = {SciPost Phys.},
  volume        = {10},
  pages         = {066},
  year          = {2021},
  doi           = {10.21468/SciPostPhys.10.3.066}
}

@article{gaoklco2026,
  author        = {B. Gao and N. Klco},
  title         = {{Finite Gaussian assistance protocols and a conic metric for extremizing spacelike vacuum entanglement}},
  journal       = {Phys. Rev. A},
  volume        = {113},
  pages         = {012430},
  year          = {2026},
  doi           = {10.1103/cm73-wdxv}
}

@article{Lupo2009Correlated,
  author        = {C. Lupo and O. V. Pilyavets and S. Mancini},
  title         = {{Capacities of lossy bosonic channel with correlated noise}},
  journal       = {New J. Phys.},
  volume        = {11},
  pages         = {063023},
  year          = {2009},
  doi           = {10.1088/1367-2630/11/6/063023}
}

@article{AdessoDatta2010,
  author        = {G. Adesso and A. Datta},
  title         = {{Quantum versus classical correlations in Gaussian states}},
  journal       = {Phys. Rev. Lett.},
  volume        = {105},
  pages         = {030501},
  year          = {2010},
  doi           = {10.1103/PhysRevLett.105.030501}
}

@article{wilde_uniform2018,
  author        = {M. M. Wilde},
  title         = {{Strong and uniform convergence in the teleportation simulation of bosonic Gaussian channels}},
  journal       = {Phys. Rev. A},
  volume        = {97},
  pages         = {062305},
  year          = {2018},
  doi           = {10.1103/PhysRevA.97.062305}
}

@article{tserkis2018,
  author        = {S. Tserkis and J. Dias and T. C. Ralph},
  title         = {{Simulation of Gaussian channels via teleportation and error correction of Gaussian states}},
  journal       = {Phys. Rev. A},
  volume        = {98},
  pages         = {052335},
  year          = {2018},
  doi           = {10.1103/PhysRevA.98.052335}
}

@article{GHG2015,
  author        = {V. Giovannetti and A. S. Holevo and R. Garc\'ia-Patr\'on},
  title         = {{A solution of Gaussian optimizer conjecture for quantum channels}},
  journal       = {Commun. Math. Phys.},
  volume        = {334},
  pages         = {1553--1571},
  year          = {2015},
  doi           = {10.1007/s00220-014-2150-6}
}

@article{GGCH2014,
  author        = {V. Giovannetti and R. Garc\'ia-Patr\'on and N. J. Cerf and A. S. Holevo},
  title         = {{Ultimate classical communication rates of quantum optical channels}},
  journal       = {Nat. Photonics},
  volume        = {8},
  pages         = {796--800},
  year          = {2014},
  doi           = {10.1038/nphoton.2014.216}
}

@article{MSW2004,
  author        = {K. Matsumoto and T. Shimono and A. Winter},
  title         = {{Remarks on additivity of the Holevo channel capacity and of the entanglement of formation}},
  journal       = {Commun. Math. Phys.},
  volume        = {246},
  pages         = {427--442},
  year          = {2004},
  doi           = {10.1007/s00220-003-0919-0}
}

@article{Caruso2008,
  author        = {F. Caruso and J. Eisert and V. Giovannetti and A. S. Holevo},
  title         = {{Multi-mode bosonic Gaussian channels}},
  journal       = {New J. Phys.},
  volume        = {10},
  pages         = {083030},
  year          = {2008},
  doi           = {10.1088/1367-2630/10/8/083030}
}

@article{Holevo2004Constraints,
  author        = {A. S. Holevo},
  title         = {{Entanglement-assisted capacities of constrained quantum channels}},
  journal       = {Theory Probab. Appl.},
  volume        = {48},
  pages         = {243--255},
  year          = {2004},
  doi           = {10.1137/S0040585X97980415}
}

@article{HolevoShirokov2006,
  author        = {A. S. Holevo and M. E. Shirokov},
  title         = {{Continuous ensembles and the capacity of infinite-dimensional quantum channels}},
  journal       = {Theory Probab. Appl.},
  volume        = {50},
  pages         = {86--98},
  year          = {2006},
  doi           = {10.1137/S0040585X97981470}
}

@misc{adesso2026,
  author        = {G. Adesso},
  title         = {{Optimality of Gaussian entanglement of formation}},
  year          = {2026},
  eprint        = {2608.01909v3},
  archivePrefix = {arXiv},
  primaryClass  = {quant-ph},
  note = {{Phys. Rev. Lett. (accepted)}},
  doi= {10.1103/5jfn-ms77}
}

@misc{ALG2026,
  author        = {G. Adesso and L. Lami and V. Giovannetti},
  title         = {{Classical capacity of single-mode bosonic Gaussian channels}},
  year          = {2026},
  eprint        = {2610.22308v1},
  archivePrefix = {arXiv},
  primaryClass  = {quant-ph}
}

@misc{Ji2026,
  author        = {S.-W. Ji},
  title         = {{Gaussian optimality of energy-constrained one-shot communication through single-mode bosonic Gaussian channels}},
  year          = {2026},
  eprint        = {2608.17239v2},
  archivePrefix = {arXiv},
  primaryClass  = {quant-ph}
}

@misc{MishraWilde2026,
  author        = {H. K. Mishra and M. M. Wilde},
  title         = {{Gaussian entanglement and quantum communication: geodesic optimization and multimode bounds}},
  year          = {2026},
  eprint        = {2610.18827v1},
  archivePrefix = {arXiv},
  primaryClass  = {quant-ph}
}

@misc{gao_geometry,
  author        = {Y. Gao},
  title         = {{The geometry of Gaussian entanglement of formation}},
  year          = {2026},
  howpublished  = {\href{https://aixiv.online/abs/2610.01476v1}{aiXiv:2610.01476v1}}
}

@misc{gao_optimality,
  author        = {Y. Gao},
  title         = {{Gaussian optimality of multimode entanglement of formation}},
  year          = {2026},
  howpublished  = {\href{https://aixiv.online/abs/2610.01682v1}{aiXiv:2610.01682v1}}
}

@misc{interpolation,
  author        = {{OpenAI}},
  title         = {{The entropy photon-number inequality}},
  year          = {2026},
  howpublished  = {OpenAI Math Release preprint},
  doi           = {10.83295/jr3v-8jya}
}

@misc{He2026frontier,
  author        = {Z. He},
  title         = {{A universal energy--entanglement capacity frontier for Gaussian optical channels}},
  year          = {2026},
  howpublished  = {Zenodo},
  doi           = {10.5281/zenodo.23479869},
  note          = {Version 1, 19 October 2026}
}

@article{Berta2013,
  author        = {M. Berta and F. G. S. L. Brand{\~a}o and M. Christandl and S. Wehner},
  title         = {{Entanglement cost of quantum channels}},
  journal       = {IEEE Trans. Inf. Theory},
  volume        = {59},
  pages         = {6779--6795},
  year          = {2013},
  doi           = {10.1109/TIT.2013.2268533}
}

@article{BraunsteinKimble1998,
  author        = {S. L. Braunstein and H. J. Kimble},
  title         = {{Teleportation of continuous quantum variables}},
  journal       = {Phys. Rev. Lett.},
  volume        = {80},
  pages         = {869--872},
  year          = {1998},
  doi           = {10.1103/PhysRevLett.80.869}
}

@article{PLOB2017,
  author        = {S. Pirandola and R. Laurenza and C. Ottaviani and L. Banchi},
  title         = {{Fundamental limits of repeaterless quantum communications}},
  journal       = {Nat. Commun.},
  volume        = {8},
  pages         = {15043},
  year          = {2017},
  doi           = {10.1038/ncomms15043}
}

@article{WTB2017,
  author        = {M. M. Wilde and M. Tomamichel and M. Berta},
  title         = {{Converse bounds for private communication over quantum channels}},
  journal       = {IEEE Trans. Inf. Theory},
  volume        = {63},
  pages         = {1792--1817},
  year          = {2017},
  doi           = {10.1109/TIT.2017.2648825}
}

@article{pirandola_uniform2018,
  author        = {S. Pirandola and R. Laurenza and S. L. Braunstein},
  title         = {{Teleportation simulation of bosonic Gaussian channels: strong and uniform convergence}},
  journal       = {Eur. Phys. J. D},
  volume        = {72},
  pages         = {162},
  year          = {2018},
  doi           = {10.1140/epjd/e2018-90253-1}
}

@article{LiuzzoScorpo2017,
  author        = {P. Liuzzo-Scorpo and A. Mari and V. Giovannetti and G. Adesso},
  title         = {{Optimal continuous variable quantum teleportation with limited resources}},
  journal       = {Phys. Rev. Lett.},
  volume        = {119},
  pages         = {120503},
  year          = {2017},
  doi           = {10.1103/PhysRevLett.119.120503},
  note          = {Erratum: \href{https://doi.org/10.1103/PhysRevLett.120.029904}{Phys. Rev. Lett. \textbf{120}, 029904 (2018)}}
}

@article{Laurenza2018,
  author        = {R. Laurenza and S. L. Braunstein and S. Pirandola},
  title         = {{Finite-resource teleportation stretching for continuous-variable systems}},
  journal       = {Sci. Rep.},
  volume        = {8},
  pages         = {15267},
  year          = {2018},
  doi           = {10.1038/s41598-018-33332-y}
}

@article{Laurenza2019,
  author        = {R. Laurenza and S. Tserkis and L. Banchi and S. L. Braunstein and T. C. Ralph and S. Pirandola},
  title         = {{Tight bounds for private communication over bosonic Gaussian channels based on teleportation simulation with optimal finite resources}},
  journal       = {Phys. Rev. A},
  volume        = {100},
  pages         = {042301},
  year          = {2019},
  doi           = {10.1103/PhysRevA.100.042301}
}

@article{BraunsteinVanLoock2005,
  author        = {S. L. Braunstein and P. van Loock},
  title         = {{Quantum information with continuous variables}},
  journal       = {Rev. Mod. Phys.},
  volume        = {77},
  pages         = {513--577},
  year          = {2005},
  doi           = {10.1103/RevModPhys.77.513}
}

@article{Weedbrook2012,
  author        = {C. Weedbrook and S. Pirandola and R. Garc{\'i}a-Patr{\'o}n and N. J. Cerf and T. C. Ralph and J. H. Shapiro and S. Lloyd},
  title         = {{Gaussian quantum information}},
  journal       = {Rev. Mod. Phys.},
  volume        = {84},
  pages         = {621--669},
  year          = {2012},
  doi           = {10.1103/RevModPhys.84.621}
}

@article{AdessoRagyLee2014,
  author        = {G. Adesso and S. Ragy and A. R. Lee},
  title         = {{Continuous variable quantum information: Gaussian states and beyond}},
  journal       = {Open Syst. Inf. Dyn.},
  volume        = {21},
  pages         = {1440001},
  year          = {2014},
  doi           = {10.1142/S1230161214400010}
}

@article{Bennett1996,
  author        = {C. H. Bennett and D. P. DiVincenzo and J. A. Smolin and W. K. Wootters},
  title         = {{Mixed-state entanglement and quantum error correction}},
  journal       = {Phys. Rev. A},
  volume        = {54},
  pages         = {3824--3851},
  year          = {1996},
  doi           = {10.1103/PhysRevA.54.3824}
}

@article{Hayden2001,
  author        = {P. M. Hayden and M. Horodecki and B. M. Terhal},
  title         = {{The asymptotic entanglement cost of preparing a quantum state}},
  journal       = {J. Phys. A: Math. Gen.},
  volume        = {34},
  pages         = {6891--6898},
  year          = {2001},
  doi           = {10.1088/0305-4470/34/35/314}
}

@article{AdessoIlluminati2005,
  author        = {G. Adesso and F. Illuminati},
  title         = {{Gaussian measures of entanglement versus negativities: Ordering of two-mode Gaussian states}},
  journal       = {Phys. Rev. A},
  volume        = {72},
  pages         = {032334},
  year          = {2005},
  doi           = {10.1103/PhysRevA.72.032334}
}

@misc{IvanSimon2008,
  author        = {J. Solomon Ivan and R. Simon},
  title         = {{Entanglement of formation for Gaussian states}},
  year          = {2008},
  doi           = {10.48550/arXiv.0808.1658},
  eprint        = {0808.1658},
  archivePrefix = {arXiv},
  primaryClass  = {quant-ph}
}

@article{SerafiniAdessoIlluminati2005,
  author        = {A. Serafini and G. Adesso and F. Illuminati},
  title         = {{Unitarily localizable entanglement of Gaussian states}},
  journal       = {Phys. Rev. A},
  volume        = {71},
  pages         = {032349},
  year          = {2005},
  doi           = {10.1103/PhysRevA.71.032349}
}

@article{SimonMukundaDutta1994,
  author        = {R. Simon and N. Mukunda and B. Dutta},
  title         = {Quantum-noise matrix for multimode systems: {$U(n)$} invariance, squeezing, and normal forms},
  journal       = {Phys. Rev. A},
  volume        = {49},
  pages         = {1567--1583},
  year          = {1994},
  doi           = {10.1103/PhysRevA.49.1567}
}

@article{EisertSimonPlenio2002,
  author        = {J. Eisert and C. Simon and M. B. Plenio},
  title         = {On the quantification of entanglement in infinite-dimensional quantum systems},
  journal       = {J. Phys. A: Math. Gen.},
  volume        = {35},
  pages         = {3911--3923},
  year          = {2002},
  doi           = {10.1088/0305-4470/35/17/307}
}

@article{HolevoShirokovWerner2005,
  author        = {A. S. Holevo and M. E. Shirokov and R. F. Werner},
  title         = {On the notion of entanglement in Hilbert spaces},
  journal       = {Russian Math. Surveys},
  volume        = {60},
  number        = {2},
  pages         = {359--360},
  year          = {2005},
  doi           = {10.1070/RM2005v060n02ABEH000830}
}

@book{Serafini2017,
  author        = {Alessio Serafini},
  title         = {Quantum Continuous Variables: A Primer of Theoretical Methods},
  year          = {2017},
  publisher     = {CRC Press},
  address       = {Boca Raton},
  doi           = {10.1201/9781315118727}
}

@article{DiGuglielmo2011,
  author        = {J. DiGuglielmo and A. Samblowski and B. Hage and C. Pineda and J. Eisert and R. Schnabel},
  title         = {{Experimental Unconditional Preparation and Detection of a Continuous Bound Entangled State of Light}},
  journal       = {Phys. Rev. Lett.},
  volume        = {107},
  pages         = {240503},
  year          = {2011},
  doi           = {10.1103/PhysRevLett.107.240503}
}

@article{Horodecki1998Bound,
  author        = {M. Horodecki and P. Horodecki and R. Horodecki},
  title         = {Mixed-State Entanglement and Distillation: Is there a ``Bound'' Entanglement in Nature?},
  journal       = {Phys. Rev. Lett.},
  volume        = {80},
  pages         = {5239--5242},
  year          = {1998},
  doi           = {10.1103/PhysRevLett.80.5239}
}

@article{Horodecki2009Entanglement,
  author        = {R. Horodecki and P. Horodecki and M. Horodecki and K. Horodecki},
  title         = {Quantum entanglement},
  journal       = {Rev. Mod. Phys.},
  volume        = {81},
  pages         = {865--942},
  year          = {2009},
  doi           = {10.1103/RevModPhys.81.865}
}

@article{Horodecki2003Rank,
  author        = {P. Horodecki and J. A. Smolin and B. M. Terhal and A. V. Thapliyal},
  title         = {Rank two bipartite bound entangled states do not exist},
  journal       = {Theor. Comput. Sci.},
  volume        = {292},
  number        = {3},
  pages         = {589--596},
  year          = {2003},
  doi           = {10.1016/S0304-3975(01)00376-0}
}

@article{Shor2004,
  author        = {Shor, Peter W.},
  title         = {Equivalence of Additivity Questions in Quantum Information Theory},
  journal       = {Commun. Math. Phys.},
  volume        = {246},
  pages         = {453--472},
  year          = {2004},
  doi           = {10.1007/s00220-003-0981-7}
}

@article{Hastings2009,
  author        = {Hastings, M. B.},
  title         = {Superadditivity of communication capacity using entangled inputs},
  journal       = {Nat. Phys.},
  volume        = {5},
  pages         = {255--257},
  year          = {2009},
  doi           = {10.1038/nphys1224}
}

@misc{KruegerWerner2005,
  author        = {Kr{\"u}ger, O. and Werner, R. F.},
  title         = {Some Open Problems in Quantum Information Theory},
  year          = {2005},
  eprint        = {quant-ph/0504166},
  archivePrefix = {arXiv},
  primaryClass  = {quant-ph},
  note          = {Problem 29: Entanglement of formation for Gaussian states},
  url           = {https://oqp.iqoqi.oeaw.ac.at/entanglement-of-formation-for-gaussian-states/}
}

@article{Walschaers2021,
  author  = {Mattia Walschaers},
  title   = {{Non-Gaussian Quantum States and Where to Find Them}},
  journal = {PRX Quantum},
  volume  = {2},
  pages   = {030204},
  year    = {2021},
  doi     = {10.1103/PRXQuantum.2.030204}
}

@article{Gwak2026,
  author  = {Geunhee Gwak and Chan Roh and Young-Do Yoon
             and M. S. Kim and Young-Sik Ra},
  title   = {{Completely characterizing multimode second-order
              nonlinear optical quantum processes}},
  journal = {Nat. Photonics},
  volume  = {20},
  pages   = {156--162},
  year    = {2026},
  doi     = {10.1038/s41566-025-01787-x}
}

@article{TakeokaGuha2014,
  author  = {Masahiro Takeoka and Saikat Guha},
  title   = {{Capacity of optical communication in loss and noise
              with general quantum Gaussian receivers}},
  journal = {Phys. Rev. A},
  volume  = {89},
  pages   = {042309},
  year    = {2014},
  doi     = {10.1103/PhysRevA.89.042309}
}

@article{Cui2025GreenMachine,
  author  = {Chaohan Cui and Jack Postlewaite and Babak N. Saif
             and Linran Fan and Saikat Guha},
  title   = {{Superadditive communication with the green machine
              as a practical demonstration of nonlocality
              without entanglement}},
  journal = {Nat. Commun.},
  volume  = {16},
  pages   = {3760},
  year    = {2025},
  doi     = {10.1038/s41467-025-59107-4}
}

@misc{GaoAdessoGEoFToolkit,
  author       = {Gao, Yuan and Adesso, Gerardo},
  title        = {{Gaussian entanglement of formation: numerical toolkit}},
  year         = {2026},
  howpublished = {Zenodo},
  note         = {Computer software},
  doi          = {10.5281/zenodo.24284176},
  url          = {https://github.com/gerardo-adesso/gaussian-eof-numerical}
}
\end{document}